\documentclass[12pt,fleqn]{article}
\usepackage{float}
\usepackage{graphicx,graphics}
\usepackage{longtable}
\usepackage{titling}
\usepackage[margin=2cm,top=1.8cm,bottom=2.5cm,centering]{geometry}

\usepackage[utf8]{inputenc} 
\usepackage[T1]{fontenc} 
\usepackage[british]{babel} 
\usepackage{subfigure}
\usepackage[final]{pdfpages} 
\usepackage{pstricks}

\usepackage[colorlinks=true,urlcolor=black,linkcolor=black]{hyperref} 

\usepackage{lmodern} 

\usepackage{amsmath} 
\usepackage{esint} 
\usepackage{amssymb} 
\usepackage{esvect} 
\usepackage{tensor}
\usepackage{mathrsfs} 

\usepackage{array}
\usepackage{cellspace} 
\usepackage{makecell}
\usepackage{multirow}

\usepackage{tikz}
\usetikzlibrary{patterns}
\usetikzlibrary{calc,arrows,positioning}
\usetikzlibrary{arrows,shadows}
\usetikzlibrary{decorations.pathmorphing}	
\usetikzlibrary{decorations.markings}

\usepackage{textcomp}
\usepackage{graphicx}
\usepackage{url}
\usepackage{appendix}
\usepackage{subfigure}
\usepackage{pgfplots}
\pgfplotsset{compat=newest}
\usepackage{verbatim} 

\usepgfplotslibrary{patchplots}
\pgfplotsset{compat=1.3}
\numberwithin{equation}{section}

\usepackage{epstopdf}
\usepackage{listings}
\usepackage{lipsum}
\usepackage{colortbl}
\usepackage{qtree}
\usepackage{color}
\usepackage{gensymb}
\usepackage{enumerate}
\usepackage{numprint}
\usepackage{caption}
\usepackage[nottoc, notlof, notlot]{tocbibind} 
\usepackage{rotating} 

\newcommand{\R}{\mathbb{R}}
\newcommand{\C}{\mathbb{C}}
\newcommand{\Z}{\mathbb{Z}}

\usepackage{hhline}

\usepackage[final]{pdfpages}
\usepackage{titlesec}
\usepackage{stmaryrd}
\usepackage{delarray}

\renewcommand{\ge}{\geqslant}
\renewcommand{\le}{\leqslant}
\def\be{\begin{equation}}
\def\ee{\end{equation}}

\newcommand\bea{\begin{eqnarray}}
\newcommand\eea{\end{eqnarray}}
\newcommand\egal{&\!\!\!=\!\!\!&}

\def\ball#1{
\pscircle[linewidth=0.7pt,linecolor=blue,](0,0){1}
\rput(0,0){\blue $#1$}
\psline[linewidth=0.7pt,linecolor=blue,](1,0)(1.5,0)
\psline[linewidth=0.7pt,linecolor=blue,](-1,0)(-1.5,0)
\psline[linewidth=0.7pt,linecolor=blue,](0,1,0)(0,1.5)
\psline[linewidth=0.7pt,linecolor=blue,](0,-1,0)(0,-1.5)
}

\def\Z{{\mathbb Z}}
\def\R{{\mathbb R}}
\def\C{{\mathbb C}}
\def\P{{\mathbb P}}
\def\E{{\mathbb E}}
\def\H{{\mathbb H}}
\def\be{\begin{equation}}
\def\ee{\end{equation}}
\def\eps{\varepsilon}

\renewcommand{\ge}{\geqslant}
\renewcommand{\le}{\leqslant}

\begin{document}     

\title{\bf Sandpile models in the large} 

\author{Philippe Ruelle}
\date{}
\maketitle

\vspace{-1.5cm}
\centerline{\em {}Institut de Recherche en Math\'ematique et Physique}
\centerline{Universit\'e catholique de Louvain, Louvain-la-Neuve, B-1348, Belgium}

\vspace{1cm}
\centerline{Invited contribution to the Research Topic}
\centerline{\em Self-Organized Criticality, Three Decades Later}
\centerline{(Frontiers in Physics)}

\vspace{1cm} 

\begin{abstract}

This contribution is a review of the deep and powerful connection between the large scale properties of critical systems and their description in terms of a field theory. Although largely applicable to many other models, the details of this connection are illustrated in the class of two-dimensional Abelian sandpile models. Bulk and boundary height variables, spanning tree related observables, boundary conditions and dissipation are all discussed in this context and found to have a proper match in the field theoretic description.

\end{abstract}

\tableofcontents

\section{Introduction}

In statistical mechanics, critical points are very special points in the space of external parameters which control the state of a system. At such a point, the system is scale invariant and its thermodynamic functions and correlations are characterized by critical exponents and power laws. In many cases, physical systems have a finite number of critical points, most often only one. Typical examples include the end-point of the liquid-gas coexistence line or the Curie point for ferromagnetic materials. In these cases, a system is brought to its critical point by tuning very precisely a few external parameters to their critical values.

In Nature however, power laws are commonplace, and can be found in a large variety of different phenomena, like avalanches, earthquakes, solar flares, dropplet formation, ... In all these cases, it is certainly not clear what parameters should be tuned, and even if they are perfectly tuned, it is unlikely that they would stay so over large periods of time. To solve this apparent paradox, Bak, Tang and Wiesenfeld suggested in the 80's that the external parameters would tune themselves dynamically: even if the system is not initially in a critical state, its own dynamics will ineluctably drive it to criticality and maintain it in that state \cite{BTW87}. This attractive idea has led to the concept of self-organized criticality (SOC). 

To support this idea, these authors proposed the sandpile model as a prototypical example of a system which shows a form of self-organized criticality. Since then, many others models showing SOC have been proposed, as abundantly illustrated in this volume, and in introductory books and reviews \cite{Ba96, Je98, Pr12, Dh06}.

The present review will be exclusively concerned with specific versions of two-dimensional sandpile models, formulated by Dhar \cite{Dh90} and called Abelian sandpile models. Even though there are among the simplest and easiest sandpile models to handle, they show a large spectrum of interesting and difficult problems which have attracted considerable attention, in both the physical and mathematical communities. From the point of view taken here (namely their scaling limit and the emerging conformal field theory), they are to our knowledge the only ones to have been studied. Yet, compared to many other equilibrium statistical models, a fair statement is that our present understanding of them is still very poor.

Our primary purpose is two-fold, namely to give the unfamiliar reader an introduction of why and how the neighbourhood of a critical point can be described by a Euclidean field theory, which, at first sight, appears to be a rather obscure statement, and also to show how this description can be worked out in practical terms. The second part will be illustrated in sandpile models, which lend themsleves very well to this kind of analysis: they are simple enough that one can follow the steps in a clear and transparent way, yet they are rich enough to show the difficulties one sometimes has to face, but also the elegance and the power of the approach. Understanding how of a field theory emerges from a stochastic lattice model enables to gain a probabilistic and intuitive view of what a field theory is in this context. 

It turns out that the field theories which appear when analyzing critical systems are conformal field theories. The simple reason for this is that their large conformal symmetry integrates the fact that critical systems have a local scale invariance. Conformal theories in two dimensions have been tremendously successful since the 80's and have led to a deep understanding of the two-dimensional critical phenomena. It is certainly not our purpose to give an introduction to conformal field theories, and we will not go very deep into its technicalities, refering to the vast literature. We restrict to their most basic features, in the hope that these will be sufficient and useful to understand how conformal theories are so well suited for our study.
   
Section \ref{asm} starts with a brief review of the Abelian sandpile models, where the most basic features of the models are recalled. Section \ref{sec3} is a general description, valid beyond the sandpile models, of what is called the scaling limit, which allows to establish the connection between the large distance regime of a critical system and the associated field theory. A brief tour of conformal theories, and specifically logarithmic conformal theories, is presented in Section \ref{sec4}. The application of the conceptual ingredients is illustrated in the next three sections. Section \ref{sec5} focuses on the bulk observables in the sandpile models, computes the first correlators and explains how these should be understood in terms field theoretic quantities. Boundary conditions and boundary observables are examined in Section \ref{sec6} as well as the way they should be thought of in conformal theories. Section \ref{sec7} discusses a dissipative variant of the sandpile models and their description by a massive field theory, and also some universality aspects of the sandpile models. The last section summarizes the present status of the conformal theory at work in sandpile models. 

The present text has some overlap with \cite{Ru13}. The latter was more concerned with the sandpile models as being described specifically by a logarithmic conformal field theory. Intended to a potentially wider readership, the present review is more devoted to the general connection between critical systems and field theories, illustrated in a specific class of models. The two are somehow complementary, and, if combined, may provide a more complete overview.

\section{Abelian sandpile models}
\label{asm}

The models we will discuss are discrete stochastic dynamical systems. Their microscopic variables are attached to the vertices of a finite connected graph $\Gamma = (V,E)$ (with $V$ the set of vertices, or sites, and $E$ the set of simple, unoriented edges), and evolve in discrete time as a random process. We label the vertices of $\Gamma$ by latin indices $i,j,\ldots$ and denote the microscopic variables by $h_i$. These are called height variables and simply give the height of the sandpile at vertex $i$ (i.e. count the number of sand grains at $i$); they are integer-valued, with $h_i \ge 1$. A height configuration $C$ is a set of heights values $\{h_i\}_{i \in V}$.

We are not quite ready to define the dynamics. For reasons that will become clear in a moment, we need to extend $\Gamma$ by adding one special vertex, noted $s$ and called the sink, as well as a number of edges connecting $s$ to some vertices in a non-empty subset $D \subset V$. Vertices in $D$ are called dissipative or open, while those in $V \setminus D$ are conservative or closed. If $\Gamma^\star = (V^\star,E^\star)$ denotes the extended graph in an obvious notation, we define $z_i$ to be the coordination number of $i$ in $\Gamma$ (the number of edges in $E$ incident to $i$, or the number of its nearest neighbours in $\Gamma$), and similarly $z_i^\star$ its coordination number in $\Gamma^\star$. Thus $z_i^\star = z_i$ if $i$ is closed, $z_i^\star > z_i$ if $i$ is open. Finally we say that a site $i$ of $V$ is stable \footnote{There is no need to keep track of the number of sand grains in the sink, and so we do not assign it a height variable.} if its height satisfies $1 \le h_i \le z_i^\star$. A height configuration is stable if all sites are stable. Clearly the number of stable configurations is equal to $\prod_{i \in V} z_i^\star$.

The discrete, stochastic dynamics of the sandpile model is defined as follows. Assume that $C_t = \{h_i\}$ is a stable configuration at time $t$. The stable configuration $C_{t+1}$ is obtained from the following two steps.

\begin{enumerate}
\item[(i)] {\it Deposition}: one grain of sand is dropped on a random site $j$ of $V$, selected with probability $p_j$, producing therefore a new configuration $C^{\rm new}$ with heights $h_i^{\rm new} = h_i + \delta_{i,j}$. If $h_j^{} \le z_j^\star$, then $C^{\rm new}$ is stable and defines $C_{t+1}$; if not, we proceed to step (ii).
\item[(ii)] {\it Relaxation}: if $h^{\rm new}_j > z_j^\star$ (it is in fact equal to $z^\star_j+1$), we let the site $j$ topple: its height is decreased by $z_j^\star$, each of its neighbours in $\Gamma$ receives one grain, and the remaining $z_j^\star-z_j^{}$ grains go to the sink. After this, one or more neighbours of $j$ in $\Gamma$ may become unstable, in which case they topple in the way explained above for the site $j$. The toppling process is pursued for all unstable sites until a stable configuration is obtained. That configuration defines $C_{t+1}$.
\end{enumerate}

It is useful to introduce the toppling matrix $\Delta$ as it will play an important role in what follows,
\be
\Delta_{i,j} = \begin{cases}
z_i^\star & {\rm for\ } i=j,\\
-1 & {\rm if\ } i {\rm \ and\ } j {\rm \ are\ neighbours\ (i.e. connected)},\\
0 & {\rm otherwise},
\end{cases}
\label{topp}
\ee
for $i,j \in V$. The sand redistribution occurring when a site $j$ topples can then be written as the update $h_i \to h_i - \Delta_{j,i}$ for all $i \in V$. The matrix $\Delta$ is like a Laplacian on $\Gamma$, with mixed boundary conditions dictated by the open and closed sites, which induce respectively Dirichlet and Neumann boundary conditions (see Section \ref{sec6}).

The above dynamics is well-defined. We see that the total number of sand grains is conserved under the toppling of a closed site whereas a non-zero number of grains are transferred to the sink under the toppling of an open site. The existence of at least one open site guarantees that the relaxation process terminates after a finite number of topplings and motivated the necessity of the extension of the graph $\Gamma$ by the sink site. Moreover if several sites are unstable during the relaxation process, the order in which they are toppled does not matter. More generally, one may define the operator $a_i$ for each $i \in V$, whose action on a stable configuration returns the stable configuration resulting from the relaxation process after the deposition of a sand grain at $i$. One can then prove that the operators $a_i$ commute \cite{Dh90}, explaining the qualifier `Abelian' used to designate the models satisfying this property.

The dynamics described above is a discrete Markov chain on a finite configuration space: at each time step, one applies the operator $a_i$ with probability $p_i$ (it is the only stochastic element of the dynamics), going from $C_t$ to $C_{t+1}=a_iC_t$. 
An important question concerns the invariant measures, since they control the behaviour of the model in the long run. 

If there is no strong reason to favour certains sites, one takes all probabilities $p_i$ equal (uniform distribution). In this case \footnote{The result holds in the more general case where $p_i \neq 0$ for every $i$.}, Dhar \cite{Dh90} has shown that there is unique invariant measure $\P_\Gamma$, which is uniform on its support. In the Markov chain terminology, the configurations in the support of $\P_\Gamma$ are called recurrent; the others are called transient. Being in the support of the unique invariant measure means that the recurrent configurations are those which are in the repeated image of the operators $a_i$. The transient ones either never appear (depending on the initial configuration) or cease to appear after some finite time. 

If indeed the unique invariant measure is uniform, the situation appears to be deceptively simple. Not so. What makes the sandpile models non-trivial, fascinating and rich is the support of the invariant measure. A generic recurrent configuration is really complicated because the height values are delicately correlated over the entire graph. In the general case, there is no simpler criterion characterizing the recurrent configurations than the following. Let $C$ be a stable configuration and let $C_F$ be its restriction to a subgraph $F \subset \Gamma$ ($F$ can be assumed to be connected). We say that $C_F$ is a forbidden subconfiguration if each vertex of $F$ has a height  smaller or equal to the number of its neighbours in $F$. It can be shown \cite{MD92} that a forbidden subconfiguration cannot be in the repeated image of the dynamics (of the operators $a_i$). It follows that a configuration is recurrent if and only if it contains no forbidden subconfiguration. The simplest example of a forbidden subconfiguration is when $F$ contains just two neighbouring vertices with height values equal 1. The criterion also implies that the maximal configuration with heights $h^{}_i = z_i^\star$ is recurrent since a vertex $i$ with height $h_i > z_i$ cannot be in a forbidden subconfiguration. It is also clearly in the image of the iterated dynamics since it can be reached from any other stable configuration by an appropriate sequence of $a_i$'s.

The characterizing condition for recurrence shows that the heights of a recurrent configuration are not at all independent. They are not only correlated locally (think of two neighbouring 1's) but also globally because asserting that a given configuration is recurrent generally requires to scan the entire graph. For instance the configuration having $h_i = z_i$ for all $i$ is not recurrent and possesses no other forbidden subconfiguration than the whole configuration itself. Moreover the recurrent status is very sensitive to local changes and can be lost or gained by the change of a single height (for the configuration just discussed, the increase by one unit of the height at a single open site makes it recurrent). However the increase of any height in a recurrent configuration preserves the recurrence.

The burning algorithm \cite{MD92} (see also the review \cite{Dh06}) provides a convenient way to test whether a given stable configuration is recurrent. In addition to provide a completely automatic procedure, more importantly it establishes a bijection between the set of recurrent configurations on $\Gamma$ and the set of rooted spanning trees on $\Gamma^\star$, rooted at the sink site $s$. Let us recall that a spanning tree is a loopless connected subgraph $(V^\star,F) \subset \Gamma^\star = (V^\star,E^\star)$ with $F \subset E^\star$. This bijection is important and useful as most of the actual calculations use the spanning tree formulation. Interestingly, there is no canonical bijection between the two sets in the sense that there are in fact many burning algorithms (the detailed definition requires a certain prescription that is largely arbitrary), each giving rise to a different bijection. This freedom in the choice of a definite algorithm, a sort of huge gauge symmetry, has remained unexploited so far.

If the notion of recurrence remains somewhat elusive in the generic case, simple arguments lead to a remarkably simple and general formula for the number of recurrent configurations \cite{Dh90}, naturally identified as the partition function $Z$ since the invariant measure is uniform,
\be
Z = \#\{{\rm recurrent\ configs}\} = \det \Delta,
\ee
for $\Delta$ the toppling matrix introduced in (\ref{topp}). It is a standard result in combinatorics (Kirchhoff's matrix-tree theorem) that $\det \Delta$ also counts the number of spanning trees on $\Gamma$ (see Section \ref{sec5.7} for a proof). The previous formula usually implies that the recurrent configurations form an exponentially small fraction of the set of stable configurations (whose number is equal to $\prod_i \Delta_{i,i}$). On a large grid in $\Z^2$ for instance, for which the density of dissipative sites goes to 0 in the infinite volume limit, the effective number of degrees of freedom per site in a recurrent configuration is roughly 3.21 (as compared to 4 in a stable configuration), meaning that $\det \Delta \simeq {\rm e}^{\frac{4G}\pi N} = (3.21...)^N$, with $N$ the total number of sites and G the Catalan constant. 

The definition of recurrence implies that all the operators $a_i$ map recurrent configurations to recurrent configurations, implying that once the dynamics has brought the sandpile into a recurrent configuration, all subsequent configurations are recurrent. Therefore the invariant measure is appropriate to study the long term behaviour of the sandpile. 

The sandpile models summarized above have raised a large number of interesting and difficult questions. In the context of this review, most if not all of them focus on the stationary regime, and study the statistical behaviour of the sandpile when it runs over the recurrent configurations. In other words, all the probabilities we are interested in are induced by the invariant measure $\P_\Gamma$. The use of $\P_\Gamma$ is what makes most of the calculations fairly hard \footnote{A notable exception concerns the linear or almost linear graphs, for which the recurrence property usually takes a simpler form and allows for a larger number of explicit results, see for instance \cite{RS92,AD95}).} because as noted earlier, that measure is non-local in terms of the (local) height variables (equivalently the recurrence criterion is non-local).  

We should remark that the measure $\P_\Gamma$ fully depends on all the minute details which are necessary in order to completely specify the sandpile model under study. Not only the graph $\Gamma$ itself, but also the number and relative positions of closed and open vertices, and the values of the local thresholds $z_i^\star$ affect the invariant measure. Many features which directly depend on these data will change if any of these parameters is modified, like the number of recurrent configurations, the structure of the sandpile group \footnote{We have mentionned that the operators $a_i$ generate an Abelian algebra. But when acting on recurrent configurations, they are invertible and therefore generate an Abelian group, called the sandpile group. The sandpile group, of order equal to $\det \Delta$, has been determined for a number of finite graphs.}, the geometric structure of the identity configuration \footnote{Recurrent configurations form an Abelian group under the sitewise addition of the heights, followed by relaxation. This group is isomorphic to the sandpile group. In particular, one of the recurrent configurations is the identity in the group, and shows remarkable geometric patterns \cite{Cr91,DRSV95,LBR02,CPS08}.}, or the average height at a given site for instance. All these features are mathematically interesting and challenging (hence interesting) but very sensitive to the underlying details. 

One should however expect that more robust features would be shared by sandpile models that are `close enough'. The same situation prevails for  other statistical models which, although having different microscopic descriptions, are considered to be essentially equivalent and grouped together to form a single universality class. Models belonging to the same universality class have identical behaviours `in the large', a point of view made technically more precise by the renormalization group analysis. 

In order to identify these common behaviours, one should not look at small scales, as these are more likely to be determined by the local details. The probability that two vertices next to each other have a height 2 for instance is not really interesting; in addition it is a pure number, different for each different model. Robust behaviours are expected to be found at large scales, as they are much less affected by the microscopic details. One convenient method to access the large distance behaviours is by taking the scaling limit. Readers familiar with the scaling limit and the ideas of the renormalization group can safely go straight to the next sections. 

\section{The scaling limit and continuum field theories}
\label{sec3}

The simple idea underlying the scaling limit is this: if we want to concentrate on the large scale behaviour of a system, let us look at it from far away ! The further away we look at the system, the larger our horizon is and the larger the distances we keep in sight. At the same time, when looking from a distance, the details get blurred and disappear: one can no longer recognize the type of graph and its connectivities are no longer visible. What we see seems to become independent of the microscopic details of the model. 

Rather than stepping back, an equivalent but more convenient way to proceed is to shrink the discrete structure (graph or grid or lattice) on which the microscopic variables live. This will involve a (real) small parameter $\varepsilon$ such that the graph can be embedded in $\varepsilon \Z^d$ (or another shrinked regular lattice). For smaller and smaller $\varepsilon$, fixing a macroscopic distance $\vec x = \varepsilon \vec m \in \eps \Z^d$ amounts to probe larger and larger scales $\vec m$ in terms of lattice units, and at the same time, allows to keep a macroscopic distance $r=|\vec x|$ under control. The scaling limit corresponds \footnote{For the scaling to be non-trivial, some external parameters may need to be appropriately scaled with $\eps$. One example of this is discussed in Section \ref{7.1}.} to take $\eps \to 0$.

We note that since the scaling limit is a way to focus on asymptotically large distances, we have to make sure that the system does have such asymptotic  distances ! Indeed the scaling limit requires that we also take the infinite volume limit, by allowing the system to remain finite but of increasing size, the growth being at least of order $1/\eps$. 

The scaling limit has interesting consequences. The first most apparent one is that the substrate of the rescaled model goes to a continuum, either $\R^d$ or a part of $\R^d$, which may be bounded \footnote{It is bounded if all the linear sizes of the finite systems in the sequence defining the infinite volume limit grow exactly like $1/\eps$.}. This is the first sign that a continuum description ought to emerge in the scaling limit. This is confirmed by a second observation: the microscopic variables --the heights in the sandpile models--, which were attached to the vertices of a graph, or a grid, should in some sense converge to variables defined on a continuum. If indeed this is expected to happen, exactly what happens is quite subtle. To realize this, one may note that all the microscopic variables attached to sites contained in a ball of radius $o(1/\eps)$ will actually collapse to the same point in the scaling limit. Thus every point in the continuum is the convergence point of an infinite number of vertices in the originial discrete setting. The infinity of microscopic variables carried by these vertices will supposedly mix and fuse to generate some kind of degree of freedom located {\it at a single point in the continuum}. What is then the nature of the emerging continuum degree of freedom at that point, and how is it related to the lattice variables supposed to collectively generate it ? The conceptual answer is provided by the renormalization group. It roughly goes as follows \footnote{Among the many books and reviews on the renormalization group in statistical mechanics, see for instance the book by Cardy \cite{Ca96}.}.

The scaling limit as explained at the beginning of this section was carried out in one stroke: all distances are scaled by $\eps$, which is then taken to 0. This limit was only designed to show how the large distance behaviours can be assessed, but is too rough to answer the question raised in the previous paragraph. The renormalization group is much better designed conceptually as it organizes the scaling limit scale by scale and keeps track, at each scale, of the degrees of freedom present in the system.

Let us suppose that we start with a statistical model defined on very large graph, or, to simplify and fix the ideas, on an infinite lattice. We fix a convenient scale $\Lambda>1$, partition the lattice into boxes of size $\Lambda$ and shrink the lattice by a factor $\Lambda$. Each box is now of linear size 1 and contains of the order of $\Lambda^d$ microscopic variables. Within each box, we associate an effective, coarse-grained degree of freedom which takes into account the overall behaviour of the microscopic variables inside the box (it could be f.i. their average value), and we then compute the sum over the microscopic variables conditioned by the values of the coarse-grained variables. The result is a statistical model for the coarse-grained variables, defined on a lattice similar to the original one. 
Once this is done (!), we iterate the process by defining a second generation of coarse-grained variables out of those of the first generation, and so on.

After the first iteration, each group of roughly $\Lambda^d$ microscopic variables has collapsed to a single coarse-grained variable of first generation; the statistical model obtained for these can be interpreted as the original model in which the fluctuations of scale smaller than $\Lambda$ have been integrated out. The second iteration yields a statistical model for the coarse-grained variables of second generation, each of which has integrated the fluctuations of $\Lambda^{2d}$ microscopic variables over scales smaller than $\Lambda^2$, and so on for the next iterations. In this way each iteration, also called renormalization, yields a model where more small scale fluctuations have been integrated out, and whose large scale behaviour should be identical to that of the original model, since the large scale fluctuations have been preserved.

The continuum degrees of freedom we were asking about are what the coarse-grained variables of higher and higher generation should converge to when the number of iterations goes to infinity. Each of them is indeed what is left of the infinite collection of the microscopic variables that were located around it. Because the coarse-grained variables of one generation are representative of those of the previous generation, the continuum degrees of freedom should similarly carry the same characteristics as the original microscopic variables. In particular the long distance correlations should be identical, at dominant order.

The continuum degrees of freedom emerging in the scaling limit are called {\it fields}. Unlike their lattice ancestors, they usually take continuous values. Fields are all what remains when the short-ranged degrees of freedom have been integrated out: they form the complete set of variables which are relevant as far as the long distance properties of the original model are concerned. It means that only the lattice degrees of freedom which have long range correlations, namely with diverging correlation lengths, will survive the scaling limit and eventually give rise to a field; all the others progressively disappear in the renormalization process.

The microscopic variables in terms of which the discrete statistical model is defined usually give rise to fields, but they are not the only ones. Any lattice observable, that is, any function of the microscopic variables, can potentially give rise to a field in the scaling limit \footnote{F.i. the energy density in the Ising model, namely the product of two neighbouring spins, gives rise to a field that is different from the one obtained from the spin variable itself. Later we will give examples of this in the sandpile models (cluster variables).}, so that one is typically left with an infinite number of different fields. Each field has its own specific properties and should be interpreted as the scaling limit of one particular lattice observable (it may also happen that different lattice observables converge to fields with the same characteristics). 

One last question must be addressed. The original statistical model was not only defined by its microscopic variables, but also by a probability measure on the configuration space. That measure, which is a joint distribution for the (non-independent) random microscopic variables, is usually given by a Gibbs measure, and written, up to normalization, as $\P(C) \sim \exp{\big(\!-\!H[C]\big)}$, where $H$ is the Hamiltonian of the system, {i.e.} some given function of the microscopic variables which determines the relative probability of a configuration $C$. What is the equivalent of the Gibbs measure for the fields ?

According to the discussion above, one starts from the original model and its Hamiltonian $H_0 \equiv H$. The first renormalization yields the coarse-grained variables of the first generation and a corresponding Hamiltonian $H_1$, computed (at least in principle) by summing $\exp{(-H_0)}$ over the microscopic variables inside the boxes. Similarly the $k$-th iteration will produce a Hamiltonian $H_k$ defining the statistical model for the coarse-grained variables of the $k$-th generation. The appropriate measure for the fields should therefore be something like the formal limit $\lim_{k \to \infty} \exp{(-H_k)}$. Physicists like to denote this formal object by $\exp{(-S)}$ where $S$, called the action, is a certain functional of the fields. 

Thus if the description of a statistical model is given, in the discrete lattice setting, in terms of a set of microscopic variables $(h^{1}_i,\, h^{2}_i, \ldots)$ and a Hamiltonian $H(h^{1}_i,\,h^{2}_i,\ldots)$, it is given in the scaling limit by a set of continuous fields $(\phi_1(\vec x),\phi_2(\vec x),\dots)$ and an action $S[\phi_1,\phi_2,\ldots]$. The pair $\{(\phi_1(\vec x),\phi_2(\vec x),\dots),S\}$ is refered to as a {\it continuum field theory} \footnote{One should add `Euclidean' field theory because it is formulated on a Euclidean space $\R^d$.}. More precisely, specifying a set of fields and their action $S$ is only {\it one} way to present a field theory; it is also the most comfortable one because it allows to compute the correlators of the various fields, at least in principle. 

Needless to say, working out the successive renormalizations along with the Hamiltonians $H_0,H_1,\ldots$ is a formidable task that is, for all practical purposes, impossible to carry out explicitely, except on extremely rare occasions (and for tailored examples). As a consequence the field theory describing the large distances of a statistical model cannot be obtained in a deductive way. 

The situation however is not hopeless. Experience, heuristic arguments or  results obtained on the lattice can often give definite hints about the nature of the seeked field theory. More importantly, and even if one has no clue of what the correct field theory is, the relevance of a trial field theory, perhaps suggested by an educated guess, can be firmly tested by comparing correlations functions. If the lattice microscopic variable $h_{i=\frac x\eps}$ (at site $i$) converge in the scaling limit to the field $\phi(x)$, it must be true that the scaling limit of the lattice correlators are equal to field theoretic correlators, namely
\be
\lim_{\eps \to 0} \: \eps^{-n\Delta} \: \langle h_{\frac {x_1}\eps} \, h_{\frac{x_2}\eps} \ldots \, h_{\frac{x_n}\eps} \rangle_{\rm lattice} = \langle \phi(x_1) \, \phi(x_2) \ldots \, \phi(x_n)\rangle_{\rm FT},
\label{match}
\ee
where the exponent $\Delta$ is determined so that the limits on the l.h.s. exist: as shown below, it will eventually be related to the scale dimension of the field $\phi$ to which the lattice variable $h_i$ converges. The previous identity must be satisfied for all $n$-point correlators, but also for any correlator of any number of lattice observables provided that for each observable $O(i)$ around site $i$ inserted in the lattice correlator, the corresponding field $\Phi(x)$ to which it converges is inserted in the field theoretic correlator,
\be
\lim_{\eps \to 0} \: \eps^{-\sum_i \Delta_i} \: \big\langle O_1({\textstyle \frac {x_1}\eps}) \, O_2({\textstyle \frac {x_2}\eps}) \, \ldots \, O_n({\textstyle \frac {x_n}\eps}) \big\rangle_{\rm lattice} = \langle \Phi_1(x_1) \, \Phi_2(x_2) \ldots \, \Phi_n(x_n) \rangle_{\rm FT}.
\label{genmatch}
\ee
So we can write the convergence of a lattice observable to a field as the formal identity,
\be
\lim_{\eps \to 0} \eps^{-\Delta} \: O({\textstyle \frac {x}\eps}) = \Phi(x),
\label{scale}
\ee
meant to be valid inside correlators.

If both types of correlators can be separately computed, the potential infinity of identities similar to the previous one put very strong constraints on the field theory proposed and allow to validate it or, on the contrary, to discard it. The more identities we are able to test, the higher the level of confidence we gain for the conjectural field theory.

At this stage we seem to be running in a vicious circle: we want to test the proposed field theory by comparing its correlators with the lattice quantities, but we cannot compute the field correlators if we do not know the field theory ! If one thinks of a field theory as being given by a set of fields and an action $S$, this is indeed a serious problem, because the action cannot be easily guessed, and even worse, there are many cases for which one has no clue as to what the action is. However the action is just one convenient (and usually not simple) way to compute correlators. One could think of other ways to determine correlators, and one of them is the presence of symmetry: enough symmetry allows to determine the correlators. It is precisely the principle underlying the conformal field theories, which therefore provides a field theoretic framework where no action is necessary. They are discussed in the next section.

Knowing the details of the field theory describing the long distance properties of a statistical model is at the same time extremely powerful and immensely complicated. On the one hand, it is indeed powerful because it captures the very essential behaviour of the statistical model without being cluttered with the many irrelevant lattice effects which make the lattice model so much more complex. On the other hand, it is also immensely complicated because every single element in the lattice model which affects the long distances must have a match in the field theory. Such elements include 
\begin{itemize}
\item of course the bulk observables as discussed above, 
\item the boundary conditions, the changes of boundary conditions, and the boundary observables, 
\item the non-local observables (like disorder lines in the Ising model), 
\item the algebra of all the observables, 
\item the specific effects arising when the lattice is embedded in topologically non-trivial geometries (cylinder, torus, ...), 
\item the symmetry, finite or other, that may be present in the model, 
\end{itemize}
and possibly many others. All this represents a huge amount of information that must be present and known in the field theory, and which can be only very rarely contemplated in full. A renown exception is when we consider critical statistical models, as we do here, which are in addition formulated on two-dimensional domains ($d=2$). 

\section{Conformal field theories}
\label{sec4}

Critical systems are primarily characterized by a scale invariance. The correlation lengths of the observables surviving the scaling limit diverge in the infinite volume limit, so that there is no intrinsic length scale left: the fluctuation patterns appear to be the same at all scales. As a consequence, the correlation functions of those observables decay algebraically rather than exponentially. The large distance 2-point correlator of a typical lattice observable $O_i$ located around site $i$ takes the following form,
\be
\langle O_i \, O_j \rangle = \frac{A}{|i-j|^{2\Delta}} + \ldots,
\label{latt}
\ee
where $A$ is a normalization, $\Delta$ is the exponent controlling the decay and the dots indicate lower order terms.

The field theory emerging in the scaling limit inherits the scale invariance. Further assuming translation and rotation symmetries, the scale invariance is enhanced to the invariance under a larger group, namely the group of conformal transformations, i.e. the coordinate transformations which preserve angles \footnote{The material recalled in this section is completely standard; useful references include \cite{DFMS97} (rather comprehensive) and \cite{He99} (more focused on critical statistical systems).}. In $d$ dimensions, the conformal transformations include the transformations mentioned above, namely the translations ($d$ real parameters), dilations (1 parameter) and rotations ($\frac{d(d-1)}2$ parameters), and the so-called special conformal transformations (or conformal inversions) which depend on an arbitrary vector $\vec b$ ($d$ additional parameters) and take the following general form
\be
\frac{\vec x'}{|\vec x'|^2} = \frac{\vec x}{|\vec x|^2} + \vec b \qquad \Longleftrightarrow \qquad {\vec x}' = \frac{\vec x + |x|^2 \, \vec b}{1 + 2 \vec b \cdot \vec x + |\vec b|^2 \, |\vec x|^2}.
\ee
Together these transformations form a finite Lie group isomorphic to SO$(d+1,1)$. They are all global conformal transformations because they are defined everywhere on $\R^d \cup \{\infty\}$ and bijective. In dimension $d>2$, a conformal transformation defined locally can be extended to a global transformation.

Typical spinless (i.e. rotationally invariant) fields transform tensorially under conformal transformations,
\be
\Phi(\vec x) \; \longrightarrow \; \Big|\frac{\partial x'^i}{\partial x^j}\Big|^{\Delta/d} \, \Phi(\vec x'),
\label{conf}
\ee
for some number $\Delta$. Fields transforming that way under global conformal transformations are called {\it quasi-primary}. Global conformal invariance then fixes the average value of a quasi-primary field,
\be
\langle \Phi(\vec x) \rangle = 0, \qquad {\rm if\ } \Delta \neq 0,
\label{1pt}
\ee
(a constant for $\Delta=0$ by translation invariance) and the 2-point correlator of two quasi-primary fields,
\be
\langle \Phi_1(\vec x_1) \, \Phi_2(\vec x_2) \rangle = 
\begin{cases}
\displaystyle \frac{A_{12}}{|\vec x_1 - \vec x_2|^{2\Delta_1}} & \text{if } \; \Delta_1 = \Delta_2,\\
\noalign{\smallskip}
0 & \text{if } \; \Delta_1 \neq \Delta_2.
\end{cases}
\label{2pt}
\ee
Specializing (\ref{conf}) to a dilation, $\vec x' = \alpha \vec x$, we have $\Phi(\vec x) \to \alpha^\Delta\, \Phi(\alpha \vec x)$ so that $\Delta$ can be identified with the dimension of the field $\Phi$ (in units of inverse length).

Global conformal invariance also completely determines the correlator $\langle \Phi_1(\vec x_1) \, \Phi_2(\vec x_2) \, \Phi_3(\vec x_3) \rangle$ of three (and not more) quasi-primary fields,
\be
\langle \Phi_1(\vec x_1) \, \Phi_2(\vec x_2) \, \Phi_3(\vec x_3) \rangle = 
\frac{A_{123}}{|\vec x_1 - \vec x_2|^{\Delta_1+\Delta_2-\Delta_3}\,|\vec x_1 - \vec x_3|^{\Delta_1+\Delta_3-\Delta_2}\,|\vec x_2 - \vec x_3|^{\Delta_2+\Delta_3-\Delta_1}}.
\label{3pt}
\ee

We see that the lattice 2-correlator (\ref{latt}) is consistent with the convergence of the observable $O_i$ to a quasi-primary field $\Phi(\vec x)$ of dimension $\Delta$ upon setting $i = \vec x/\eps$ since the matching identity (\ref{match}) is satisfied,
\be
\lim_{\eps \to 0} \: \eps^{-2\Delta} \; \langle O_{\vec x_1/\eps} \, O_{\vec x_2/\eps} \rangle_{\rm lattice} = \frac{A}{|\vec x_1 - \vec x_2|^{2\Delta}} = \langle \Phi(\vec x_1) \, \Phi(\vec x_2) \rangle_{\rm FT}.
\label{scalim}
\ee
We note that all subdominant terms in the lattice correlator (\ref{latt}) drop out when taking the limit $\eps \to 0$, confirming once more that a field theory captures the large distance behaviour of a critical lattice model.

What has been just recalled is valid in any dimension $d \ge 2$ but is only the beginning of the story for $d=2$. The global conformal group discussed above remains, but is more conveniently presented in complex coordinates as the SL$(2,\C)$ group of M\"obius transformations $w = \frac{az+b}{cz+d}$, for $a,b,c,d \in \C$ satisfying $ad-bc = 1$. 

The two-dimensional world has however many more conformal transformations in store. Indeed it is a well-known fact that any {\it analytic map} $w(z)$ of the complex plane is conformal. Surely an analytic function requires an infinite number of parameter to fix it (f.i. the coefficients of its Laurent expansion in some neighbourhood), so that the conformal `group' is certainly infinite-dimensional. The term {\it group} is not really appropriate because the composition of analytic maps is generally not defined everywhere on the complex plane: unless it is a M\"obius transformation, an analytic map is either not defined everywhere or else its image is not the whole complex plane. For instance the map $w = \frac{L}{2\pi{\rm i}} \log z$ maps the complex plane to a cylinder of circumference $L$. The discussion of two-dimensional conformal group is thus usually carried out at the level of its algebra, for which infinitesimal transformations of the form $w = z + \epsilon \, z^{n+1}$ are considered. The corresponding generators satisfy the famous infinite-dimensional {\it Virasoro algebra},
\be
[L_m,L_n] = (m-n) L_{m+n} + \frac c{12} m(m^2-1) \delta_{m+n,0}, \qquad m,n \in \Z,
\ee
a central extension of the Witt algebra. The real number $c$ is the central charge, and is one of the most important data of a two-dimensional conformal field theory (CFT). The modes $L_0,L_{\pm 1}$, whose algebra is unaffected by the central charge, are the infinitesimal generators of the M\"obius group, with $L_{-1}$ and $L_0$ corresponding to translations and dilations respectively. As it turns out, a second commuting copy of the Virasoro algebra, with modes $\overline L_n$, can formally be considered for the conformal transformations of the antiholomorphic variable $\overline z$.

It is not our purpose to give an introduction to CFT, but one can easily conceive the huge difference between a finite symmetry algebra and an infinite one. A field theory that is to be invariant under an infinite algebra is immensely more constrained, and therefore much more rigid, leaving the hope that one should be able to say a lot more about it. It is indeed the case.

For one thing, the field content of a CFT must be organized into representations of the Virasoro algebra, which are all infinite dimensional, and this opens up the possibility that an infinite number of fields be in fact accomodated in a finite number of representations (such CFT are called rational). In this respect, the {\it primary fields} are particularly important. They are the strengthened version of quasi-primary fields in the sense that they transform tensorially under any conformal transformation. A primary field is an eigenfield of $L_0$ and $\overline L_0$ with real eigenvalues $h$ and $\overline h$, and, more importantly, is annihiliated by all positive modes $L_{n>0},\overline L_{n>0}$. It is in particular characterized by a total weight $\Delta = h+\overline h$ (its eigenvalue under $L_0+\overline L_0$, the real dilation generator) and is of course quasi-primary. The action of any string of negative Virasoro modes $L_{n<0},\overline L_{n<0}$ on a primary field produces infinitely many new fields, called {\it descendant fields}, which include all derivatives of the primary field, since $L_{-1}=\partial_z$ and $\overline L_{-1}=\partial_{\overline z}$ act as derivatives on any field. All of them are eigenfields of $L_0$ and $\overline L_0$. Together they form a highest weight representation of the Virasoro algebra whose structure is similar to highest weight representations of simple Lie algebras, the primary field playing the role of the highest weight state. 

Like in higher dimension, the forms of the 1-, 2- and 3-point of quasi-primary fields are completely fixed by their invariance under M\"obius transformations. They are more easily written in complex coordinates ($z_{ij}=z_i-z_j$),
\bea
&&\langle \Phi(z,\overline z) \rangle = A \, \delta_{h,0\phantom{\bar h}} \! \delta_{\bar h,0},\\
&&\langle \Phi_1(z_1,\overline z_1) \, \Phi_2(z_2,\overline z_2) \rangle = \frac{A_{12}}{z_{12}^{h_1+h_2} \, \overline z_{12}^{\,\bar h_1+\bar h_2}} \: \delta_{h_1,h_2\phantom{\bar h}} \!\! \delta_{\bar h_1,\bar h_2},\\
&&\langle \Phi_1(z_1,\overline z_1) \, \Phi_2(z_2,\overline z_2) \, \Phi_3(z_3,\overline z_3) \rangle = \frac{A_{123}}{z_{12}^{h_1+h_2-h_3}\,z_{13}^{h_1+h_3-h_2}\,z_{23}^{h_2+h_3-h_1}\,\overline z_{12}^{\,\bar h_1+\bar h_2-\bar h_3}\,\overline z_{13}^{\,\bar h_1+\bar h_3-\bar h_2}\,\overline z_{23}^{\,\bar h_2+\bar h_3-\bar h_1}}.\nonumber\\
\eea
These forms suggest that the conformal weights $h_i,\overline h_i$ are positive, so that the correlators decrease with the separation distances, as seems natural from a physical point of view. We will nonetheless encounter physical fields with negative weights, for which the correlators have a different meaning.

Occasionally we will consider {\it chiral correlators} for which we only retain the dependence in the $z_i$ variables of the full correlators (equivalently the action of the holomorphic modes $L_n$). Chiral correlators are appropriate for observables living on a boundary, like the real line bordering the upper-half plane, since a boundary is one-dimensional. In this case, only one copy of the Virasoro algebra remains, so that the fields are characterized by a single conformal weight.  Chiral correlators are also useful to compute the correlators of bulk variables on surfaces with boundaries, see Section \ref{sec6}.

The precise structure of a Virasoro highest weight representation $(c,\Delta)$ based on a primary field of weight $\Delta$ is crucial. In the good cases, it determines the properties of the primary field (and of its descendants) by fixing its correlators with itself or with other fields. The 2-point correlator of a primary has the form (\ref{2pt}) since it is quasi-primary, and the same is true for the 3-point correlator. To go beyond, the global conformal invariance is not enough \footnote{A general 3-point correlator $\langle \Phi_1(z_1) \, \Phi_2(z_2) \, \Phi_3(z_3) \rangle$ is a function of three complex numbers; if all three fields are quasi-primary, that function can be determined by trading $z_1,z_2,z_3$ for the three complex parameters of a general M\"obius transformation.}. It turns out to be often the case that the structure of a Virasoro highest weight representation implies that the correlators $\langle \Phi(z,\overline z) \ldots \;\rangle$ involving the primary field $\Phi$ obey differential equations. Four-point functions can be routinely computed in this way. All correlators can then be determined, at least in principle, without knowing anything of a possible Lagrangian realization of the underlying field theory (through its action). 

The miracle of 2d CFTs can be paraphrased in the following way: to completely solve a CFT, i.e. compute all its correlation functions, and thereby to know everything there is to know of the large distance limit of a critical model, it is sufficient to know enough of the Virasoro representations making up that CFT. This methodology has been immensely successful since the mid-80's and has led to a profound understanding of the many aspects of critical models listed at the end of the previous section. The Ising model is the prominent example of a model that can be treated that way, but the same is true of more general statistical models involving local interactions between the microscopic variables. 

More recently, models showing some form of non-locality have been examined at the conformal light. Sandpile models are in this class, since, as we have seen earlier, the height variables are in strong interaction over the entire domain to form global recurrent configurations. Other models with non-local interactions and/or non-local degrees of freedom include percolation, critical polymers and more general loop models. It may sound surprising but the conclusion seems to be that the conformal approach is still relevant. However the CFTs underlying these models are more complex, essentially because the representations of the Virasoro algebra that appear have a far more complicated structure. These special CFTs are called {\it logarithmic conformal field theories} (LCFT). What follows is a very basic  introduction to the salient features of LCFT; various reviews and applications may be found in the special issue \cite{GRR13}. Let us also mention \cite{Fl03} which reviews the extension to LCFT of the calculational tools used in CFT.

For the highest weight representations discussed above, the operators $L_0,\overline L_0$ are diagonalizable. LCFTs have the distinct feature to include Virasoro representations for which $L_0$ and $\overline L_0$ are no longer diagonalizable, but instead contain (infinitely many) Jordan blocks of finite rank. To have a rough idea of what these representations look like, one can think of a highest weight representation for which the highest weight is not a single primary field, but a pair of fields $(\Phi,\Psi)$, of which only $\Phi$ is primary. The action of $L_0$ on them would be typical of a rank 2 Jordan cell, namely
\be
L_0 \Phi = h \Phi, \qquad L_0 \Psi = h \Psi + \lambda \Phi,
\label{jordan}
\ee
where $\Psi$ is called the {\it logarithmic partner} of the primary field $\Phi$, and a similar action of $\overline L_0$ (with $\overline h$). Under the action of the negative Virasoro modes, the Jordan block structure will propagate among the descendant fields. The presence of Jordan blocks is a sort of minimal ingredient to make a representation logarithmic; many mathematical complications can and do arise, see for instance \cite{KR09}. Higher rank Jordan blocks can also appear.

A immediate consequence of the presence of Jordan blocks explains the use of the word `logarithmic': the correlators of fields in a LCFT contain logarithmic terms in addition to the power laws encountered before. For instance the 2-point correlators of the logarithmic pair $\{\Phi,\Psi\}$, both of weights $(h,\overline h)$, read
\begin{subequations}
\bea
\langle \Phi(z_1,\overline z_1)\, \Phi(z_2,\overline z_2) \rangle &\!\!\!=\!\!\!& 0, \qquad \langle \Phi(z_1,\overline z_1)\, \Psi(z_2,\overline z_2) \rangle = \frac{B}{(z_1 - z_2)^{2h}(\overline z_1 - \overline z_2)^{2\bar h}}, \\
\langle \Psi(z_1,\overline z_1)\, \Psi(z_2,\overline z_2) \rangle &\!\!\!=\!\!\!& \frac{C - 2 \lambda B \log{|z_1 - z_2|^2}}{(z_1 - z_2)^{2h}(\overline z_1 - \overline z_2)^{2\bar h}}.
\eea
\label{2ptlog}
\end{subequations}
$\!\!$For rank $r$ Jordan blocks, the 2-point correlators would involve up to $(r-1)$-th powers of logarithms. The parameter $\lambda$ is not intrinsic as it can be absorbed in the normalization of $\Phi$ or of $\Psi$; likewise, the logarithmic partner $\Psi$ is defined up to a multiple of $\Phi$ without affecting the defining relations (\ref{jordan}). The chiral version of the above 2-point functions reads
\begin{subequations}
\bea
\langle \Phi(z_1)\, \Phi(z_2) \rangle &\!\!\!=\!\!\!& 0, \qquad \langle \Phi(z_1)\, \Psi(z_2) \rangle = \frac{B}{(z_1 - z_2)^{2h}}, \\
\langle \Psi(z_1)\, \Psi(z_2) \rangle &\!\!\!=\!\!\!& \frac{C - 2 \lambda B \log{(z_1 - z_2)}}{(z_1 - z_2)^{2h}},
\eea
\label{2ptlogch}
\end{subequations}

It should not be too surprising that Jordan blocks and logarithms go hand in hand. Under dilation by a factor $\alpha$, a logarithmic term transforms inhomogeneously $\log z \rightarrow \log z + \log \alpha$, reflecting the inhomogeneous action of the dilation generator $L_0$ on $\Psi$. Under a finite dilation $w=\alpha z$, the transformation laws of $\Phi$ and $\Psi$ read 
\be
\Phi'(w,\overline w) = |\alpha|^{-\Delta} \, \Phi(z,\overline z), \qquad \Psi'(w,\overline w) = |\alpha|^{-\Delta} \,  \{\Psi(z,\overline z) - \lambda \log|\alpha|^2 \, \Phi(z,\overline z) \}.
\ee
One may check that the form of the correlators (\ref{2ptlog}) is indeed invariant under the replacement $\Phi(z,\overline z) \rightarrow \Phi'(w,\overline w)$ and $\Psi(z,\overline z) \rightarrow \Psi'(w,\overline w)$. Let us also note that the scaling (\ref{scale}) must be redefined for the lattice observables described by logarithmic fields since it involves a dilation by a factor $1/\eps$, to which the field responds by an inhomogeneous term.

Despite all the efforts spent, LCFTs are generally much less understood than their non-logarithmic cousins, altough a number of general features are known. On the statistical side, few models have been thoroughly studied, as their non-local features make it hard to carry out exact calculations on the lattice. On the field-theoretic side, it is not known what a generic LCFT looks like. The simplest of all (but non-trivial) and probably the only LCFT to be fully under control is the symplectic fermion theory with central charge $c=-2$, also called the triplet theory. It has been introduced in \cite{Gu93} and then investigated in greater detail in \cite{GK99,GR06}. It has the following Lagrangian realization in terms of a pair of free, massless, Grassmanian scalar fields $\theta,\tilde \theta$,
\be
S = \frac 1{\pi} \int {\rm d}z {\rm d}\overline z \; \partial \theta \overline\partial \tilde \theta\,, \qquad\qquad \partial = \partial_z, \: \overline\partial = \partial_{\overline z}.
\ee
Several fields in this theory form logarithmic pairs, like the identity ${\mathbb I}$ and the composite field $\theta \tilde\theta$. We note that (\ref{2ptlog}) then implies the somewhat unusual relation $\langle {\mathbb I} \rangle = 0$, which indeed follows, using the rules of integration over Grassmanian variables, from the fact that the above action does not depend on the constant modes of $\theta$ and $\tilde\theta$. Since this is a free scalar theory, all correlators of fields that are local (i.e. product of derivatives) in $\theta,\tilde\theta$ are polynomials in the derivatives of the Green function (the kernel of the inverse Laplacian $-4\partial\overline\partial$) given in complex coordinates by $G(z,w) = -\log|z-w|$. 

To finish, let us note that the statistical models which have a non-diagonalizable transfer matrix (when there is a proper one) are the natural candidates for being described by LCFTs in their scaling regime. Indeed such a transfer matrix gives rise to a non-diagonalizable Hamiltonian, which itself is the lattice version of the field-theoretic operator $L_0+\overline L_0$. As said above, the non-diagonalizability of $L_0,\overline L_0$ is the hallmark of LCFTs. The logarithmic minimal models form an infinite series of such lattice models \cite{PRZ06}.

The rest of this review is devoted to discussing the variables of the 2d sandpile models which have been successfully (i.e. with enough confidence) identified in the corresponding LCFT. These elements reveal {\it some} facets of the field theory at work in sandpile models: the big and complete picture is well out of reach for the moment.

%
%
%
%
%

\section{Bulk variables}
\label{sec5}

The height variables are certainly the first and most natural variables to look at, as they are the microscopic variables in terms of which the models are defined. The introduction we gave in Section \ref{asm} was for the Abelian sandpile on an arbitrary graph. If large distance properties should be rather robust against local modifications of a graph, they are not expected to be the same on a graph with a high degree of connectivity (the extreme example being the complete graphs), a regular graph with a moderate degree of connectivity or a graph with a strong hierarchical structure (like Cayley trees). Most of the results reviewed here are obtained when the graph is a rectangular portion of the square lattice $\Z^2$; varying the size of the grid is an easy way to approach the infinite volume limit and this choice ensures that conservative sites away from the boundary have height variables taking the same number of values (namely, 4). The triangular and honeycomb lattices, for which the number of height values is respectively 6 and 3, will be briefly discussed as well in order to address universality issues, see Section \ref{7.2}.

In most cases, the only dissipative sites will be located on the boundary \footnote{For rectangular grids $\Gamma \subset \Z^2$, the notion of boundary is clear: when $\Gamma$ is embedded in $\Z^2$, the boundary sites are those which are connected to sites of $\Z^2$ not in $\Gamma$.}, except when we discuss the insertion of isolated dissipation. With one exception, we will exclusively consider open and closed boundary conditions, by which we mean that whole stretches of boundary sites are either dissipative or conservative respectively. The choice of boundary conditions has clearly an effect at finite volume, but also in the infinite volume limit if some of the boundaries are kept at finite distance (for example on the upper half-plane or on a strip of finite width, see Section \ref{sec6}). 

On a finite grid $\Gamma$, the heights assigned to the vertices form stable configurations, but only the recurrent ones have a non-zero (and uniform) weight with respect to the invariant measure $\P_\Gamma$. So far, we have no clear idea of what a generic recurrent configuration looks like. Answers to questions like ``What is the proportion of sites having height 1, height 2, ... ?'' can certainly help figure out. Also the heights must be correlated within a recurrent configuration. Can one characterize these correlations ? Are they exponential or power-lawed ? The computation of multisite height probabilities answers these questions and helps understand the statistics of recurrent configurations. 

To be definite, let us consider $\Gamma$ to be an $L \times M$ rectangular grid in $\Z^2$, with open boundary conditions: the non-boundary sites are conservative and have maximal height value equal to $z_i^\star = z_i = 4$, whereas the boundary sites have maximal height value chosen to be $z_i^\star = 4 > z_i$ (boundary and corner sites dissipate 1 resp. 2 grains of sand under toppling; both types are connected to the sink). Thus the toppling matrix is four times the identity minus the adjacency matrix of the grid, and the height at every site takes values in $\{1,2,3,4\}$. In this section, all boundaries are sent off to infinity in the scaling limit, so that the domain converges to $\R^2$; in that limit, all multisite probabilities are fully invariant under translations. 

\subsection{One-site height probabilities}
\label{5.1}

As a warm up for what has to come, we ask the following: what is the probability $\P_\Gamma(h_i=a)$ that, in a recurrent configuration, a given site $i$ has height equal to $a$, between 1 and 4 ? Because we are interested in the infinite volume limit of these numbers, we take $i$ to be deep in the middle of the grid, well away from the boundaries.

If we pause for a while and ponder over that simple question, we feel a bit at a loss on how to handle it 
because the only mean we have is the general criterion of recurrence, namely the non-existence of forbidden subconfigurations. Let us start with the height 1.

Since the total number of recurrent configuration is equal to $\det \Delta_\Gamma$ (see Section \ref{asm}), we can write
\be
\P_\Gamma(h_i = 1) = \frac{\#\{{\rm recurrent\ configs\ with\ }h_i=1\}}{\det \Delta_\Gamma},
\ee
For $h_i=1$ to be in a recurrent configuration $C$, the height of none of its neighbours N, E, S or W can be equal to 1 (as they would form a forbidden subconfiguration). Following the clever trick proposed in \cite{MD91}, we consider a new grid $\tilde \Gamma_i$ by deleting from $\Gamma$ the vertex $i$ and the four edges incident to it. We also define from $C$ a new configuration $\tilde C$ on $\tilde\Gamma_i$ by setting 
\be
\tilde h_j = 
\begin{cases}
h_j & {\rm for\ }j \not\in \{i,{\rm N,E,S,W}\},\\
h_j-1 \ge 1 & {\rm for\ }j \in \{{\rm N,E,S,W}\}.
\end{cases}
\ee
Looking back at the criterion of recurrence for an arbitrary graph, it is not difficult to see that a configuration $C$ with $h_i=1$ is recurrent on $\Gamma$ if and only if $\tilde C$ is recurrent on $\tilde \Gamma_i$. We thus obtain
\be
\P_\Gamma(h_i = 1) = \frac{\det \Delta_{\tilde\Gamma_i}}{\det \Delta_\Gamma} = \frac{\det [\Delta_{\tilde\Gamma_i} \oplus 1_{ii}]}{\det \Delta_\Gamma},
\ee
where the matrix in the numerator has been extended by a one-dimensional diagonal block labelled by the vertex $i$, without changing the value of the determinant. One then can write
\be
\Delta_{\tilde\Gamma_i} \oplus 1_{ii} = \Delta_\Gamma + B(i), \qquad 
\ee
with $B(i)$ the defect matrix given by
\be
B(i)_{k,k'} = {\textstyle \begin{pmatrix}
-3 & 1 & 1 & 1 & 1 \\
1 & -1 & 0 & 0 & 0 \\
1 & 0 & -1 & 0 & 0 \\
1 & 0 & 0 & -1 & 0 \\
1 & 0 & 0 & 0 & -1
\end{pmatrix}}, \qquad k,k' \in \{i,{\rm N,E,S,W}\},
\label{defect}
\ee
and $B(i)$ is zero everywhere else. We obtain  
\be
\P_\Gamma(h_i = 1) = \frac{\det [\Delta_\Gamma + B(i)]}{\det \Delta_\Gamma} = \det[{\mathbb I} + \Delta_\Gamma^{-1} \, B(i)].
\label{p1}
\ee
It reduces to the computation of a finite determinant since $B(i)$ has finite rank. In the infinite volume limit (both $L,M \to \infty$), this probability converges to a constant $\P_1$ (by translation invariance). As the matrix $\Delta_\Gamma$ becomes the discrete Laplacian on $\Z^2$ in that limit \footnote{The reader will legitimately point out that the Laplacian on $\Z^2$ has a zero mode and is therefore not invertible. A closer look at the determinants (\ref{p1}) however reveals that they only depend on differences of the inverse matrix entries, which are perfectly well-defined.}, standard results yield \cite{MD91}
\be
\P_1 \equiv \lim_{|\Gamma| \to \infty} \: \P_\Gamma(h_i = 1) = \frac{2(\pi-2)}{\pi^3} \simeq 0.073\,63.
\ee
It also means that a recurrent configuration has an average of about 7\% of sites with a height equal to 1.

What about higher heights ? We know for sure that the inequalities $\P_4 > \P_3 > \P_2 > \P_1$ hold because adding one grain of sand to a recurrent configuration, at a site where $h_i = a$, yields a recurrent configuration if $a<4$. However to actually compute these numbers, can one use the same trick as for the height 1 ? The answer is definitely negative: no local modification of $\Gamma$ like what we did above will allow to compute the corresponding probabilities. To undertand this, we turn to the description in terms of spanning trees. 

As was briefly mentioned in Section 2, the burning algorithm yields a one-to-one correspondence between a recurrent configuration and a spanning tree rooted at the sink site $s$ and growing into the interior of $\Gamma$. In a given spanning tree $\mathcal T$, a site $j$ is called a {\it predecessor} of $i$ if the unique path in $\mathcal T$ from $j$ to the root passes through $i$. Let us also denote by $X_k(i)$ the fraction of all spanning trees for which the site $i$ has $k$ predecessors {\it among its nearest neighbours}, for $0 \le k \le 3$. A careful analysis of the burning algorithm shows the following \cite{Pr94},
\be
\P_\Gamma(h_i = a) = \P_\Gamma(h_i = a-1) + \frac{X_{a-1}(i)}{5-a}, \qquad 1 \le a \le 4.
\ee

For $a=1$, we see that $\P_\Gamma(h_i = 1)$ is related to $X_0(i)$, namely the fraction of spanning trees on $\Gamma$ for which the site $i$ is a leaf. All such trees can be obtained from arbitrary spanning trees on $\tilde \Gamma_i$ by adding one edge between one neighbour of $i$ and $i$ itself (four different possibilities). Thus both points of view coincide and lead to the same local modification $\Gamma \rightarrow \tilde \Gamma_i$.

The next case is $\P_\Gamma(h_i = 2)$, related to $X_1(i)$. Here the situation is dramatically different because the condition that $i$ has only one predecessor among its nearest neighbours is highly non-local. The reason for this is that there are two manners for a neighbour of $i$ to be a predecessor of $i$ in a given tree. The first one is that the tree includes the edge between the two sites so that the neighbour of $i$ is directly connected to $i$. In the second manner, the tree contains a potentially long chain of edges that forms a path between the two sites. The first one is a local connection and is easy to check, the second one is non-local and more difficult. The same remark applies to the fractions $X_2(i)$ and $X_3(i)$, and make the calculation of the corresponding probabilities much more complicated. 

In fact this first natural and simple looking question we have raised, namely the value of $\P(h_i=a)$, turned into a fairly long warming up exercise, as it took about twenty years before the completely explicit probabilities could be found. By using a rather heavy graph-theoretical technology, Priezzhev \cite{Pr94} obtained the first expressions for $\P_2,\P_3$ and $\P_4$, but these were given in the form of multivariate integrals. The problem was reconsidered in \cite{JPR06}, where the following explicit values were conjectured, 
\begin{subequations}
\bea
\P_2 \egal {1 \over 4} - {1 \over 2\pi} - {3 \over \pi^2} + {12 \over \pi^3} \simeq 0.173\,90,\\
\noalign{\smallskip}
\P_3 \egal {3 \over 8} + {1 \over \pi} - {12 \over \pi^3} \simeq 0.306\,29,\\
\noalign{\smallskip}
\P_4 \egal {3 \over 8} - {1 \over 2\pi} + {1 \over \pi^2} + {4 \over \pi^3} \simeq 0.446\,17.
\eea
\label{234}
\end{subequations}
A few years later, three independent proofs were given. The first one was based on a relation with the probability of a loop-erased random walk (LERW) to visit a fixed nearest neighbour of its starting point, which was then computed in terms of dimer arrangements \cite{PPR11}. The second proof also used the relation with LERW passage probabilities but within a much more general approach \cite{KW15}. Finally the third one \cite{CS12} carried out the direct computation of the multiple integrals left open in \cite{Pr94}. Let us mention that the technique developed in \cite{KW15} to enumerate so-called cycle-rooted groves (which generalize spanning trees to spanning forests with marked points) currently provides by far the most efficient way to compute height probabilities, reducing the calculation of $\P_2,\P_3$ to just a few lines, see \cite{PR17}. Most of the height correlators presented below have been computed using this technique. Also noteworthy in this context is the work \cite{KW16} which presents a direct and elementary derivation of the average height $\langle h \rangle = \sum_a a \, \P_a$ on planar lattices (from the formulas above, it is equal to $\frac {25}8$ on $\Z^2$) without computing the individual height probabilities. 

Ironically, the four numbers $\P_a$ are not very useful for a comparison with a field theory, because they will have to be subtracted in correlators (see below). And indeed some of the correlators have been determined exactly before the 1-site probabilities $\P_a$ were found.

Even though the explicit expressions for the $\P_a$'s have the same level of simplicity, the far larger complexity of the combinatorial problem posed by the calculation of $\P_{a\ge 2}$ hints at a striking difference of nature between the height 1 and the higher heights: the height 1 is essentially local, the others are non-local. This will soon be confirmed.

\subsection{Height cluster probabilities}
\label{sec5.2}

Cluster height probabilities are a rather obvious generalization of one-site probabilities, by which we ask for the probability that a specific connected subconfiguration occurs in recurrent configurations, away from the boundaries and in the infinite volume limit. Examples of height clusters are shown below.

\be
\begin{pspicture}[shift=0](-0.5,-0.2)(60,0.8)
\psset{xunit=0.28cm}
\psset{yunit=0.28cm}
\psset{runit=0.28cm}
\rput(0,1.5){\ball1}
\rput(6,0){\ball2}
\rput(9,0){\ball2}
\rput(9,3){\ball1}
\rput(15,0){\ball2}
\rput(18,0){\ball3}
\rput(18,3){\ball1}
\rput(21,0){\ball2}
\rput(21,3){\ball3}
\rput(24,3){\ball1}
\rput(32,1.5){\ball2}
\rput(38,1.5){\ball2}
\rput(41,1.5){\ball4}
\rput(44,1.5){\ball3}
\rput(50,0){\ball2}
\rput(53,0){\ball4}
\rput(53,3){\ball3}
\rput(56,3){\ball2}
\end{pspicture}
\label{clusters}
\ee


The three clusters on the left belong to the family of {\it weakly allowed subconfigurations}, or minimal height clusters, first introduced in \cite{MD91}, and which contains the cluster made of a single height equal to 1. They are minimal subconfigurations in the sense that if one decreases any of its heights by 1, the clusters become (or contain) forbidden subconfigurations. As was done in the previous subsection for the single height 1, their occurrence probabilities around position $i$ can be computed by cutting off appropriate lattice sites and edges. They take the form of finite determinants $\P_S(i) = \det[{\mathbb I} + \Delta^{-1}_\Gamma B_S(i)]$ where the defect matrix $B_S(i)$ depends on the cluster $S$ considered \cite{MD91,MR01}. 

The three clusters on the right of (\ref{clusters}) are not minimal and generalize the simple cluster made of a single height larger or equal to 2. Their level of complexity is comparable to the latter and are best computed using the methods of \cite{KW15}. Explicit calculations become fairly tedious as the size of the cluster increases.

\subsection{Height correlations}
\label{sec5.3}

In terms of the subtracted height variables \footnote{In order to make contact with fields, we slightly change the notation $h_i \to h(i)$ for the height at site $i$.},
\be
h_a(i) \equiv \delta_{h(i),a} - \P_a,
\ee
the $n$-point correlation functions are given by
\be
\sigma_{a_1,a_2,\ldots,a_n}(i_1,i_2,\ldots,i_n) = {\mathbb E}\big[h_{a_1}(i_1)\, h_{a_2}(i_2) \ldots h_{a_n}(i_n)\big].
\ee
These are the functions we are primarily interested in for a future comparison with a conformal field theory. To make the comparison sensible, we have to take the infinite volume limit {\it and} the limit of large separations $|i_k - i_\ell| \to +\infty$. In addition, to avoid the boundary effects --they will be studied later on--, all insertion points $i_k$ are to stay (infinitely) far from the boundaries. In practice, one first replaces $\Delta_\Gamma$ by the Laplacian $\Delta$ on $\Z^2$, and then expand the Green matrix ({\it i.e.} the inverse Laplacian) for large separations. 

The computation of correlations of heights 1 (or indeed any weakly allowed subconfigurations, see below) poses no particular problem. The argument used in Section \ref{5.1} leading to consider new configurations $\tilde C$ on a locally modified lattice $\tilde\Gamma$ is simply repeated for the neighbourhood of each cluster. Thus the probability to find a height $h(i_1)=1$ at site $i_1$, a height $h(i_2)=1$ at site $i_2$, and so on, is equal to 
\be
\P_\Gamma\big(h(i_1)=1,\,h(i_2)=1,\ldots\big) = \frac{\det[\Delta_\Gamma + B(i_1) + B(i_2) + \ldots]}{\det \Delta_\Gamma} = \det\Big[{\mathbb I} + \Delta^{-1}_\Gamma \big\{B(i_1) + B(i_2) + \ldots\big\}\Big].
\label{multi}
\ee
The correlators $\sigma_{1,1,\ldots,1}$ are obtained by taking appropriate subtractions and the limits discussed above.

The first few $n$-point correlators can be easily computed for arbitrary configurations of insertion points \cite{MR01,PR17}. By construction, the 1-point function vanishes, $\sigma_1(i_1) = 0$ (the relation (\ref{1pt}) is indeed the main motivation for the subtraction). The 2-point function is found to be ($i_1-i_2 = \vec r = r {\rm e}^{i\varphi}$)
\be
\sigma_{1,1}(i_1,i_2) = - \frac{\P_1^2}{2r^4} - \frac{4(\pi-2)[1 + (\pi-2) \cos{4\varphi}]}{\pi^6 \, r^6}  + \ldots
\label{11}
\ee
where the dots stand for lower order terms. 

This first result is instructive for several reasons. First, for large separation distances, the dominant term indicates that the correlation decay is algebraic, which shows that the model is critical and makes room for a conformal field theoretic description. Second, choosing the scale dimension $\Delta=2$, the scaling limit (\ref{scalim}) indeed retains the first term only, the form of which has the expected form (note that the second term, like all other subdominant ones, has only the lattice rotation invariance, and is therefore not expected to survive the scaling limit). And third, the dominant term is negative, indicating an anticorrelation between the heights 1. This is consistent with the fact that the presence of many heights 1 in a configuration makes it more likely to be non-recurrent. Interestingly, the calculation can be carried out in $\Z^d$, with the result that the correlation decays like $r^{-2d}$, giving a dimension $\Delta=d$ \cite{MD91}.

The mixed correlator of a height 1 and a height 2, 3 or 4 is harder. They have first been obtained in \cite{PGPR08,PGPR10} by using classical graph-theoretic techniques, and then reconsidered and extended in \cite{PR17} using the results of \cite{KW15}. Whatever the method used, one has to evaluate the fractions $\tilde X_k(i_1)$ of spanning trees, as defined in Section \ref{sec5.2}, but on a lattice modified around the $i_2$ where the height 1 is located. This modification affects the toppling (Laplacian) matrix and its inverse, and consequently the whole computation, heavily based on these two matrices. The result for a height 1 and a height 2 reads, at dominant order,
\be
\sigma_{2,1}(i_1,i_2) = - \frac{\P_1^2}{2r^4} \left\{\log{r} + \Big( \gamma + \frac 32 \log{2} + \frac{16-5\pi}{2(\pi-2)} \Big) \right\}  + \ldots
\label{21}
\ee
where $\gamma = 0.577\,216...$ is the Euler constant. The first subdominant correction is of order $r^{-6}$ and contains a non-trivial angular dependence, like in (\ref{11}), but also a $\log{r}$ term \cite{PR17}. The expressions of $\sigma_{3,1}$ and $\sigma_{4,1}$ are similar, with different coefficients. 

The expressions $\sigma_{a,1}$ for $a>1$ definitely establish the logarithmic character of the CFT underlying the sandpile model. The forms (\ref{11}) and (\ref{21}) are strongly reminiscent of those in (\ref{2ptlog}), but do not quite match. If in the scaling limit, the heights 1 and 2 were to converge to a logarithmic pair $\{h_1(z),h_2(z)\}$, one would think that $\sigma_{1,1}$ and $\sigma_{2,1}$ ought to go over to the 2-point functions $\langle h_1(z_1) h_1(z_2) \rangle$ and $\langle h_2(z_1) h_1(z_2) \rangle$ respectively. However, conformal invariance implies that the former vanishes identically, whereas the latter is not logarithmic. We could think of computing $\sigma_{2,2}$ to see what comes out, but large distance correlators with several heights strictly larger than 1 are far beyond our present computational capabilities. Let us add that the calculation of $\sigma_{a,b}(i_1,i_2)$ for $a,b > 1$ does not merely reduce to the evaluation of numbers like $X_{a-1,b-1}(i_1,i_2)$ which would generalize the numbers $X_{a-1}(i)$ defined earlier and enumerate the spanning trees with fixed numbers of predecessors among the nearest neighbours of $i_1$ resp. $i_2$. Indeed the possibility that neighbours of $i_1$ are predecessors of $i_2$, or vice-versa, substantially complicates the matter. Details on how to perform the correct counting have been given in \cite{PR17}.

To reconcile the previous lattice results and the LCFT predictions, we pause for a while to examine the effects of a seemingly unrelated observable. 

\subsection{Isolated dissipation}
\label{sec5.4}

In the previous section, the calculation of height probabilities started on a finite grid $\Gamma$, where the only dissipative sites are boundary sites. We did not pay too much attention to exactly which boundary sites are dissipative; in fact, since the infinite volume limit sends the boundaries off to infinity, there is no need to know precisely which boundary conditions are used (this is what we meant when we said that $\Delta_\Gamma$ becomes the Laplacian on $\Z^2$). 
We show now that the situation changes if we make some of the bulk sites dissipative \cite{PR04}. It is not difficult to understand why this is so in terms of spanning trees. We remember that dissipative sites are sites that are connected to the sink, the root of the trees, from which the branches of the tree are growing. Therefore the existence of dissipative sites in the bulk make it possible that branches grow from the middle of the grid, thereby affecting the macroscopic structure of the spanning trees.

To make a bulk site $i_1$ dissipative, one simply has to connect it to the sink. In the notations of Section \ref{asm}, this amounts to increase the value $z_{i_1}^\star$, for instance from $z_{i_1}=4$ (on $\Z^2$) to 5 (a higher value would not make much difference). In turn this changes by 1 the diagonal entry $(\Delta_\Gamma)_{i_1,i_1}$ of the toppling matrix, that is, $\Delta_\Gamma \to \Delta_\Gamma + D_{i_1}$, with $(D_{i_1})_{i,j}=\delta_{i,i_1} \delta_{j,i_1}$. More generally, the new toppling matrix $\tilde \Delta_n \equiv \Delta_\Gamma + D_{i_1} + D_{i_2} + \ldots + D_{i_n}$ defines a new model in which several bulk sites $i_k$ are dissipative. As a consequence, the height variables at these sites take values in the set $\{1,2,3,4,5\}$.

A simple and natural way to evaluate the effect of inserting isolated dissipation is to consider the change in the number of recurrent configurations, by computing the ratio $\det \tilde\Delta_n/\det \Delta_\Gamma$, first at finite volume, then in the infinite volume limit.

We start by inserting dissipation at single site $i$, far from the boundaries. The ratio is easy to compute since the defect matrix $D_i$ has rank 1,
\be
\frac{\det \tilde\Delta_1}{\det\Delta_\Gamma} = \det({\mathbb I} + \Delta_\Gamma^{-1} D_i) = 1 + (\Delta_\Gamma^{-1})_{i,i}.
\label{1diss}
\ee
It is a finite number at finite volume, but diverges in the infinite volume limit, no matter where the site $i$ is located. The divergence reflects the fact that the extra value $h_i=5$ allows enormously more recurrent configurations in the modified model \footnote{We note that the inverse ratio $\det\Delta_\Gamma/\det \tilde\Delta_1$ is equal to  ${\rm Prob[all\ }h_j \le 4] = {\rm Prob}[h_i \le 4] = 1 - {\rm Prob}[h_i=5]$ where the probabilities are evaluated in the modified model. The divergence mentioned in the text therefore implies that $h_i=5$ with probability 1.}. 

The {\it same} divergence is present in the ratios $\det \tilde\Delta_n/\det \Delta_\Gamma$, which suggests to change the normalization and compare the effect of inserting $n$ dissipative sites with respect to the situation where there is only one dissipative site, that is, to consider instead the ratios $\det \tilde\Delta_n/\det \tilde\Delta_1$, perfectly well-defined. Let us also remark that in the infinite volume limit, the denominator does not depend on the location of the (only) dissipative site, so that the ratios are fully symmetric in the insertion points $i_k$ and translation invariant.

The first two ratios read, with $r=|i_1-i_2|$ for $n=2$,
\be
\frac{\det \tilde\Delta_1}{\det \tilde\Delta_1} = 1, \qquad \frac{\det \tilde\Delta_2}{\det \tilde\Delta_1} = \frac 1\pi \log{r} + 2\gamma_0 + {\mathcal O}(r^{-2}),
\ee
where $\gamma_0 = \frac 1{2\pi}(\gamma + \frac 32 \log{2})+1$. If we denote by $\omega(z,\overline z)$ the field that describes, in the scaling limit, the insertion of dissipation at a bulk site, the previous two equations would imply 
\be
\langle \omega(z,\overline z) \rangle = 1, \qquad \langle \omega(z_1,\overline z_1) \, \omega(z_2,\overline z_2) \rangle = \frac 1\pi \log|z_1 - z_2| + 2\gamma_0.
\ee
Interestingly, they exactly match the last two equations of (\ref{2ptlog}), with the logarithmic pair $\{\Phi,\Psi\}$ identified with $\{{\mathbb I},\omega\}$, both fields having the weights $h=\overline h=0$ (the identity field is primary). Moreover the logarithmic term in the 2-point correlation fixes the coefficient $\lambda$ of the logarithmic pair $({\mathbb I},\omega)$ equal to $\lambda = -\frac 1{4\pi}$ so that $L_0 \, \omega = \overline L_0 \, \omega = -\frac 1{4\pi}{\mathbb I}$. The relation $\langle {\mathbb I} \rangle = 0$, as noted for the free symplectic fermion theory, is here understood as being given by the inverse of the divergent quantity in (\ref{1diss}). 

The lattice calculation of $\det \tilde\Delta_3/\det \tilde\Delta_1$, corresponding to the insertions of three dissipative sites, is not difficult and yields the following 3-point correlation, with $z_{ij}\equiv z_i-z_j$,
\be
\langle \omega(1) \, \omega(2) \, \omega(3) \rangle = 3\gamma_0^2 + \frac{\gamma_0}{2\pi} \log{|z_{12}z_{13}z_{23}|^2} + \frac 1{16\pi^2} \left[ \log{|z_{12}|^2} \: \log\Big|\frac{z_{13}z_{23}}{z_{12}}\Big|^2 + {\rm cyclic} \right].
\label{3om}
\ee
It is fully consistent with the general 3-point correlators of fields in a logarithmic pair \cite{Fl03}. Many additional checks have been carried out \cite{PR04} which all confirm the consistency of the above field assignment. It has been shown \cite{JPR06} that the bulk dissipation field can be realized in terms of symplectic free fermions as 
\be
\omega(z,\overline z) = \frac1{2\pi} \theta \tilde \theta + \gamma_0,
\ee
in the sense that the correlators of this composite field, computed in the symplectic fermion theory, reproduce the above expressions.

\subsection{Height correlations cont'd}
\label{sec5.5}

The multisite height probabilities computed in Section \ref{sec5.3} were obtained by taking the limit over a sequence of grids of increasing size. Because of the dissipation along the boundaries, the probabilities are well-defined for each finite grid, and properly converge. On the field-theoretic side, the CFT supposedly describing the scaling limit is defined right away on the infinite continuum, and does not know about the dissipation of the finite systems. To make the CFT connect with the lattice description, we have to insert by hand the required dissipation in the correlators. Since on the lattice side, the boundary dissipation is pushed off to infinity when we take the infinite volume limit, the previous section suggests that we insert the additional field $\omega(\infty)$ in the correlators. Thus the proposal, first made in \cite{JPR06}, is that a lattice $n$-point height correlator is described in the scaling limit by an $(n+1)$-point field correlator,
\be
\sigma_{a_1,a_2,\ldots,a_n}(i_1,i_2,\ldots,i_n) \stackrel{{\rm scalim}}{\longrightarrow} \langle h_{a_1}(z_1)\, h_{a_2}(z_2) \ldots h_{a_n}(z_n) \, \omega(\infty) \rangle.
\ee

It turns out that the proposed field correlations exactly reproduce the form of the lattice results obtained in Section \ref{sec5.3}. If $\{\Phi,\Psi\}$ are fields of weights $h = \overline h$ forming a logarithmic pair such that $(L_0 - h) \Psi = (\overline L_0 - h)\Psi = \lambda \Phi$, one finds with $\Delta=2h$ \cite{JPR06}
\begin{subequations}
\bea
\hspace{-5mm} \langle \Phi(z_1,\overline z_1)\, \Phi(z_2,\overline z_2) \, \omega(\infty) \rangle &\!\!\!=\!\!\!& \frac{A}{|z_{12}|^{2\Delta}}, \quad \langle \Phi(z_1,\overline z_1)\, \Psi(z_2,\overline z_2) \, \omega(\infty) \rangle = \frac{B - \lambda A \log|z_{12}|^2}{|z_{12}|^{2\Delta}}, \\
\hspace{-5mm} \langle \Psi(z_1,\overline z_1)\, \Psi(z_2,\overline z_2) \, \omega(\infty) \rangle &\!\!\!=\!\!\!& \frac{C - 2\lambda B \log|z_{12}|^2 + \lambda^2 A \log^2{|z_{12}|^2}}{|z_{12}|^{2\Delta}}.
\eea
\end{subequations}
Comparing with (\ref{2ptlog}), we see that the insertion of dissipation at infinity through $\omega(\infty)$ allows a non-zero value of $A$, and cures the problem encountered in Section \ref{sec5.3}. 

From the dominant terms in (\ref{11}) and (\ref{21}) for the lattice correlations $\sigma_{1,1}(i_1,i_2)$ and $\sigma_{2,1}(i_1,i_2)$, we infer that the (subtracted) bulk lattice height 1 and height 2 variables converge to fields, $h_1(z)$ and $h_2(z)$, that form a logarithmic pair of weight $\Delta=2$. Moreover if we assign them the same normalization as their lattice companions ($A=-\frac{\P_1^2}2$), we find the parameter of the logarithmic pair $(h_1,h_2)$ equal to $\lambda = -\frac 12$. The explicit results for $\sigma_{3,1}(i_1,i_2)$ and $\sigma_{4,1}(i_1,i_2)$ \cite{PGPR10} show that the fields $h_3(z)$ and $h_4(z)$ are also logarithmic partners of $h_1(z)$ albeit of different normalizations and for different values of $\lambda$. As noted in Section \ref{sec4}, it means that they can be written as linear combinations \footnote{The four height fields $h_a(z)$ also satisfy the trivial identity $h_1(z) + h_2(z) + h_3(z) + h_4(z) = 0$.} of $h_1(z)$ and $h_2(z)$ with known coefficients,
\be
h_a(z) = \alpha_a h_2(z) + \beta_a h_1(z), \qquad \alpha_1 = 0, \; \alpha_2 = 1, \; \alpha_3 = \frac{8-\pi}{2(\pi-2)}, \; \alpha_4 = -\frac{\pi+4}{2(\pi-2)},
\label{ha}
\ee
and other values for the $\beta_a$ \cite{JPR06}. These field assignments predict that the lattice correlation of heights larger or equal to 2 behave asymptotically as
\be
\sigma_{a,b}(i_1,i_2) \simeq -\alpha_a \alpha_b \frac{\P_1^2}{2} \: \frac{\log^2 r}{r^4} + {\mathcal O}\Big(\frac{\log r}{r^4}\Big), \qquad a,b \ge 2.
\label{ab}
\ee
Because $\alpha_4$ is the only negative coefficient among the $\alpha_a$, the height variables are all anticorrelated, except the height 4 which has a positive correlation with the other three heights. Numerical simulations have successfully confirmed the behaviour (\ref{ab}) \cite{JPR06}. A lattice proof however remains one of the greatest challenges in the sandpile models.

The lattice 2-point correlators discussed above correspond to 3-point functions in the CFT. They are therefore completely generic, depending only on the weights of the fields involved and a few assumptions about their global conformal transformations. Higher correlators are not generic and depend on finer details of the nature of the fields and of the specific CFT at work, in particular its central charge. In this regard, the first hint for the value of the central charge was given in \cite{MD92} by looking at the finite-size corrections of the partition function (i.e. the number of recurrent configurations); the analysis yields the value $c=-2$. 

The simplest higher correlators to consider on the lattice are the 3- and 4-point height 1 correlators. They have been computed in \cite{MR01} with the following results. Since the height 1 variable has weight $\Delta=2$ in the scaling limit, one would expect the dominant contribution to the 3-point correlator to be homogeneous of degree $-6$ in the seperation distances. Surprisingly the first non-zero term has degree $-8$,
\be
\sigma_{1,1,1}(i_1,i_2,i_3) = 0 + \ldots
\ee
implying that its scaling limit, corresponding to the CFT 4-point function $\langle h_1(z_1) \, h_1(z_2) \, h_1(z_3) \, \omega(\infty) \rangle$, vanishes identically. 

The lattice 4-point correlator has the expected dominant degree $-8$,
\bea
\hspace{-0.cm} \sigma_{1,1,1,1}(i_1,i_2,i_3,i_4) \egal {\P_1^4 \over 8} \, \left\{{1 \over |z_{12}\,z_{34}|^4} + {1 \over |z_{13}\,z_{24}|^4} + {1 \over |z_{14}\,z_{23}|^4} \right. \nonumber\\ 
&& \hspace{-1cm}\left. - {1 \over (z_{12}\,z_{34}\,\overline z_{13} \,\overline z_{24})^2} - {1 \over (z_{13}\,z_{24}\,\overline z_{14} \,\overline z_{23})^2} - {1 \over (z_{14}\,z_{23}\,\overline z_{12} \,\overline z_{34})^2} 
+ {\rm c.c.}\right\} + \ldots 
\label{latt4pt}
\eea
and is much more instructive: it is precisely the expression we obtain for the 5-point CFT correlation function $\langle h_1(z_1) \, h_1(z_2) \, h_1(z_3) \, h_1(z_4) \, \omega(\infty) \rangle$ if we assume that the height 1 field $h_1(z)$ is a primary field of dimensions $(h,\overline h)=(1,1)$ in a CFT with central charge $c=-2$, and that it satisfies a certain degeneracy condition at level 2. This last condition furnishes a differential equation \cite{DFMS97}, from which the correlator can be fully determined, the result being exactly the function (\ref{latt4pt}) ! From this result, one can actually infer that the previous correlator $\langle h_1(z_1) \, h_1(z_2) \, h_1(z_3) \, \omega(\infty) \rangle$ must vanish if it is to be symmetrical in the three insertion points \cite{PR17}.

The lattice 3-point correlators $\sigma_{a,1,1}(i_1,i_2,i_3)$, $a \ge 2$, have been computed more recently in \cite{PR17}. For simplicity, the three points were assumed to be aligned horizontally in the plane, with real separations $x_{ij}$. The following result was obtained to dominant order,
\be
\sigma_{a,1,1}(i_1,i_2,i_3) = \alpha_a \, \frac{\P_1^3}8 \: \frac{1}{x_{21}^3 x_{31}^3} + \ldots, \qquad (a \ge 2),
\label{a11}
\ee
where the coefficients $\alpha_a$ are those given in (\ref{ha}). In addition to being very simple, this expression is  surprisingly non-logarithmic. Because its scaling limit should be given by $\langle h_a(z_1) \, h_1(z_2) \, h_1(z_3)$ $\, \omega(\infty) \rangle$, it is particularly important to understand it from the CFT point of view. Indeed the computation of such a correlator requires additional information on the height fields $h_{a\ge 2}$ as logarithmic partners of $h_1$. Since $h_3$ and $h_4$ can be regarded as linear combinations of $h_1$ and $h_2$, it is sufficient to consider $h_2$.

Inspired by the conformal representations appearing in the bosonic sector of the symplectic theory \cite{GK99}, the following proposal has been made in \cite{JPR06} regarding the conformal nature of $h_2(z,\overline z)$.  A more complete account will be presented in Section \ref{sec8}.

The field $h_2(z,\overline z)$ is not primary since it transforms into $h_1(z,\overline z)$ under dilations. It is also not quasi-primary because its $L_1$ and $\overline L_1$ transforms generate two new fields, $\rho(z,\overline z)$ and $\overline \rho(z,\overline z)$ respectively, with weights $(0,1)$ and $(1,0)$. Moreover the field $\rho(z,\overline z)$ is left primary and its $\overline L_1$ transform is equal to $\kappa {\mathbb I}$; likewise $\overline \rho(z,\overline z)$ is right primary and its $L_1$ transform is also equal to $\kappa {\mathbb I}$. All this results in the following transformation law of $h_2(z,\overline z)$ under a general conformal transformation $z \to w(z)$ and $\overline z \to \overline w(\overline z)$,
\bea
h_2(z,\overline z) \egal \Big|\frac{{\rm d}w}{{\rm d}z}\Big|^2 \, \Big[h_2(w,\overline w) + \log\Big|\frac{{\rm d}w}{{\rm d}z}\Big|^2 \: h_1(w,\overline w) \Big] + \frac12 \Big(\frac{{\rm d}^2 w}{{\rm d}z^2}\Big/\frac{{\rm d}w}{{\rm d}z}\Big) \, \frac{{\rm d}\overline w}{{\rm d}\overline z} \, \rho(w,\overline w) \nonumber\\
\noalign{\medskip}
&& \hspace{4mm} + \; \frac12 \frac{{\rm d}w}{{\rm d}z} \, \Big(\frac{{\rm d}^2 \overline w}{{\rm d}\overline z^2}\Big/\frac{{\rm d} \overline w}{{\rm d}\overline z}\Big)\, \overline\rho(w,\overline w) + \frac{\kappa}4 \Big(\frac{{\rm d}^2 w}{{\rm d}z^2}\Big/\frac{{\rm d}w}{{\rm d}z}\Big)\Big(\frac{{\rm d}^2 \overline w}{{\rm d}\overline z^2}\Big/\frac{{\rm d} \overline w}{{\rm d}\overline z}\Big), \quad \kappa = - \frac{\P_1}4.
\label{finitepsi}
\eea

This conformal transformation law of $h_2$ is sufficient to compute correlators involving $h_2$, but substantially complicates the calculations. Using this transformation and the left and right level 2 degeneracy of $h_1(z,\overline z)$, the required correlator can be nonetheless determined. The result reads \cite{PR17}
\be
\langle h_2(z_1) \, h_1(z_2) \, h_1(z_3) \, \omega(\infty) \rangle = \frac{\P_1^3}{16}  \, \frac1{|z_{12}z_{13}|^2} \, \Big[ \frac1{z_{13}\overline z_{12}} + \frac1{z_{12}\overline z_{13}} \Big].
\ee
When the tree points $z_i$ are aligned horizontally, it exactly reproduces the lattice result (\ref{a11}) for $a=2$. The cases $a=3,4$ follow by multiplying by the proper coefficient $\alpha_a$ since $\langle h_1(z_1) \, h_1(z_2) \, h_1(z_3) $ $\omega(\infty) \rangle=0$.

\subsection{Minimal height cluster correlations}
\label{sec5.6}

The calculation of occurrence probabilities of minimal subconfigurations have been briefly discussed in Section \ref{sec5.2}. Their correlations can be computed very much like those of heights 1 by using a defect matrix. The calculation of mixed 2-point correlators for about a dozen different minimal subconfigurations has been reported in \cite{MR01}. It turns out that each such cluster $S$ can be specified by a triplet $(a,b_1,b_2)$ of real numbers.

We define as before subtracted variables 
\be
h_S(i) = \delta_{S(i)} - \P_S,
\ee
where $\delta_{S(i)}$ denotes the event ``a minimal subconfiguration $S$ is found around site $i$'', and $\P_S(i) = \E\big[\delta_{S(i)}\big]$ is the probability of such an event. The mixed correlator of two such variables takes the form
\be
\sigma_{S,S'}(i_1,i_2) = \E\big[h_S(i_1)\,h_{S'}(i_2) \big] = -\frac1{2r^4} \Big\{aa' + (b^{}_1b'_1 - b^{}_2b'_2) \cos{4\varphi} \Big\} + \ldots
\label{SSp}
\ee
We see that the dominant contribution retains an angular dependence, which in this case is not surprising since the minimal clusters  are generally not rotationally invariant. As a matter of illustration, the cluster reduced to a single height 1 is characterized by the triplet $(a,b_1,b_2) = (\P_1,0,0)$ while for the second one in (\ref{clusters}), one has
\be
S = \hspace{-2mm} 
\begin{pspicture}[shift=0](-0.5,0.2)(0.5,0.8)
\psset{xunit=0.23cm}
\psset{yunit=0.23cm}
\psset{runit=0.23cm}
\rput(0,0){\ball2}
\rput(3,0){\ball2}
\rput(3,3){\ball1}
\end{pspicture}
\hspace{.6truecm} :\; a = \frac{(4-\pi)(72-19\pi)(3\pi-8)^2}{18\pi^5},\; b_1 = 0,\; b_2 = -\frac{(4-\pi)(32-9\pi)(3\pi-8)^3}{18\pi^6}.
\ee

In the scaling limit, the subtracted cluster variables give rise to the fields $h_S(z)$, whose mixed correlators are given by the terms displayed in (\ref{SSp}). Interestingly it has been observed \cite{MR01} that these fields have a realization in terms of the symplectic free fermions discussed at the end of Section \ref{sec4}. Indeed one may check that the explicit fields given by
\be
h_S(z,\overline z) = - \Big\{ a \big(\partial \theta \, \overline\partial \tilde \theta + \overline \partial \theta \, \partial \tilde \theta\big) + (b_1 + {\rm i} b_2) \, \partial \theta \, \partial \tilde \theta + (b_1 - {\rm i} b_2)  \, \overline \partial \theta \, \overline\partial \tilde \theta \Big\},
\label{real}
\ee
reproduce the above 2-point correlators, as well as the higher order correlators computed in \cite{MR01}, provided the dissipation field $\omega$, proportional to $\theta \tilde\theta$, is inserted in the correlators, as explained earlier. In particular, the height 1 field $h_1(z,\overline z)$ is recovered upon setting $a=\P_1$ and $b_1=b_2=0$. Let us note a generic field $h_S(z,\overline z)$ is a linear combination of three fields with different conformal weights, namely $(1,1)$, $(2,0)$ and $(0,2)$, and with therefore different conformal transformations; the last two are responsible for inducing an angular dependence in the correlators. The field realization (\ref{real}) has been proved in a much greater generality in \cite{Je05a}: any lattice observable based on a conservative local bond modification \footnote{The qualifier `conservative' means that the defect matrix that implements the bond modifications has zero row and column sums. The defect matrix used in Section \ref{5.1} to compute the height 1 probability does not have this property. It can however be replaced by another one that does have it \cite{MR01}. An example of a non-conservative bond modification is given in Section \ref{sec5.7}.} converges in the scaling limit to a field of the form (\ref{real}).

On general grounds, this should not be surprising. On the one hand, the multisite probabilities for minimal clusters can be computed by using defect matrices which implement the local bond modifications. On the other hand, defect matrices always yield contributions that are given by finite determinants of discrete Green matrix entries. In the limit of large separations, the determinants converge to polynomial expressions in the Green function and its derivatives. It is therefore not a complete surprise that the associated fields can be constructed out from the symplectic free fermions $\theta,\tilde\theta$. Indeed, because the 2-point correlators of $\theta,\tilde\theta$ are given by the Green function, the correlators of any fields that are local in $\theta,\tilde\theta$ and their derivatives are necessarily polynomials in the Green function and its derivatives (Wick's theorem). We expect this observation to extend to all the observables that correspond to local perturbation of the toppling matrix. Isolated dissipation and minimal cluster variables are among them; the arrow variables discussed in the next section are in this class too. These general remarks apply to the massive extension of the sandpile model, see Section \ref{7.1}.

What about the height 2, 3 and 4 variables ? Can they also be accomodated in the free symplectic fermion theory ? As explained earlier, these three variables cannot be handled with finite rank perturbations of the toppling matrix, because they involve non-local constraints on the nearest neighbours (some of them should not be predecessors). Using the technique developed in \cite{KW15}, the 1-site probabilities $\P_{a \ge 2}$ can be efficiently computed. Surprisingly, the details show that the explicit values are given in terms of a few entries of the lattice Green matrix (at short distances), which explains why the values of $\P_{a\ge 2}$ quoted in (\ref{234}) are not much more complicated than for $\P_1$. This is no longer the case for large distance correlations $\sigma_{a\ge 2,1}(i_1,i_2)$. The analysis \cite{PR17} shows that those correlators are expressed in terms of sums of product of Green matrix entries over a path connecting $i_1$ to $i_2$, and thus in terms of quantities that are not local in the Green matrix. It supports the view that the height fields $h_{a \ge 2}$ do not belong to the free symplectic fermion theory. A detailed analysis of this question has been carried out in \cite{JPR06}, and has reached the same conclusion. Section \ref{sec8} below summarizes this  somewhat strange situation.

\subsection{Spanning tree related variables}
\label{sec5.7}

A recurrent configuration of the sandpile model can be specified as a set of height values or as a spanning tree; the former has the local heights as natural variables, the latter has local connectivities as natural variables, namely the existence or absence of specific bonds in the spanning tree.

We recall that a spanning tree is a connected subgraph with no loop which contains all vertices, including the sink. The latter is chosen to be the root of the tree, implying that there is a unique path connecting any vertex to the root, and therefore any vertex to any other vertex. A rooted spanning tree can then naturally be oriented by deciding that the edges of the tree all point towards the root. As a consequence, in any rooted spanning tree, there is exactly one outgoing edge at each vertex but the root; there may however be more (or less) ingoing edges (a vertex with no ingoing edge is a leaf). A site $j$ is a then a predecessor of $i$ if the unique path from $j$ to $i$ is consistently oriented (equivalently if the unique path form $j$ to $i$ does not pass through the root). As we have seen, the question of being predecessor is a non-local problem, even if $i$ and $j$ are close to each other, even nearest neighbours \cite{PP11}.

Connectivities between neighbouring sites can be handled in much the same way as heights 1 or minimal height cluster variables. To see this, we must first understand why the determinant of the toppling matrix on a graph counts the number of spanning trees on that graph. In the perspective of this section, we can generalize the matrix by assigning arbitrary weights to the oriented edges of the graph $\Gamma^\star=(V^\star,E^\star)$. We define $x_{ij}$ as the weight carried by the edge from vertex $i$ to vertex $j$ ($x_{ij}=0$ means that there is no edge from $i$ to $j$), and we set, for $i,j \in V$,
\be
\Delta_{i,j} = \begin{cases}
\; y_i = x_{i\star} + \sum_{j\neq i}\, x_{ij} & {\rm for\ } i=j,\\
-x_{ij} & {\rm for\ } i \neq j,
\end{cases}
\ee
In the context of the sandpile model, the difference $x_{i\star} = y_i - \sum_{j\neq i}\, x_{ij}$ can be viewed as the weight of the oriented edge from $i$ to the sink $\star$, so that the conservative vertices have this difference equal to 0 (no connection to the root). 

If $N = |V|$ is the number of vertices in the graph, let us write the determinant of $\Delta$ as a sum over the permutations $\sigma$ of the symmetric group $S_N$, which we partition according to the number $k$ of proper cycles they contain, that is, the cycles of length strictly larger than 1,
\be
\det \Delta = \sum_{\sigma \in S_N} \, \varepsilon_\sigma \, \Delta_{1,\sigma(1)} \Delta_{2,\sigma(2)} \ldots \Delta_{N,\sigma(N)} = \sum_{k=0}^{[N/2]} (-1)^k \sum_{\sigma{\rm\ has\ }k \atop {\rm proper\ cycles}}\, \Delta^+_{1,\sigma(1)} \ldots \Delta^+_{N,\sigma(N)}
\ee
where the matrix $\Delta^+$ is $\Delta$ without the minus signs in the non-diagonal part. The second equality follows by combining the signs in the non-diagonal entries of $\Delta$ with the parity of $\sigma$: every cycle of length $\ell \ge 2$ in a permutation $\sigma$ brings a sign $(-1)^{\ell-1}$ coming from the parity $\varepsilon_\sigma$ and another sign $(-1)^{\ell}$ from the product of non-diagonal entries of $\Delta$, resulting in an overall sign $-1$ per proper cycle. A cycle of length $\ell=1$, corresponding to a vertex left invariant by $\sigma$, brings no sign.

The term $k=0$ is simply equal to $\prod_i y_i$, as the only permutation with no proper cycle is the identity. In combinatorial terms, the product $\prod_i y_i$ is the weighted sum over all configurations of $N$ arrows, where each vertex has exactly one arrow pointing to one of the other vertices or to the root, each configuration being weighted by the product of the weights carried by the arrows. The generic term $k \neq 0$, apart for the sign $(-1)^k$, is a weighted sum of arrow configurations which contain at least $k$ oriented loops. Indeed the $k$ cycles contained in a fixed $\sigma$ give rise to $k$ loops, and the arrows attached to the vertices left invariant by $\sigma$ are unconstrained and possibly form more loops. 

By using the inclusion-exclusion principle, one can see that the above alternating sum has the effect to subtract from the term $k=0$ the weights of all the arrow configurations which contain at least one loop \cite{Pr94,IPR07}. Thus $\det \Delta$ is the sum over the oriented spanning trees on $\Gamma^\star$, each tree being weighted by the product of the weights of the oriented edges present in the tree (Kirchhoff's theorem). These oriented trees are also rooted spanning trees because one vertex at least must have its arrow oriented to the sink (a configuration of $N$ arrows on $N$ vertices necessarily contains a loop). When all weights $x_{ij}$ are equal to 0 or 1, $\det \Delta$ is simply the number of spanning trees.

Let us come back to the question of local connectivities on a rectangular grid in $\Z^2$ and compute the probability that the outgoing arrow from the site $i$ is oriented to its right neighbour. This amounts to compute the fraction of spanning trees with such an arrow, namely 
\be
\P_\to(i) = \P({\rm right\ arrow\ at\ }i) = \frac{\{\#{\rm \ trees\ with\ right\ arrow\ at\ }i\}}{\det \Delta},
\ee
where $\Delta$ is the discrete Laplacian. We take $i$ to be a conservative, non-boundary site.

According to the general discussion above, in order to force an arrow from $i$ to its right neighbour E, one could simply set to 0 the weights between $i$ and its three neighbours S, W and N. It is however computationally more efficient to set the weight from $i$ to its right neighbour to $x$, and take the limit $x \to +\infty$ so as to give the edges to the other three neighbours a relative weight equal to 0. This implies that we define a new matrix $\tilde \Delta$ which coincides with $\Delta$ except on two entries, namely $\tilde \Delta_{i,i+\hat e_1} = -x$ and $\tilde \Delta_{i,i} = x+3$. As before we write $\tilde \Delta = \Delta + B_\to(i)$ for the defect matrix $B_\to(i)$ which is zero everywhere except on the two sites $i,i+\hat e_1$, where it reduces to $\footnotesize \big({x-1 \atop 0}\;{1-x \atop 0}\big)$. Since $x$ is in any case large, we can simply set that part of $B_\to(i)$ to $\footnotesize \big({x \atop 0}\;{-x \atop 0}\big)$. From Kirchhoff's theorem, we obtain
\be
\P_\to(i) = \lim_{x \to +\infty} \: \frac 1x \: \det[{\mathbb I} + \Delta^{-1} B_\to(i)],
\ee
which reduces to a 2-by-2 determinant. In the infinite volume limit, one finds $\P_\to(i) = \frac 14$, as expected. If we want to have the arrow at $i$ oriented to its neighbour $j$ other then E, we use similar defect matrices $B_\uparrow,\, B_{\downarrow}, \, B_{\leftarrow}$, which look the same as $B_\to$ but with the non-zero 2-by-2 block $\footnotesize \big({x \atop 0}\;{-x \atop 0}\big)$ placed on the sites $i,j$. The four orientations of the arrow at $i$ yield the same result $\frac 14$.

Multipoint arrow probabilities can be computed in the now usual way, placing appropriate defect matrices at the different sites. For instance the probability to find a right arrow at two sites $i_1$ and $i_2$ is
\be
\sigma_{\to,\to}(i_1,i_2) = \lim_{x \to +\infty} \: \frac 1{x^2} \: \det\big[{\mathbb I} + \Delta^{-1} \{B_\to(i_1) + B_\to(i_2)\}\big].
\ee
In the infinite volume limit and for a large distance, the following two-point probabilities are found at dominant order,
\begin{subequations}
\bea
\sigma_{\to,\to}(i_1,i_2) - \frac 1{16} \egal \frac 1{16\pi^2} \: \frac{(z + \overline z)^2}{|z|^4} + \ldots \\
\sigma_{\uparrow,\uparrow}(i_1,i_2) - \frac 1{16} \egal - \frac 1{16\pi^2} \: \frac{(z - \overline z)^2}{|z|^4} + \ldots \\
\sigma_{\to,\uparrow}(i_1,i_2) - \frac 1{16} \egal - \frac {\rm i}{16\pi^2} \: \frac{z^2 - \overline z^2}{|z|^4} + \ldots
\eea
\end{subequations}

Looking for symplectic fermion realizations of fields $\rho_{\to}$ and $\rho_{\uparrow}$ which reproduce these two-point correlators, one quickly sees that these have to include two parts with respective weights $(1,0)$ and $(0,1)$, leading in a natural way to the following forms,
\be
\rho_{\to}(z,\overline z) = \frac1{2\pi} \big(\theta \, \partial \tilde \theta + \theta \, \overline\partial \tilde \theta\big), \qquad \rho_{\uparrow}(z,\overline z) = \frac{\rm i}{2\pi} \big(\theta \, \partial \tilde \theta - \theta \, \overline\partial \tilde \theta\big),
\ee
in agreement with the fact that $\rho_{\uparrow}(z,\overline z)$ is formally the rotated form of $\rho_{\to}(z,\overline z)$ (under $z \to -{\rm i}z$). The first two correlators are indeed related by rotation. This observation also suggests that the other two orientations are described by fields which are the opposite of the previous two, $\rho_{\leftarrow}(z,\overline z) = -\rho_{\to}(z,\overline z)$ and $\rho_{\downarrow}(z,\overline z) = -\rho_{\uparrow}(z,\overline z)$. Explicit calculations confirm it.

In a similar way, probabilities that edges belong to a random spanning tree, irrespective of their orientation, can be computed. The probability that a single, fixed edge belongs to a tree is the sum of the probabilities to find it in either of the two possible orientations, is thus equal to $\frac 12$.

Likewise the probability to find $m$ edges in a tree is the sum of the probabilities to find them in all possible orientations, and so is the sum of $2^m$ probabilities of $m$ oriented edges. That sum can however be obtained in one go by replacing the block $\footnotesize \big({x \atop 0}\;{-x \atop 0}\big)$ used for the oriented edges by $\footnotesize \big({x \atop -x}\;{-x \atop x}\big)$ and dividing as above the determinant by $x^m$. Because two arrows with opposite orientations can not occupy the same edge (as they would form a loop), the summation over the $2^m$ terms is correctly realized. 

Correlations of unoriented edges should decay faster than that of oriented edges, because the sum of the two orientations is zero, in view of the relations $\rho_{\leftarrow} = -\rho_{\to}$ and $\rho_{\downarrow} = -\rho_{\uparrow}$, at least at the order that was dominant for the oriented edges ($r^{-2}$). Indeed explicit calculations yield a $r^{-4}$ decay,
\begin{subequations}
\bea
\sigma_{\leftrightarrow,\leftrightarrow}(i_1,i_2) - \frac 1{4} \egal \sigma_{\updownarrow,\updownarrow}(i_1,i_2) - \frac 1{4} = -\frac1{16\pi^2} \: \frac{(z^2 + \overline z^2)^2}{|z|^8} + \ldots \\
\sigma_{\leftrightarrow,\updownarrow}(i_1,i_2) - \frac 1{4} \egal \frac1{16\pi^2} \: \frac{(z^2 - \overline z^2)^2}{|z|^8} + \ldots
\eea
\end{subequations}

From these correlators, the associated fields $\phi_\leftrightarrow$ and $\phi_\updownarrow$ must have components with conformal weights $(2,0)$, $(1,1)$ and $(0,2)$. One finds the same form as the fields associated to the minimal clusters, in agreement with a previous remark since the defect matrix $\footnotesize \big({x \atop -x}\;{-x \atop x}\big)$ is conservative. More precisely, the following fermionic expressions reproduce the above correlations,
\begin{subequations}
\bea
\phi_\leftrightarrow(z,\overline z) \egal \frac1{2\pi} \big(\partial \theta \, \overline\partial \tilde \theta + \overline \partial \theta \, \partial \tilde \theta + \partial \theta \, \partial \tilde \theta + \overline \partial \theta \, \overline\partial \tilde \theta \big) \:= \:(\partial + \overline\partial) \rho_{\to} - \frac 1{2\pi} \big(\theta \partial^2 \tilde\theta + \theta \, \overline\partial^2 \tilde\theta\big),\\
\phi_\updownarrow(z,\overline z) \egal \frac1{2\pi} \big(\partial \theta \, \overline\partial \tilde \theta + \overline \partial \theta \, \partial \tilde \theta - \partial \theta \, \partial \tilde \theta - \overline \partial \theta \, \overline\partial \tilde \theta \big) \: = \:{\rm i}(\partial - \overline\partial) \rho_{\uparrow} - \frac 1{2\pi} \big(\theta \partial^2 \tilde\theta + \theta \, \overline\partial^2 \tilde\theta\big).
\eea
\end{subequations}
In fact, given that an unoriented edge is a sum of two oriented edges with opposite orientation, or, from what we said above, a difference of two oriented edges with the same orientation, one would expect that the fields describing a horizontal resp. vertical unoriented edge are proportional to the horizontal resp. vertical derivative of the fields describing the oriented edges, namely $\phi_\leftrightarrow \sim (\partial + \overline\partial) \rho_{\to}$ and $\phi_\updownarrow \sim {\rm i}(\partial - \overline\partial) \rho_{\uparrow}$. It turns out not to be quite the case.

\section{Boundaries, boundary conditions and boundary variables}
\label{sec6}

Formulating the sandpile model on a surface with boundaries is important to see how they affect the statistics of the model. The multisite correlations discussed in the previous section are likely to be modified by the presence of a boundary, and by the associated boundary conditions. Moreover the microscopic variables on a boundary or very close to it will surely have a different behaviour from their bulk versions. In the field theoretic description, the boundary fields have to be properly identifed, and the way changes of boundary conditions are implemented must be clarified. All this adds to the known set of bulk fields a number of boundary related fields, and offers the opportunity to further test the consistency of their identification by computing mixed correlations combining both types of variables. 

Surfaces with boundaries arise in the thermodynamic limit when some of the boundaries of the finite system are not sent off to infinity, unlike the situation considered in the previous section. The simplest case is when only one boundary of the rectangular grid is kept at finite distance, leading to a domain converging to the upper-half plane $\H = \{(x,y) \in \R^2 : y \ge 0\}$. There is only one boundary to care about, and the invariance under horizontal translations is preserved. 

From our earlier discussion of conservative versus dissipative sites, we have already defined two possible boundary conditions: open and closed. Let us recall that the boundary condition is open resp. closed if the boundary sites are dissipative resp. conservative. As before, a boundary open site has $z^\star_i = 4$ while a boundary closed site has $z^\star_i = z_i = 3$. In the scaling limit, it endows $\H$ with a homogeneous boundary condition, open or closed. We also can (and will) consider inhomogeneous boundary conditions, by alternating open and closed stretches on a single boundary. 

In terms of height variables, the open boundary condition is equivalent \footnote{Indeed the burning algorithm implies that the boundary sites, all with a height equal to 4, will burn at the first step of the burning process. The sites on the next layer all have $z^\star_i = 4$, corresponding to the open condition.} to fix all the boundary heights to 4, whereas the closed condition amounts to constrain the boundary heights not to take the value 4. The fixed boundary condition with boundary heights equal to 1 is not possible (two neighbouring 1's form a forbidden subconfiguration); the fixed boundary conditions with boundary heights equal to 2 and/or 3 should be possible but seem to be difficult to handle in practice. 

Two more boundary conditions, defined in the spanning tree description and previously called windy boundary conditions will be discussed at the end of this section (as we will see, they are not so far from the possibility just mentioned, namely that of having height variables being equal to 2 or 3). No other boundary condition has been considered so far though it would be very surprising that no other exist \footnote{Of course we talk here of no other universality class of boundary conditions. Many boundary conditions may differ in the way they are microscopically defined on the lattice and nevertheless renormalize to the same continuum boundary condition in the scaling limit.}.

\subsection{Bulk variables with homogeneous open or closed boundary}
\label{sec6.1}

In this section, we would like to reconsider the multisite height probabilities, but on a domain with a boundary, the upper-half plane (UHP) being the simplest case. The principle underlying the calculations on the UHP stays the same as on the full plane. The most essential difference is that the toppling matrix becomes in the thermodynamic limit the Laplacian matrix on the discrete UHP with the appropriate boundary condition, open or closed. In this section, we consider homogeneous boundary conditions only.

To be specific, we choose the boundary row of sites to be located on the horizontal line $y=1$, so that the discrete UHP we consider is $\{(x,y) \in \Z^2 \;|\; y \ge 1\}$. For either boundary condition, the Laplacian matrix $\Delta^{\rm op}$ or $\Delta^{\rm cl}$ is minus the adjacency matrix of the discrete UHP plus a diagonal matrix, everywhere equal to 4 for the open condition, equal to 4 and 3 respectively for the bulk and boundary sites for the closed condition. By the method of images, the Green matrices $G^{\rm op/cl} = (\Delta^{\rm op/cl})^{-1}$ can be easily computed in terms of that on the full plane,
\begin{subequations}
\bea
G^{\rm op}_{(x_1,y_1),(x_2,y_2)} \egal G_{(x_1,y_1),(x_2,y_2)} - G_{(x_1,y_1),(x_2,-y_2)}, \label{dir} \\
G^{\rm cl}_{(x_1,y_1),(x_2,y_2)} \egal G_{(x_1,y_1),(x_2,y_2)} + G_{(x_1,y_1),(x_2,1-y_2)}, \label{neu}
\eea
\end{subequations}
for $y_1,y_2 > 0$. As anticipated, we verify that $G^{\rm op}$ satisfies the Dirichlet condition, namely it is odd under the reflection through the line $y=0$ and therefore vanishes on it, and that $G^{\rm cl}$ satisfies the Neumann condition, namely it is even under the reflection through the line $y=\frac 12$, inducing a vanishing normal derivative in the scaling limit. The calculations of the previous section can then be generalized to the UHP geometry by using these Green matrices. For the height 1 and for the cluster variables, one merely has to use the appropriate Green matrix. For higher heights, the presence of a boundary makes the calculations more complicated because the combinatorics involved is heavier.

\begin{table}[t]
\centering
\renewcommand{\arraystretch}{1.8}
\tabcolsep11pt
\begin{tabular}{|c||cccc|}
\hline
 & $a=1$ & $a=2$ & $a=3$ & $a=4$  \\
\hline\hline
$c_a$ & $\frac{\P_1}4 = \frac{\pi-2}{2\pi^3}$ & $\frac{\pi-2}{2\pi^3} \, \widetilde\gamma + \frac{34-11\pi}{8\pi^3}$ & $\frac{8-\pi}{4\pi^3} \, \widetilde\gamma + \frac{-88+5\pi+2\pi^2}{16\pi^3}$ & $-\frac{\pi+4}{4\pi^3} \, \widetilde\gamma + \frac{36+9\pi-2\pi^2}{16\pi^3}$ \\
\hline
$d_a$ & 0 & $\frac{\P_1}4 = \frac{\pi-2}{2\pi^3}$ & $\frac{8-\pi}{4\pi^3}$ & $-\frac{\pi+4}{4\pi^3}$ \\
\hline
\end{tabular}
\caption{Numerical coefficients for one-site height probabilities on the UHP, with $\widetilde \gamma = \gamma + \frac 52 \log{2}$. They satisfy the relations $\sum_a c_a = \sum_a d_a = 0$.}
\label{tab1}
\end{table}

The simplest case is the 1-site height probability $\P_1^{\rm op/cl}(y)$ to find a height equal to 1 at a distance $y$ from the boundary. It can be computed by using the formula (\ref{p1}) where $\Delta^{-1}_\Gamma$ is replaced, in the infinite volume limit, by one of the two Green matrices given above. This was historically the first calculations of boundary effects in sandpile models \cite{BIP93}, with the result
\be
\sigma_1^{\rm op}(y) = \P_1^{\rm op}(y) - \P_1 = \frac{\P_1}{4y^2} + \ldots, \qquad \sigma_1^{\rm cl}(y) = \P_1^{\rm cl}(y) - \P_1  = -\frac{\P_1}{4y^2} + \ldots
\label{s1}
\ee

The analogous results for higher heights were obtained somewhat later \cite{PR05a,JPR06} and were the first to firmly establish their logarithmic nature. They take the following form, valid for $a \ge 1$,
\be
\sigma_a^{\rm op}(y) = \frac{1}{y^2}\Big( c_a + \frac{d_a}2 + d_a \log{y} \Big) + \ldots, \qquad \sigma_a^{\rm cl}(y) = -\frac{1}{y^2}\Big( c_a + d_a \log{y} \Big) + \ldots
\label{sa}
\ee
up to terms of order ${\mathcal O}(y^{-4}\log{y})$, which have since been explicitly computed \cite{PR17}, as they enter the calculations of $\sigma_{a1}^{\rm op/cl}$ given below. The coefficients are explicitly known and are collected in Table \ref{tab1}. One may check that the relations (\ref{ha}) expressing $h_3$ and $h_4$ linearly in terms of $h_1,h_2$ are confirmed. The distinctive change of sign between the two boundary conditions, the fact that for fixed $a$, both are controlled by the same constants $c_a$ and $d_a$ and the equality $d_2=c_1$ are striking. As will be explained below and in one of the next sections, all three features will follow from the CFT picture. 

Let us mention that these lattice calculations have been carried out on another lattice realization of the UHP, namely on the diagonal upper-half plane $\{(x,y) \in \Z^2 \;|\; y > x\}$, for which the method of images allows to explicitly compute the Green matrices for the two boundary conditions. As expected, the dominant terms are exactly the same as above in terms of the Euclidean distance between the height 1 and the diagonal boundary, while the subdominant terms are different \cite{PR17}.

The 2-site height correlators in the bulk of the UHP, at sites $i_1=(x_1,y_1)$ and $i_2=(x_2,y_2)$, and which involve the same subtractions as before,
\be
\sigma_{a,1}^{\rm op/cl}(x;y_1,y_2) = \P^{\rm op/cl}_{a,1}(i_1,i_2) - \P^{\rm op/cl}_a(i_1) \, \P^{}_1 - \P^{}_a \, \P^{\rm op/cl}_1(i_2) + \P^{}_a \,\P^{}_1,
\ee
have been computed in \cite{PR17} when the two sites are far from the boundary and far from each other, again using the technique developped in \cite{KW15}. They depend on three real variables, the horizontal distance $x=x_1-x_2$ between the two sites and their vertical positions $y_1,y_2$. For simplicity however, the lattice calculations have been carried out for two vertically aligned sites, that is, for $x=0$.

Defining the two bivariate functions,
\be
P(u,v) = \frac1{8u^2v^2} - \frac 1{(u-v)^4} - \frac 1{(u+v)^4}\,, \qquad Q(u,v) = \frac 1{(u-v)^4} - \frac 1{(u+v)^4},
\ee
the results for $a=1,2$ take the following form, at dominant order,
\begin{subequations}
\bea
&& \hspace{-7mm} \sigma_{1,1}^{\rm op}(0;y_1,y_2) = \sigma_{1,1}^{\rm cl}(0;y_1,y_2) = \frac{\P_1^2}2 \, P(y_1,y_2) + \ldots, \\
&& \hspace{-7mm} \sigma_{2,1}^{\rm op/cl}(0;y_1,y_2) = \frac{\P_1^2}2 \Big[P(y_1,y_2) \Big(\log y_1+\gamma+\frac{5}{2}\log 2\Big) + Q(y_1,y_2)\log\Big|\frac{y_2+y_1}{y_2-y_1}\Big|\Big] \nonumber\\
&& \hspace{4cm} + \:\frac{H^{\rm op/cl}(y_1^2,y_2^2)}{y_1^2 \, y_2^2 \, (y_1^2-y_2^2)^4} +\ldots,
\eea
\end{subequations}
where $H^{\rm op}(u,v)$ and $H^{\rm cl}(u,v)$ are homogeneous polynomials of degree 4 in $u,v$, with explicitly known coefficients. The results for $a=3,4$ take the same form with different coefficients and confirm once more the linear relations (\ref{ha}).

Let us discuss these results in the CFT picture, using what we already know about the height fields $h_a(z,\overline z)$. For a homogeneous boundary condition like here, boundary CFT prescribes to compute bulk correlators on the UHP by viewing a field $\phi(z,\overline z)$ with conformal weights $(h,\overline h)$ as the product $\phi_{h^{}}(z) \phi_{\overline h}(\overline z)$ of two chiral fields of weight $h$ and $\overline h$ respectively, located at the points $z$ and $\overline z$ (the latter being in the lower-half plane) \cite{Ca84}. A correlation function of $n$ bulk (non-chiral) fields on the UHP can then be computed as a correlation of $2n$ chiral fields on the full plane; the correlation appropriate for the boundary condition under consideration is accordingly selected in the solution space of these $2n$-correlators. 

The above prescription must however be adapted in the case of logarithmic fields because the chiral factorization is not consistent with the non-diagonal action of $L_0$. Indeed let us consider a logarithmic pair $(\Phi(z,\overline z),\Psi(z,\overline z))$. If we factorize the logarithmic partner as $\Psi(z,\overline z) = \psi_h^{\phantom{2}}(z) \psi_{\overline h}(\overline z)$, we find from the action of $L_0$, 
\be
L_0 \Psi = (L_0 \psi_h^{\phantom{2}}) \psi_{\overline h} = (h \psi_h^{\phantom{2}} + \lambda \, \phi_h^{\phantom{2}}) \psi_{\overline h} = h \Psi + \lambda \, \phi_h^{\phantom{2}} \psi_{\overline h},
\ee
that the chiral factorization of the primary partner is $\Phi = \phi_h^{\phantom{2}} \psi_{\overline h}$. The same argument with $\overline L_0$ shows that an equally good factorization is $\Phi = \psi_h^{\phantom{2}} \phi_{\overline h}$.

Let us first see how this works for $\sigma_a^{\rm op}(y)$. Their dominant terms, made explicit in (\ref{s1}) and (\ref{sa}), should correspond to $\langle h_a(z,\overline z) \rangle_{\rm op}$. Note that, unlike the correlations on the plane discussed in Section \ref{sec5}, we do {\it not} insert dissipation at infinity since the whole boundary is open and therefore dissipative. From the prescription recalled above, these 1-point functions should have the general form of chiral 2-point functions. If $\psi$ and $\phi$ denote chiral versions of the height 2 and height 1 fields respectively, with $h = \overline h = 1$, the chiral factorizations of the height fields $h_1$ and $h_2$ read $h_1(z,\overline z) = \psi(z) \phi(\overline z)$ and $h_2(z,\overline z) = \psi(z) \psi(\overline z)$. The CFT formalism gives the general forms (\ref{2ptlogch}),
\be
\langle h_1(z,\overline z) \rangle_{\rm op} = \langle \phi(z) \psi(\overline z) \rangle = \frac B{(z-\overline z)^2}, \quad
\langle h_2(z,\overline z) \rangle_{\rm op} = \langle \psi(z) \psi(\overline z) \rangle = \frac{C - 2 \lambda B \log{(z - \overline z)}}{(z - \overline z)^{2}}.
\ee
With the value $\lambda = -\frac 12$ noted in Section \ref{sec5}, these forms exactly reproduce the lattice results (\ref{sa}), including the relation $d_2 = c_1 = B$.

For $\sigma_a^{\rm cl}(y)$ and since the closed boundary is not dissipative, we insert by hand dissipation at infinity, so that $\sigma_a^{\rm op}(y)$ should be given by $\langle h_a(z,\overline z) \omega(\infty) \rangle_{\rm cl}$. Using the same chiral factorization as above leads to a 3-point function. The selection of a physically sensible solution leads to the same general form as for the open boundary condition \cite{JPR06}.

The conformal calculations required to account for $\sigma_{a,1}^{\rm op}$ are only technically more involved. The needed chiral correlators are $\langle \phi(z_1)\psi(\overline z_1) \phi(z_2) \psi(\overline z_2) \rangle$ for $a=1$ and $\langle \psi(z_1)\psi(\overline z_1) \phi(z_2) \psi(\overline z_2) \rangle$ for $a=2$. Both can be computed from the primary nature of the chiral field $\phi$, as established in Section \ref{sec5}, by solving a second order differential equation and selecting the appropriate solution. As the details are given in \cite{PR17}, we merely give the results, valid for any relative positions of the two heights, $z_1=(x_1,y_1)$ and $z_2=(x_2,y_2)$,
\be
\langle h_1(z_1,\overline z_1) h_1(z_2,\overline z_2) \rangle_{\rm op} = \frac{\P_1^2}2 \Big\{\frac2{(z_1-\overline z_1)^2(z_2-\overline z_2)^2} - \frac1{|z_1-z_2|^4} - \frac1{|z_1-\overline z_2|^4} \Big\},
\ee
and
\bea
\langle h_2(z_1,\overline z_1) h_1(z_2,\overline z_2) \rangle_{\rm op} \egal \frac{\P_1}{32y_1^2y_2^2} \, \frac{t^4-2t^3+4t-2}{(1-t)^2} \, \Bigg[ \frac{3(3\pi-10)}{2\pi^3} - \P_1 \Big(\log{y_1} + \gamma + \frac52 \log{2}\Big) \Bigg] \nonumber\\
&& \hspace{-3cm} + \: \frac{\P_1^2}{64y_1^2y_2^2} \, \Bigg[\frac{t^3(t-2)}{(1-t)^2} \, \Big(\log{(1-t)} + \frac{y_1}{2y_2}\Big) - \frac{t^2}{1-t} \Bigg], \qquad t = -\frac{4y_1y_2}{|z_1-z_2|^2}.
\label{h2h1}
\eea
One can check that setting $x_1=x_2$ exactly reproduces the lattice results $\sigma^{\rm op}_{1,1}(0;y_1,y_2)$ and $\sigma^{\rm op}_{2,1}(0;y_1,y_2)$ reported above. 

The analogous calculation for the closed boundary has been carried in the case $a=1$, yielding the same expression as for the open boundary. No calculation however has been successful for $a=2$ as it involves a non-trivial 5-point chiral correlator (in this case the dissipation field $\omega$ must be added). 

Similar calculations with isolated bulk dissipation instead of height variables have been considered in \cite{PR04}; it was found in all cases that the CFT predictions compare successfully with the lattice results. 
 
\subsection{Changing the boundary condition}
\label{sec6.2}

We have considered so far two different boundary conditions, the open and closed conditions. This allows to address a fundamentally new issue, namely how to think of a change of boundary condition, both on the lattice and in the emerging field theory. Like in the previous section, we consider the UHP.

We have seen that the calculation of correlations on the UHP, of height or dissipation variables, involves the use of the appropriate Laplacian (toppling) matrix and its inverse. On the lattice, the way we can change the boundary condition at a boundary site $i$ is thus fairly clear: since an open boundary site has $\Delta_{i,i} = z_i^\star = 4$ and a closed one has $\Delta_{i,i} = z_i^\star = 3$,  we simply lower by 1 the diagonal entry $\Delta_{i,i}$ to close an open site, and we increase it by 1 to open a closed site (as we did in Section \ref{sec5.4} to introduce dissipation at bulk sites). We do it either way for $n$ consecutive boundary sites to change the boundary condition on a interval $I$ of length $n$, that is, we do the following change on the toppling matrix $\Delta \to \Delta \pm D_I$, where $D_I$ implements the diagonal shifts described above.

Let us examine the effect of closing $n$ consecutive sites in an otherwise open boundary. We decide to measure this effect as in Section \ref{sec5.4}, namely by comparing the number of recurrent configurations before and after the closing of $n$ sites. So we want to compute the ratio $Z_{\rm op}(n)/Z_{\rm op}$. At finite volume, the two partition functions can be computed as determinants of the corresponding toppling matrices on rectangular grids, with say four open boundaries in the case of $Z_{\rm op}$, and with $n$ closed sites inserted on the lower boundary for $Z_{\rm op}(n)$. As usually, we can readily write the infinite volume limit of the ratio as
\be
\frac{Z_{\rm op}(n)}{Z_{\rm op}} = \frac{\det \Delta^{\rm op}(n)}{\det \Delta^{\rm op}} = \frac{\det [\Delta^{\rm op} - D_{I_n}]}{\det \Delta^{\rm op}} = \det [{\mathbb I} - G^{\rm op} D_{I_n}] = \det [{\mathbb I} - G^{\rm op}]_{i,j \in I_n},
\label{59}
\ee
where $(D_{I_n})_{i,j} = \delta_{i,j}$ for $i,j \in I_n$, is zero elsewhere, and $I_n = \{(\ell,1) \;:\; 1 \le \ell \le n\}$ is the set of sites being closed. Using the relation (\ref{dir}) expressing $G^{\rm op}$ in terms of the Green matrix on the full plane $\Z^2$, the matrix in the determinant reads
\be
\Big({\mathbb I} - G^{\rm op}\Big)_{i,j \in I} = \Big(\delta_{\ell,\ell'} - G_{(\ell,1),(\ell',1)} + G_{(\ell,1),(\ell',-1)}\Big)_{1 \le \ell,\ell' \le n}.
\ee

By the horizontal translation invariance of $G^{\rm op}$, this is a Toeplitz matrix of the form $a_{\ell - \ell'}$. Using standard results on the Green matrix on the plane, one finds that the entries $a_m$ are the Fourier coefficients of the following symbol, which has a so-called Fisher-Hartwig singularity,
\be
\sigma^{\rm op}(k) = \sqrt{1 - \cos k} \cdot \big\{ \sqrt{3 - \cos k} - \sqrt{1 - \cos k}\big\}.
\ee
For large $n$, the asymptotics of such Toeplitz determinants is well-known (see f.i. \cite{DIK13}), and leads to the following result \cite{Ru02},
\be
\frac{Z_{\rm op}(n)}{Z_{\rm op}} \simeq A \, n^{1/4} \, {\rm e}^{-\frac{2{\rm G}}\pi n}, \qquad n \gg 1,
\label{op}
\ee   
with G$= 0.915965...$ the Calatan constant. The proportionality constant $A$ is explicitly known but is unimportant here.

What if we consider the opposite situation, in which we open $n$ consecutive sites of a closed boundary ? Reasoning as above, we quickly get the corresponding ratio,
\be
\frac{Z_{\rm cl}(n)}{Z_{\rm cl}} = \frac{\det \Delta^{\rm cl}(n)}{\det \Delta^{\rm cl}} = \frac{\det [\Delta^{\rm cl} + D_{I_n}]}{\det \Delta^{\rm cl}} = \det [{\mathbb I} + G^{\rm cl}]_{i,j \in I_n},
\ee
which is also a Toeplitz determinant. However this one is infinite --each entry is infinite-- for the same reason we have pointed out in Section \ref{sec5.4}. Adopting the same point of view, we similarly evaluate the effect of opening $n$ sites with respect to the situation where only one site is open. One therefore considers instead the ratio $\frac{Z_{\rm cl}(n)}{Z_{\rm cl}(1)}$, which one can write as 
\be
\frac{Z_{\rm cl}(n)}{Z_{\rm cl}(1)} = \frac 1{b_0} \det \big(b_{\ell - \ell'} \big)_{1 \le \ell,\ell' \le n},
\label{ncl}
\ee
where the entries $b_m$ are the Fourier coefficients of a symbol $\sigma^{\rm cl}(k)$ given by
\be
\sigma^{\rm cl}(k) = \frac 12 (1 - \cos k)^\alpha \cdot \big\{ \sqrt{3 - \cos k} + \sqrt{1 - \cos k}\big\}, \qquad \alpha = -\frac 12.
\ee
Its Fourier coefficients are well-defined for $\alpha > -\frac 12$, diverge in the limit $\alpha \to -\frac 12$ but nonetheless keep the ratio (\ref{ncl}) finite. Remarkably, for large $n$, it takes the form \cite{Ru02}
\be
\frac{Z_{\rm cl}(n)}{Z_{\rm cl}(1)} \simeq A \, n^{1/4} \, {\rm e}^{\frac{2{\rm G}}\pi (n-1)}, \qquad n \gg 1,
\label{cl}
\ee
for the same constant $A$ as above.

Before discussing the CFT side, let us remark that the exponential factors in (\ref{op}) and (\ref{cl}) are expected. On a finite $N \times N$ grid, all four partition functions (numerators and denominators) are asymptotically dominated by the bulk free energy, given by ${\rm e}^{\frac{4{\rm G}}\pi N^2}$ as mentioned in Section \ref{asm}. These terms drop out in the ratios. The next correction is related to the boundary free energy $f_{\rm b}$ (per site) and takes the form ${\rm e}^{4N f_{\rm b}}$ in case the boundary condition b is the same at all boundary sites. For the partition functions considered above, the boundary conditions only differ on the lower edge of the grid, so that for large $N \gg n \gg 1$ the ratios are asymptotic to
\be
\frac{Z_{\rm op}(n)}{Z_{\rm op}} \simeq \frac{{\rm e}^{(N-n) f_{\rm op} + n f_{\rm cl}}}{{\rm e}^{N f_{\rm op}}} = {\rm e}^{-n (f_{\rm op} - f_{\rm cl})}\,, \qquad \frac{Z_{\rm cl}(n)}{Z_{\rm cl}(1)} \simeq \frac{{\rm e}^{(N-n) f_{\rm cl} + n f_{\rm op}}}{{\rm e}^{(N-1) f_{\rm cl} + f_{\rm op}}} = {\rm e}^{(n-1) (f_{\rm op} - f_{\rm cl})}.
\ee
The free energies $f_{\rm op}$ and $f_{\rm cl}$ represent (the logarithm of) the effective number of values taken by the boundary heights in the set of recurrent configurations. The number of possible values taken by the height at an open boundary site is 4, and is 3 at a closed site. If these numbers values get effectively reduced in the set of recurrent configurations, one should expect that the number of values at an open boundary site remains larger than that at a closed site, implying $f_{\rm op} - f_{\rm cl} > 0$. An explicit calculation \cite{Ru02} confirms this and yields $f_{\rm op} - f_{\rm cl} = \frac{2{\rm G}}\pi$, in agreement with the above results. To fix the ideas, the effective number of values taken by a boundary height is ${\rm e}^{f_{\rm op}}=3.70$ at an open site, and ${\rm e}^{f_{\rm cl}}=2.07$ at a closed site.

In the CFT approach, a change of boundary condition at $x$, from condition a to condition b, is implemented by the insertion in the correlators of a specific field $\phi^{\rm a,b}(x)$. Such boundary condition changing fields \footnote{In the correspondence between statistical system and field theory, the boundary condition changing fields are somehow special. They describe the effects of a change of boundary condition but are not associated to a lattice observable, unlike the height fields $h_a$ for instance.} are usually expected to be chiral primary fields, and satisfy $\phi^{\rm a,b}(x) = \phi^{\rm b,a}(x)$ when the boundary conditions a and b do not carry an intrinsic orientation (see Section \ref{sec6.4} for counterexamples). The insertion of the product $\phi^{\rm a,b}(x_1) \, \phi^{\rm b,a}(x_2)$ accounts for the change at $x_1$ from condition a to condition b, and then back from b to a at $x_2$, but does {\it not} account for the exponential terms related to the difference of boundary free energies of condition a versus condition b, namely the terms we have just discussed in the previous paragraph. These are clearly non-universal, i.e. depend on the specific model under consideration, and cannot be accounted for by the underlying CFT, which itself applies to all the models in the universality class to which the sandpile model belongs. 

It follows that the effect of changing the boundary condition given above, in which we omit the exponential terms, should correspond to the 2-point function $\langle \phi^{\rm op,cl}(0) \, \phi^{\rm cl,op}(n) \rangle = \langle \phi^{\rm cl,op}(0) \, \phi^{\rm op,cl}(n) \rangle$. The two are indeed equal on the lattice and asymptotic to $n^{\frac 14}$, and from this, we infer that the boundary condition changing field $\phi^{\rm op,cl}(x) = \phi^{\rm cl,op}(x)$ is a chiral conformal field of weight $h = -\frac 18$, with a correlator given by
\be
\langle \phi^{\rm op,cl}(x_1) \, \phi^{\rm cl,op}(x_2) \rangle_{\rm op} = A \, |x_1 - x_2|^{1/4}.
\label{clop}
\ee
For physical reasons, we might worry about having a correlator that actually increases with the distance, suggesting somehow the existence of a strange interaction that would get stronger at larger distances. There is nothing strange however, as it does not really correspond to the physical correlation of two observables. As said above, the field $\phi^{\rm op,cl}$ is expected to be primary. As usually, this conjecture can be put to the test: the consequences of this statement must have a match in the lattice properties of the model. 

One of the strongest consequence of the primary nature of $\phi^{\rm op,cl}$ and the assumed structure of the conformal module that contains it, is that any correlator where this field appears must satisfy a second-order partial differential equation \footnote{Again the technical assumption is that the field $\phi^{\rm op,cl}$ is degenerate at level 2, similarly to the height 1 field $h_1$, see Section \ref{sec5.5}.}, the precise form of which depends on the other fields involved. A first and simple test is to look at a 4-point function \footnote{The CFT is really defined on the UHP plus the point at infinity. The boundary must therefore be thought of as the real line plus the two points $\pm \infty$ identified, and forming a loop closing at infinity. Any change of boundary condition thus involves an even number of insertions of $\phi^{\rm op,cl}$. For instance $\langle \phi^{\rm op,cl}(0) \, \phi^{\rm cl,op}(\infty) \rangle$ changes the boundary condition from open to closed on the positive real axis.}, for instance $\langle \phi^{\rm op,cl}(x_1)\,\phi^{\rm cl,op}(x_2)\,\phi^{\rm op,cl}(x_3)\,\phi^{\rm cl,op}(x_4) \rangle$, which should describe the effect of closing the sites on two disjoint interval $[x_1,x_2]$ and $[x_3,x_4]$ in the otherwise open boundary of the UHP. Using the global conformal invariance, one can reduce the partial differential equation to a second-order ordinary differential equation. In the two-dimensional solution space, we select the only solution which reduces to the product $\langle \phi^{\rm op,cl}(x_1)\,\phi^{\rm cl,op}(x_2)\rangle \langle \phi^{\rm op,cl}(x_3)\,\phi^{\rm cl,op}(x_4) \rangle$ when the two intervals are infinitely distant. This unique solution solution reads (with $x_{ij}=x_i-x_j$)
\be
\langle \phi^{\rm op,cl}(x_1)\,\phi^{\rm cl,op}(x_2)\,\phi^{\rm op,cl}(x_3)\,\phi^{\rm cl,op}(x_4) \rangle_{\rm op} = \frac{2A^2}\pi (x_{12} \, x_{34})^{1/4} \, (1-t)^{1/4} \, K(t), \quad t \equiv \frac{x_{12}x_{34}}{x_{13}x_{24}},
\label{ellip}
\ee
where $K(t) = \int_0^{\frac \pi 2} \frac{{\rm d}\theta}{\sqrt{1 - t \sin^2{\theta}}}$ is the complete elliptic integral.

To compare with a lattice calculation, we take $x_i$ integers, with $x_{21}, \, x_{32}$ and $x_{43}$ all large, and try to compute the determinant in (\ref{59}) with $I = I_1 \cup I_2$ the union of the two intervals $[x_1,x_2]$ and $[x_3,x_4]$. This determinant is no longer Toeplitz, which makes it difficult to compute its asymptotics analytically. Dividing it by the prefactor $(x_{12} \, x_{34})^{1/4}$, it can however be evaluated numerically as a function of $t$ by varying the lengths of the intervals and their separation distance. The agreement with (\ref{ellip}) is more than satisfactory \cite{Ru02}.

The opposite situation --two open intervals in a closed boundary-- has also been considered. The appropriate 4-point function can be obtained from (\ref{ellip}) by making a simple cyclic permutation $(x_1,x_2,x_3,x_4) \to (x_4,x_1,x_2,x_3)$, with the result that $K(x)$ gets replaced by $K(1-x)$. An equally successful agreement was observed \cite{PR04}. Many other crosschecks have been done, confirming that the open/closed boundary condition changing field is indeed a primary field with conformal weight $h=-\frac 18$. One of them, particularly convincing, is presented in the next section.

\subsection{Bulk variables with inhomogeneous boundary}
\label{sec6.3}

In Section \ref{sec6.1}, we have computed the lattice 1- and 2-site height probabilities on the UHP, with either the open or the closed boundary condition. Here we would like to revisit these results in the light of what we have learned of the boundary condition changing field, in terms of which one should be able to relate the probabilities for the two boundary conditions. In particular, we would like to understand the 1-site probabilities $\sigma_a^{\rm op}(y)$ and $\sigma_a^{\rm cl}(y)$, 
\be
\sigma_a^{\rm op}(y) = \frac{1}{y^2}\Big( c_a + \frac{d_a}2 + d_a \log{y} \Big) + \ldots, \qquad \sigma_a^{\rm cl}(y) = -\frac{1}{y^2}\Big( c_a + d_a \log{y} \Big) + \ldots
\ee

One can do this by computing, on the CFT side, a more general probability. Namely we look for the probability to find a height equal to $a$ at a distance $y$ from the boundary, when the boundary condition is mixed, namely open everywhere except on the interval $[x_1,x_2]$ where the condition is closed. The two homogeneous open and closed conditions can be recovered in the limits $x_1 \to x_2$ and $x_1 \to -\infty,\, x_2 \to \infty$. Let us denote by $\langle h_a(z,\overline z) \rangle_{\rm mix}$ the corresponding quantity in the CFT, given by
\be
\langle h_a(z,\overline z) \rangle_{\rm mix} = \frac{\langle \phi^{\rm op,cl}(x_1) \, \phi^{\rm cl,op}(x_2) \, h_a(z,\overline z) \rangle_{\rm op}}{\langle \phi^{\rm op,cl}(x_1) \, \phi^{\rm cl,op}(x_2)\rangle_{\rm op}},
\label{ratios}
\ee
where the division by $\langle \phi^{\rm op,cl}(x_1) \, \phi^{\rm cl,op}(x_2)\rangle_{\rm op}$ comes from the fact that we want to evaluate the probability to have a height 1 in front of a mixed boundary condition, and not the combined effects of having a height 1 and the closing the boundary between $x_1$ and $x_2$. The denominator is known from (\ref{clop}).

To compute the numerator, we represent the height fields in terms of the chiral fields as $h_1(z,\overline z) = \phi(z) \psi(\overline z)$ and $h_2(z,\overline z) = \psi(z) \psi(\overline z)$ (as usually, considering the heights $a=1,2$ is enough) and write the differential equation statisfied by the two ensuing 4-point correlators, as a consequence of the primary nature of $\phi^{\rm op,cl}$. Because $\psi$ is the chiral logarithmic partner of $\phi$, the general solution for $\langle \phi^{\rm op,cl}(x_1) \, \phi^{\rm cl,op}(x_2) \, \psi(z) \, \psi(\overline z) \rangle_{\rm op}$ in fact depends on that of $\langle \phi^{\rm op,cl}(x_1) \, \phi^{\rm cl,op}(x_2) \, \phi(z) \, \psi(\overline z) \rangle_{\rm op}$. 

All calculations done, one finds that they depend on two integration constants $c_2$ and $d_2$ in such a way that the ratios (\ref{ratios}) take the following forms where $y = {\rm Re\ }z$ \cite{JPR06},
\begin{subequations}
\bea
\langle h_1(z,\overline z) \rangle_{\rm mix} \!\egal\! \frac{d_2}{2y^2} \: \frac{1+t}{\sqrt{t}}, \hspace{5cm} t = \frac{(x_1-\overline z)(x_2-z)}{(x_1-z)(x_2-\overline z)},\\
\langle h_2(z,\overline z) \rangle_{\rm mix} \!\egal\! \frac{1}{2y^2} \: \frac{1+t}{\sqrt{t}} \:\left\{c_2 + \frac{d_2}8 \frac{(1+\sqrt{t})^2}{\sqrt{t}}  + d_2 \log{y} - d_2 \frac{{\rm i}y(1-t)}{(x_1-x_2)} \Big[\frac {1}{1+t} - \frac {1}{2\sqrt{t}}\Big]\right\}.\nonumber\\
\label{h2mix}
\eea
\label{mix}
\end{subequations}
Although $t$ is complex, both expressions are real on account of $t^* = 1/t$.

Let us now discuss the above two limits $x_1 \to x_2$ and $x_1 \to -\infty,\, x_2 \to \infty$. For convenience, we set $x_1=-x_2$ and examine the limits $x_2 \to 0^+$ and $x_2 \to +\infty$. To compute the two limits, the important thing to notice is that the complex variable $t$, now equal to
\be
t = \frac{(x_2+\overline z)(x_2-z)}{(x_2+z)(x_2-\overline z)},
\ee
has complex norm equal to 1 and loops anticlockwise around the origin as $x_2$ varies from $0^+$ to $+\infty$, starting from $1+0{\rm i}$ to $1-0{\rm i}$. It follows that $t$ itself goes to 1 in both limits but $\sqrt{t}$ goes to $+1$ when $x_2 \to 0^+$ and goes to $-1$ when $x_2 \to +\infty$. The actual limits yield
\begin{subequations}
\bea
\langle h_1(z,\overline z) \rangle_{\rm op} \egal \lim_{x_2 \to 0^+} \langle h_1(z,\overline z) \rangle_{\rm mix} = \frac{d_2}{y^2}, \qquad \langle h_1(z,\overline z) \rangle_{\rm cl} = \lim_{x_2 \to +\infty} \langle h_1(z,\overline z) \rangle_{\rm mix} = -\frac{d_2}{y^2},\\
\langle h_2(z,\overline z) \rangle_{\rm op} \egal \frac{1}{y^2}\Big( c_2 + \frac{d_2}2 + d_2 \log{y} \Big), \qquad 
\langle h_2(z,\overline z) \rangle_{\rm cl} = -\frac{1}{y^2}\Big( c_2 + d_2 \log{y} \Big),
\eea
\end{subequations}
in complete agreement with the lattice results: the change of the overall sign between the open and closed boundary conditions, the specific dependence on the two coefficients $c_2$ and $d_2$, and the equality $c_1=d_2$ are all accounted for ! The conformal approach however cannot fix the two coefficients $c_2$ and $d_2$; these must be determined by lattice calculations. 

The expressions (\ref{mix}) can also be tested in situations where the boundary condition along the real axis is no longer homogeneous. A particularly instructive case is when the boundary condition is closed on the negative part of the real axis, and open on the positive part, corresponding to the limits $x_1 \to -\infty$ and $x_2 \to 0$. The conformal transformation $w = \frac L\pi \log z$ can be used to map the UHP onto an infinite strip of width $L$, with open boundary condition on the left side, closed on the right side. The conformal transformation rules of the fields involved being known, the expressions (\ref{mix}) can be transformed to the strip and compared with numerical simulations on a truncated (and large) strip (exact calculations on the lattice are not available). It was found \cite{JPR06} that the conformal predictions and the numerical plots match remarkably well, thereby confirming once more all the field identifications made so far.

\subsection{Wind on the boundary}
\label{sec6.4}

The open and closed boundary conditions are very natural as the very definition of the sandpile model uses dissipative and conservative sites. One may wonder what other type of boundary condition could be thought of. Perhaps we could think of alternating open and closed boundary sites; we expect however that such a boundary condition would flow to the open condition in the scaling limit, as numerical experiments confirm. We have already commented on the possibility to uniformly fix the boundary heights. Fixing the boundary heights to 2 or to 3, or even to 2 or 3, seems difficult. The two boundary conditions, different from open and closed, which have been considered in \cite{Ru07}, are in fact closely related, but not quite identical, to the third possibility. They are fixed boundary conditions but in the language of spanning trees. 

We recall that in a rooted spanning tree, there is exactly one outgoing arrow at each vertex. The two new boundary conditions, noted $\leftarrow$ and $\to$, force the outgoing arrows at the boundary sites to be uniformly left or uniformly right \footnote{In contrast, the outgoing arrow of a closed boundary site of the UHP can point left, up or right, while that of an open boundary site can point in any of the four directions, a down arrow pointing to the root.}. In terms of height values, either condition means that none of the boundary sites has a height 1 (because each boundary site has an ingoing arrow) or a height 4 (because the burning algorithm would imply that the arrow is pointing down, towards the root). The converse is however not true: recurrent configurations with height values equal to 2 or 3 on the boundary do not necessarily have boundary arrows uniformly oriented.

The way the orientation of an edge can be forced has been briefly discussed in Section \ref{sec5.7}. This allows to evaluate the effects of inserting a stretch of left or right arrows into an open or a closed boundary, similarly to what we did in Section \ref{sec6.2}. We refer the reader to \cite{Ru07} for the details of the analysis, and restrict here to a summary of the results. 

An obvious but unusual feature of the boundary conditions $\leftarrow$ and $\to$ is that they are intrinsically oriented. It implies that the boundary condition changing field $\phi^{{\rm a},\to}$ turning the boundary condition from ${\rm a}$ to $\to$ may not be the same as the field $\phi^{\to,{\rm a}}$ implementing the opposite change. With ${\rm a},{\rm b} \in \{{\rm op},{\rm cl},\leftarrow,\to\}$, this makes potentially twelve distinct fields $\phi^{{\rm a},{\rm b}}$ (the fields $\phi^{{\rm a},{\rm a}}$ are just the identity). We already know $\phi^{\rm op,cl} = \phi^{\rm cl,op}$, and likewise, if ${\rm a} \in \{{\rm op},{\rm cl}\}$ is unoriented and ${\rm b} \in \{\leftarrow,\to\}$ is oriented, we expect the identifications $\phi^{{\rm a},\to} = \phi^{\leftarrow,{\rm a}}$ and $\phi^{{\rm a},\leftarrow} = \phi^{\to,{\rm a}}$ on the basis of a left-right reflection symmetry. These identifications have been confirmed and reduce the number of distinct fields to seven. 

There is an additional subtlety for the field that changes the orientation from $\to$ to $\leftarrow$. Indeed the right and left arrows $\circ \!\!\!\to \!\bullet \!\leftarrow \!\!\!\circ$ point to the same boundary site (in black), and whether that site is open or closed may be relevant. Indeed if it is open, the flow of arrows, which eventually terminates at the root, can go directly to the root; if it is closed, it must necessarily go upwards into the bulk of the UHP. In the two cases, the macroscopic configurations of arrows are different. Thus we should distinguish two different fields, $\phi^{\to\stackrel{\rm op}{,}\leftarrow}$ and $\phi^{\to\stackrel{\rm cl}{,}\leftarrow}$. The detailed analysis confirm that they are distinct fields as their conformal weights are different.

We therefore have eight distinct boundary condition changing fields. A mix of analytical calculations and numerical simulations has been used to determine the conformal weights of these eight fields. The results are given in Table \ref{tab2}.

\renewcommand{\arraystretch}{1.7}
\begin{table}[t]
\tabcolsep6pt
\hfill 
\begin{center}
\begin{tabular}{|c||c|c|c|c|}
\hline
$h[\phi^{{\rm a},{\rm b}}]$ & open & closed & $\to$ & $\leftarrow$ \\
\hline\hline
open & & $-{1 \over 8}$ & $0$ & $0$\\
\hline
closed & $-{1 \over 8}$ & & $-{1 \over 8}$ & ${3 \over 8}$\\
\hline
$\to$ & $0$ & ${3 \over 8}$ & & 
$\renewcommand{\arraystretch}{1.2}
\begin{array}{l}
{}0  \ \ ({\rm center\  op})\\
{}1  \ \ ({\rm center\  cl})
\end{array}$ \\
\hline
$\leftarrow$ & $0$ & $-{1 \over 8}$ & $0$ &  \\
\hline
\end{tabular}
\end{center}
\caption{Conformal weights of the fields $\phi^{{\rm a},{\rm b}}$ which implement a change of boundary condition from a (row label) to b (column).}
\label{tab2}
\end{table}

The more delicate question of the exact nature of all these fields has been addressed by considering the fusion of the representations to which they belong. Loosely speaking, the fusion rules implement the composition law $\phi^{{\rm a},{\rm b}} \star \phi^{{\rm b},{\rm c}} \simeq \phi^{{\rm a},{\rm c}}$ of boundary condition changing fields in the limit where the insertion points coincide. The ensuing consistency conditions suggest that all of them are primary fields, except two, which could belong to logarithmic representations (i.e. reducible indecomposable with Jordan cells). Also the fields of weight 0 are non-trivial, that is, not equal to the identity (they are found to be degenerate at level 3). Relying on these proposals, various 4-point correlators have been computed and successfully compared with numerical simulations. We refer to \cite{Ru07} for more details on these specific points.

\subsection{Boundary height variables}
\label{sec6.5}

The boundary condition changing fields are not the only ones to live on a boundary. The lattice model includes observables in the bulk as well as on the boundaries. Those in the bulk have been discussed at length and give rise in the scaling limit to non-chiral fields $\Phi(z,\overline z)$, characterized by a pair of conformal weights $(h,\overline h)$; those on the boundaries give rise to boundary, chiral fields $\Phi^{\rm a}(x)$, characterized by a single conformal dimension $h_{\rm a}$. In general, the nature of the boundary field associated wih a boundary observable and its conformal weight depend on the boundary condition.

In the Abelian sandpile model, only the boundary fields arising from the height variables and from the insertion of isolated dissipation have been studied on the UHP. In both case, only open and closed boundaries have been considered.

The case of isolated dissipation is simpler and has been examined in details in \cite{PR04}, where isolated dissipation has been considered on a closed boundary only. The calculation proceeds much like those for the bulk, reviewed in Section \ref{sec5.4}, for which the same regularization is used. The results are similar: the dissipation field $\omega^{\rm cl}(x)$ turns out to be a chiral field with conformal weight $h_{\rm cl}=0$, and is a logarithmic partner of the identity. The multipoint correlators involve various combinations of logarithms like for its bulk version. On an open boundary, already dissipative, the dissipation field $\omega^{\rm op}(x)$ is expected to be a descendant of the identity. Isolated dissipation is the simplest observable that can be associated and computed in terms of a local defect matrix. This, from what we have said in Section \ref{sec5.6} of the minimal clusters, suggests that both $\omega^{\rm cl}$ and $\omega^{\rm op}$ can be realized as local fields in the symplectic fermions. It was indeed shown that $\omega^{\rm cl} \sim \theta\tilde\theta$ \cite{PR04} and $\omega^{\rm op} \sim \partial\theta\partial\tilde\theta$ \cite{Je05b} reproduce all known correlations. We note that the latter is proportional to the boundary stress-energy tensor $T(x)$ of the symplectic theory, a non-primary chiral field of weight $h_{\rm op}=2$, and a descendant of the identity since $T(x) \sim (L_{-2}\, \mathbb I)(x)$.

Boundary height variables are more complicated than dissipation but simpler than the bulk height variables. The first results have been derived by Ivashkevich in \cite{Iv94}, where the one- and two-site height probabilities on open and closed boundaries were obtained. The probabilities involving heights 1 only are no more complicated than in the bulk and can be easily obtained by using a defect matrix. As could be expected, probabilities for higher heights are more difficult. 

On a boundary, heights larger or equal to 2 are characterized as in the bulk, namely in terms of the number of predecessors among their nearest neighbours. So it leads essentially to the same problems of computing non-local contributions. Both in the bulk and on a boundary, one can write linear identities expressing combinations of non-local contributions in terms of local ones, themselves calculable with a defect matrix. In turn, the non-local contributions can be used to calculate probabilities. In the bulk, the linear system is underdetermined and cannot be inverted to provide the required non-local contributions, and then the probabilities themselves. The main observation made in \cite{Iv94} was that on a boundary, the linear system can be inverted, and therefore allows to compute the height probabilities and correlations in terms of local contributions only. The following results were obtained.

The 1-site height probabilities on the boundary, open and closed, of the infinite UHP were computed exactly. For comparison purposes, we reproduce here their numerical values (the exact values can be found in \cite{Iv94}) and recall those in the bulk, as given in Section \ref{5.1},
\bea
&&\P_1 = 0.073\,63, \quad \P_2 = 0.173\,90, \quad \P_3 = 0.306\,29, \quad \P_4 = 0.446\,17,\\
&&\P_1^{\rm op} = 0.103\,82, \quad \P_2^{\rm op} = 0.216\,57, \quad \P_3^{\rm op} = 0.316\,23, \quad \P_4^{\rm op} = 0.363\,38,\\
&&\P_1^{\rm cl} = 0.113\,38, \quad \P_2^{\rm cl} = 0.216\,571, \quad \P_3^{\rm cl} = 0.316\,225.
\eea
On the open boundary, for which the comparison makes more sense, lower heights are thus more likely. 

Mixed 2-site correlators $\tau_{a,b}^{\rm op}(x_1,x_2)$ and $\tau_{a,b}^{\rm cl}(x_1,x_2)$ on an open or closed boundary were also computed in \cite{Iv94}; all of them were found to decay like $|x_1-x_2|^{-4}$. Although logarithmic conformal field theory was in its infancy at the time, it indicates in hindsight that unlike their bulk cousins, boundary height fields are not logarithmic. This is also in agreement with the fact explained above that boundary height correlations can be fully computed in terms of local contributions.

The decay of the 2-site correlators strongly suggest that all boundary height fields, whatever the boundary condition, have a conformal dimension equal to $h_{\rm op} = h_{\rm cl} = 2$. But like for the other observables discussed so far, we are interested to know the precise nature of the associated fields. Since the multisite boundary height probabilities appear to be calculable in terms of local contributions using defect matrices, it suggests again to look for field candidates constructed out from the symplectic free fermions $\theta,\tilde\theta$. This was done independently in \cite{Je05b} and in \cite{PR05b}, following however two different approaches: the former computed various 3-point correlators whereas the latter considered 2-point correlators only but in the massive extension of the sandpile model (see Section \ref{7.1}). The massive extension indeed allows to distinguish more efficiently different fields which would otherwise have the same 2-point correlators in the non-massive (critical) limit.

The results are as follows. The four height fields on an open boundary are all proportional to a single field,
\be
h_a^{\rm op}(x) = O_a \: \partial\theta \, \partial \tilde\theta, \qquad 1 \le a \le 4,
\ee
with explicit normalization constants $O_a$ and where the $\theta,\tilde\theta$ fields satisfy the Dirichlet boundary condition. Thus on an open boundary, the four height fields and the dissipation field turn out to be all proportional to each other. On a closed boundary, the three height fields are distinct and given by
\be
h_1^{\rm cl}(x) = C_1 \: \partial\theta \, \partial \tilde\theta, \quad h_2^{\rm cl}(x) = C_2 \: \partial\theta \, \partial \tilde\theta + \frac 1{2\pi} \, \theta\,\partial\partial\tilde\theta, \quad h_3^{\rm cl}(x) = C_3 \: \partial\theta \, \partial \tilde\theta - \frac 1{2\pi} \, \theta\,\partial\partial\tilde\theta,
\ee
where the $\theta,\tilde\theta$ fields now satisfy the Neumann boundary condition. In both cases, the boundary condition means that the correlators are computed using the Wick theorem with the Wick contractions given by the Green functions $G^{\rm op}$ or $G^{\rm cl}$, see (\ref{dir}) and (\ref{neu}). Let us point out that, for both boundary conditions, the 3-site correlations of three heights 1 do not vanish, unlike their bulk version.

The question of the nature of the height fields on the windy boundary conditions discussed in the previous section is definitely interesting, but has not been considered so far.

\subsection{Duality}
\label{sec6.6}

This long section on boundaries has been largely devoted to a discussion of the open and closed boundary conditions, the best known and most studied ones. To finish, it is worth pointing out that a duality exists between these two boundary conditions, which has not been fully investigated nor exploited. This duality follows from a duality relation for planar graphs, well-known in graph theory, and acquires in the framework of the sandpile model an interesting flavour. It has been considered and discussed in \cite{IPRH05,IPR07} in the dimer model, intrinsically related, like the sandpile model, to spanning trees.

\begin{figure}[t]
\begin{pspicture}[shift=0](-1.2,-1.5)(-0.1,3)
\psset{xunit=0.9cm}
\psset{yunit=0.9cm}
\psset{runit=0.9cm}
\psline[linecolor=blue,linewidth=0.7pt](0,-1)(0,2)(2,2)(2,-1)
\psline[linecolor=blue,linewidth=0.7pt](1,-1)(1,2)
\psline[linecolor=blue,linewidth=0.7pt](-1,2)(3,2)
\psline[linecolor=blue,linewidth=0.7pt](-1,1)(3,1)
\psline[linecolor=blue,linewidth=0.7pt](-1,0)(3,0)
\psline[linecolor=blue,linestyle=dashed,dash=.05 .05,linewidth=0.7pt](-1,2)(-1,0)
\psline[linecolor=blue,linestyle=dashed,dash=.05 .05,linewidth=0.7pt](0,-1)(2,-1)
\psline[linecolor=blue,linestyle=dashed,dash=.05 .05,linewidth=0.7pt](3,0)(3,2)
\psarc[linecolor=blue,linestyle=dashed,dash=.05 .05,linewidth=0.7pt](0,0){1}{180}{270}
\psarc[linecolor=blue,linestyle=dashed,dash=.05 .05,linewidth=0.7pt](2,0){1}{270}{360}
\pscircle[linewidth=0pt,linecolor=blue,fillstyle=solid,fillcolor=blue](0,0){0.07}
\pscircle[linewidth=0pt,linecolor=blue,fillstyle=solid,fillcolor=blue](0,1){0.07}
\pscircle[linewidth=0pt,linecolor=blue,fillstyle=solid,fillcolor=blue](0,2){0.07}
\pscircle[linewidth=0pt,linecolor=blue,fillstyle=solid,fillcolor=blue](1,2){0.07}
\pscircle[linewidth=0pt,linecolor=blue,fillstyle=solid,fillcolor=blue](1,1){0.07}
\pscircle[linewidth=0pt,linecolor=blue,fillstyle=solid,fillcolor=blue](1,0){0.07}
\pscircle[linewidth=0pt,linecolor=blue,fillstyle=solid,fillcolor=blue](2,0){0.07}
\pscircle[linewidth=0pt,linecolor=blue,fillstyle=solid,fillcolor=blue](2,1){0.07}
\pscircle[linewidth=0pt,linecolor=blue,fillstyle=solid,fillcolor=blue](2,2){0.07}
\pscircle[linewidth=0.7pt,linecolor=blue,fillstyle=solid,fillcolor=white](-1,2){0.07}
\pscircle[linewidth=0.7pt,linecolor=blue,fillstyle=solid,fillcolor=white](-1,1){0.07}
\pscircle[linewidth=0.7pt,linecolor=blue,fillstyle=solid,fillcolor=white](-1,0){0.07}
\pscircle[linewidth=0.7pt,linecolor=blue,fillstyle=solid,fillcolor=white](0,-1){0.07}
\pscircle[linewidth=0.7pt,linecolor=blue,fillstyle=solid,fillcolor=white](1,-1){0.07}
\pscircle[linewidth=0.7pt,linecolor=blue,fillstyle=solid,fillcolor=white](2,-1){0.07}
\pscircle[linewidth=0.7pt,linecolor=blue,fillstyle=solid,fillcolor=white](3,0){0.07}
\pscircle[linewidth=0.7pt,linecolor=blue,fillstyle=solid,fillcolor=white](3,1){0.07}
\pscircle[linewidth=0.7pt,linecolor=blue,fillstyle=solid,fillcolor=white](3,2){0.07}
\end{pspicture}
\begin{pspicture}[shift=0](-4.7,-1.3)(-4.8,3)
\psset{xunit=0.9cm}
\psset{yunit=0.9cm}
\psset{runit=0.9cm}
\psline[linecolor=blue,linewidth=0.7pt](0,-1)(0,2)(2,2)(2,-1)
\psline[linecolor=blue,linewidth=0.7pt](1,-1)(1,2)
\psline[linecolor=blue,linewidth=0.7pt](-1,2)(3,2)
\psline[linecolor=blue,linewidth=0.7pt](-1,1)(3,1)
\psline[linecolor=blue,linewidth=0.7pt](-1,0)(3,0)
\psline[linecolor=blue,linestyle=dashed,dash=.05 .05,linewidth=0.7pt](-1,2)(-1,0)
\psline[linecolor=blue,linestyle=dashed,dash=.05 .05,linewidth=0.7pt](0,-1)(2,-1)
\psline[linecolor=blue,linestyle=dashed,dash=.05 .05,linewidth=0.7pt](3,0)(3,2)
\psarc[linecolor=blue,linestyle=dashed,dash=.05 .05,linewidth=0.7pt](0,0){1}{180}{270}
\psarc[linecolor=blue,linestyle=dashed,dash=.05 .05,linewidth=0.7pt](2,0){1}{270}{360}
\pscircle[linewidth=0pt,linecolor=blue,fillstyle=solid,fillcolor=blue](0,0){0.07}
\pscircle[linewidth=0pt,linecolor=blue,fillstyle=solid,fillcolor=blue](0,1){0.07}
\pscircle[linewidth=0pt,linecolor=blue,fillstyle=solid,fillcolor=blue](0,2){0.07}
\pscircle[linewidth=0pt,linecolor=blue,fillstyle=solid,fillcolor=blue](1,2){0.07}
\pscircle[linewidth=0pt,linecolor=blue,fillstyle=solid,fillcolor=blue](1,1){0.07}
\pscircle[linewidth=0pt,linecolor=blue,fillstyle=solid,fillcolor=blue](1,0){0.07}
\pscircle[linewidth=0pt,linecolor=blue,fillstyle=solid,fillcolor=blue](2,0){0.07}
\pscircle[linewidth=0pt,linecolor=blue,fillstyle=solid,fillcolor=blue](2,1){0.07}
\pscircle[linewidth=0pt,linecolor=blue,fillstyle=solid,fillcolor=blue](2,2){0.07}
\pscircle[linewidth=0.7pt,linecolor=blue,fillstyle=solid,fillcolor=white](-1,2){0.07}
\pscircle[linewidth=0.7pt,linecolor=blue,fillstyle=solid,fillcolor=white](-1,1){0.07}
\pscircle[linewidth=0.7pt,linecolor=blue,fillstyle=solid,fillcolor=white](-1,0){0.07}
\pscircle[linewidth=0.7pt,linecolor=blue,fillstyle=solid,fillcolor=white](0,-1){0.07}
\pscircle[linewidth=0.7pt,linecolor=blue,fillstyle=solid,fillcolor=white](1,-1){0.07}
\pscircle[linewidth=0.7pt,linecolor=blue,fillstyle=solid,fillcolor=white](2,-1){0.07}
\pscircle[linewidth=0.7pt,linecolor=blue,fillstyle=solid,fillcolor=white](3,0){0.07}
\pscircle[linewidth=0.7pt,linecolor=blue,fillstyle=solid,fillcolor=white](3,1){0.07}
\pscircle[linewidth=0.7pt,linecolor=blue,fillstyle=solid,fillcolor=white](3,2){0.07}
\psline[linecolor=red,linewidth=0.7pt](-0.5,2.5)(-0.5,-0.5)(2.5,-0.5)(2.5,2.5)
\psline[linecolor=red,linewidth=0.7pt](-0.5,1.5)(2.5,1.5)
\psline[linecolor=red,linewidth=0.7pt](-0.5,0.5)(2.5,0.5)
\psline[linecolor=red,linewidth=0.7pt](0.5,-0.5)(0.5,2.5)
\psline[linecolor=red,linewidth=0.7pt](1.5,-0.5)(1.5,2.5)
\psline[linecolor=red,linestyle=dashed,dash=.05 .05,linewidth=0.7pt](-0.5,2.5)(2.5,2.5)
\pscircle[linewidth=0pt,linecolor=red,fillstyle=solid,fillcolor=red](-0.5,1.5){0.07}
\pscircle[linewidth=0pt,linecolor=red,fillstyle=solid,fillcolor=red](-0.5,0.5){0.07}
\pscircle[linewidth=0pt,linecolor=red,fillstyle=solid,fillcolor=red](-0.5,-0.5){0.07}
\pscircle[linewidth=0pt,linecolor=red,fillstyle=solid,fillcolor=red](0.5,1.5){0.07}
\pscircle[linewidth=0pt,linecolor=red,fillstyle=solid,fillcolor=red](0.5,0.5){0.07}
\pscircle[linewidth=0pt,linecolor=red,fillstyle=solid,fillcolor=red](0.5,-0.5){0.07}
\pscircle[linewidth=0pt,linecolor=red,fillstyle=solid,fillcolor=red](1.5,1.5){0.07}
\pscircle[linewidth=0pt,linecolor=red,fillstyle=solid,fillcolor=red](1.5,0.5){0.07}
\pscircle[linewidth=0pt,linecolor=red,fillstyle=solid,fillcolor=red](1.5,-0.5){0.07}
\pscircle[linewidth=0pt,linecolor=red,fillstyle=solid,fillcolor=red](2.5,1.5){0.07}
\pscircle[linewidth=0pt,linecolor=red,fillstyle=solid,fillcolor=red](2.5,0.5){0.07}
\pscircle[linewidth=0pt,linecolor=red,fillstyle=solid,fillcolor=red](2.5,-0.5){0.07}
\pscircle[linewidth=0.7pt,linecolor=red,fillstyle=solid,fillcolor=white](-0.5,2.5){0.07}
\pscircle[linewidth=0.7pt,linecolor=red,fillstyle=solid,fillcolor=white](0.5,2.5){0.07}
\pscircle[linewidth=0.7pt,linecolor=red,fillstyle=solid,fillcolor=white](1.5,2.5){0.07}
\pscircle[linewidth=0.7pt,linecolor=red,fillstyle=solid,fillcolor=white](2.5,2.5){0.07}
\end{pspicture}
\begin{pspicture}[shift=0](-8.9,-1.1)(-9,1,3)
\psset{xunit=0.9cm}
\psset{yunit=0.9cm}
\psset{runit=0.9cm}
\psline[linecolor=red,linewidth=0.7pt](-0.5,2.5)(-0.5,-0.5)(2.5,-0.5)(2.5,2.5)
\psline[linecolor=red,linewidth=0.7pt](-0.5,1.5)(2.5,1.5)
\psline[linecolor=red,linewidth=0.7pt](-0.5,0.5)(2.5,0.5)
\psline[linecolor=red,linewidth=0.7pt](0.5,-0.5)(0.5,2.5)
\psline[linecolor=red,linewidth=0.7pt](1.5,-0.5)(1.5,2.5)
\psline[linecolor=red,linestyle=dashed,dash=.05 .05,linewidth=0.7pt](-0.5,2.5)(2.5,2.5)
\pscircle[linewidth=0pt,linecolor=red,fillstyle=solid,fillcolor=red](-0.5,1.5){0.07}
\pscircle[linewidth=0pt,linecolor=red,fillstyle=solid,fillcolor=red](-0.5,0.5){0.07}
\pscircle[linewidth=0pt,linecolor=red,fillstyle=solid,fillcolor=red](-0.5,-0.5){0.07}
\pscircle[linewidth=0pt,linecolor=red,fillstyle=solid,fillcolor=red](0.5,1.5){0.07}
\pscircle[linewidth=0pt,linecolor=red,fillstyle=solid,fillcolor=red](0.5,0.5){0.07}
\pscircle[linewidth=0pt,linecolor=red,fillstyle=solid,fillcolor=red](0.5,-0.5){0.07}
\pscircle[linewidth=0pt,linecolor=red,fillstyle=solid,fillcolor=red](1.5,1.5){0.07}
\pscircle[linewidth=0pt,linecolor=red,fillstyle=solid,fillcolor=red](1.5,0.5){0.07}
\pscircle[linewidth=0pt,linecolor=red,fillstyle=solid,fillcolor=red](1.5,-0.5){0.07}
\pscircle[linewidth=0pt,linecolor=red,fillstyle=solid,fillcolor=red](2.5,1.5){0.07}
\pscircle[linewidth=0pt,linecolor=red,fillstyle=solid,fillcolor=red](2.5,0.5){0.07}
\pscircle[linewidth=0pt,linecolor=red,fillstyle=solid,fillcolor=red](2.5,-0.5){0.07}
\pscircle[linewidth=0.7pt,linecolor=red,fillstyle=solid,fillcolor=white](-0.5,2.5){0.07}
\pscircle[linewidth=0.7pt,linecolor=red,fillstyle=solid,fillcolor=white](0.5,2.5){0.07}
\pscircle[linewidth=0.7pt,linecolor=red,fillstyle=solid,fillcolor=white](1.5,2.5){0.07}
\pscircle[linewidth=0.7pt,linecolor=red,fillstyle=solid,fillcolor=white](2.5,2.5){0.07}
\end{pspicture}
\begin{pspicture}[shift=0](-13.1,-1.3)(-13.2,3)
\psset{xunit=0.9cm}
\psset{yunit=0.9cm}
\psset{runit=0.9cm}
\psline[linecolor=blue,linewidth=0.5pt](0,-1)(0,2)(2,2)(2,-1)
\psline[linecolor=blue,linewidth=0.5pt](1,-1)(1,2)
\psline[linecolor=blue,linewidth=0.5pt](-1,2)(3,2)
\psline[linecolor=blue,linewidth=0.5pt](-1,1)(3,1)
\psline[linecolor=blue,linewidth=0.5pt](-1,0)(3,0)
\psline[linecolor=blue,linewidth=1.8pt](-1,0)(1,0)(1,1)(2,1)(2,2)(1,2)
\psline[linecolor=blue,linewidth=1.8pt](1,1)(0,1)(0,2)
\psline[linecolor=blue,linewidth=1.8pt](2,0)(2,-1)
\psline[linecolor=blue,linestyle=dashed,dash=.05 .05,linewidth=0.5pt](-1,2)(-1,0)
\psline[linecolor=blue,linestyle=dashed,dash=.05 .05,linewidth=0.5pt](0,-1)(2,-1)
\psline[linecolor=blue,linestyle=dashed,dash=.05 .05,linewidth=0.5pt](3,0)(3,2)
\psarc[linecolor=blue,linestyle=dashed,dash=.05 .05,linewidth=0.5pt](0,0){1}{180}{270}
\psarc[linecolor=blue,linestyle=dashed,dash=.05 .05,linewidth=0.5pt](2,0){1}{270}{360}
\psline[linecolor=red,linewidth=0.5pt](-0.5,2.5)(-0.5,-0.5)(2.5,-0.5)(2.5,2.5)
\psline[linecolor=red,linewidth=0.5pt](-0.5,1.5)(2.5,1.5)
\psline[linecolor=red,linewidth=0.5pt](-0.5,0.5)(2.5,0.5)
\psline[linecolor=red,linewidth=0.5pt](0.5,-0.5)(0.5,2.5)
\psline[linecolor=red,linewidth=0.5pt](1.5,-0.5)(1.5,2.5)
\psline[linecolor=red,linewidth=1.8pt](-0.5,-0.5)(1.5,-0.5)(1.5,0.5)(2.5,0.5)(2.5,2.5)
\psline[linecolor=red,linewidth=1.8pt](2.5,-0.5)(2.5,0.5)
\psline[linecolor=red,linewidth=1.8pt](0.5,0.5)(-0.5,0.5)(-0.5,2.5)
\psline[linecolor=red,linewidth=1.8pt](1.5,1.5)(0.5,1.5)(0.5,2.5)
\psline[linecolor=red,linestyle=dashed,dash=.05 .05,linewidth=0.5pt](-0.5,2.5)(2.5,2.5)
\end{pspicture}
\caption{The drawing codes for the three figures are as follows. The open circle stands for the sink vertex, while the solid circles stand for the non-sink vertices. The solid lines represent true edges of the extended graphs, unlike the dashed lines connecting the open circles which indicate that these should be identified as the unique sink vertex. Let us note that the corner vertices which lie at the intersection of an open and a closed boundary have a single edge to the sink; those at the intersection of two open boundaries have two such edges. (a) The left figure shows the extended graph of a $3 \times 3$ grid with open boundary conditions on the left, lower and right boundaries, and closed boundary condition on the upper boundary. (b) The second pannel shows how the dual of the blue graph, in red, is constructed. The red sink is the vertex associated to the outer face of the blue graph. (c) For a better readability, the dual graph alone is reproduced on the third pannel. (d) Two dual spanning trees are drawn on the far right.}
\label{fig1}
\end{figure}
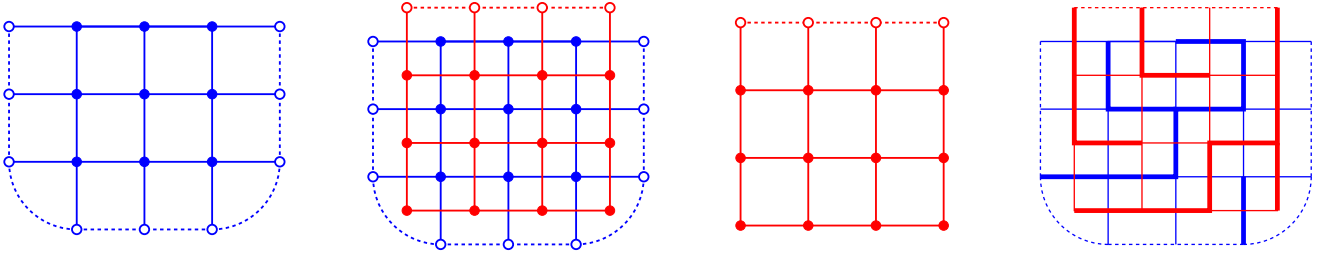

Let us consider a rectangular portion of $\Z^2$, that is, the graph $\Gamma$ made of a rectangular array of vertices, in which two adjacent vertices are linked by a single edge. The boundary conditions chosen for the boundary vertices determine the extended graph $\Gamma^\star$, obtained from $\Gamma$ by adding the sink vertex and the edges connecting the open boundary sites to the sink. The graph $\Gamma^\star$ corresponding to a $3 \times 3$ grid with three open edges and one closed egde is shown in Figure \ref{fig1}. 

Once the graph $\Gamma^\star$ is embedded in the plane \footnote{This requires $\Gamma^\star$ to be planar, and therefore excludes that some of the bulk vertices {\it and} some of the boundary vertices be open (dissipative) at the same time, except in a few very special cases.}, the faces of $\Gamma^\star$ are the connected components of its complementary in the plane (for a finite graph, there is thus a large outer face, encircling the graph). The definition of the dual graph $(\Gamma^\star)^*$ is standard: the vertices of $(\Gamma^\star)^*$ are associated to the faces of $\Gamma^\star$, and two such vertices are connected if their corresponding two faces are separated from each other by an edge of $\Gamma^\star$. The dual graph of the example above is also shown is Figure \ref{fig1}.

By comparing the two graphs, one immediately notices that the boundary conditions are exchanged: if a boundary is homogeneously open resp. closed in $\Gamma^\star$, it becomes homogeneously closed resp. open in $(\Gamma^\star)^*$. In addition, the dual graph $(\Gamma^\star)^*$ is the extension $(\Gamma^*)^\star$ by a sink of a dual rectangular grid $\Gamma^*$, of size slightly different from the original grid $\Gamma$. 

A classical result states that the number of spanning trees on $\Gamma^\star$ is equal to the number of spanning trees on its dual $(\Gamma^\star)^*$. In fact, for every spanning tree $\mathcal T$ on $\Gamma^\star$, there is a unique dual spanning tree ${\mathcal T}^*$ on $(\Gamma^\star)^*$ such that the two are perfectly interdigitating: the edges of ${\mathcal T}^*$ are exactly those of $(\Gamma^\star)^*$ which cross the edges of $\Gamma^\star$ not used in $\mathcal T$, and vice-versa. An example of this is given in Figure \ref{fig1}.

This dual picture implies that the recurrent configurations for the sandpile model defined on $\Gamma^\star$ can be isomorphically described by those on $(\Gamma^*)^\star$. As far as the counting goes, the equality of their partition functions can be explicitly written for rectangular grids. If $\Gamma$ is an $L_1 \times L_2$ rectangular grid with $k$ of its four boundaries being open, the other $4-k$ being closed, the dual $\Gamma^*$ is an  $L'_1 \times L'_2$ rectangular grid with swapped boundary conditions, and the following identity holds,
\be
Z_{[k{\rm \ op}\,|\,(4-k){\rm \ cl}]}\,(L_1,L_2) = Z_{[(4-k){\rm \ op}\,|\,k{\rm \ cl}]}\,(L'_1,L'_2).
\ee
The dimensions are related as follows: $L_i'=L_i+1$ resp $L_i-1$ if the opposites sides of length $L_i$ of $\Gamma$ are both open resp. closed, and $L'_i=L_i$ otherwise. If $k=4$, the dual rectangle has all its boundaries closed with a single boundary site open.

The isomorphism of the two descriptions may be hard to formulate in concrete terms for the height variables as it is defined for the associated trees. Its practical utility remains to be seen.

\section{More developments}
\label{sec7}

We would like to add a few more considerations about two further features of the sandpile model, namely the dissipative sandpile model and some aspects of universality.

\subsection{The massive sandpile model}
\label{7.1}

In the standard sandpile we have studied so far, the sites in the bulk of grid, that is, the vast majority of sites, are conservative. This meant that when such a site topples, it loses a certain number of sand grains which are all redistributed to its nearest neighbours. Sand moves in the grid but remains conserved. Dissipative sites must be present for the dynamics of the model to be well-defined; however the dissipative sites were located most of the time on the boundaries. 

The mostly conservative nature of the model is what drives it dynamically to a critical state: when enough sand is stored in the system, large avalanches become likely and span macroscopic parts of it, inducing strong correlations between distant heights. In the long run, the system enters a critical state described by the invariant measure $\P$, characterized by infinite correlation lengths in the infinite volume limit, and algebraic decays of the correlation functions. The field theory emerging in the scaling limit is conformal, and consequently massless.

From the above point of view, a natural way to take the sandpile model out of criticality is to introduce a fair amount of dissipation so as to make the range of the avalanches shorter. It is not completely clear what {\it a fair amount} means, as there are several ways to introduce dissipation. In the most common version, every site is made dissipative, with a dissipation rate controlled by an external parameter. In this case, it has been argued that indeed criticality is broken, resulting in an exponential decay of the correlations \cite{GLJ97,TK00,MR01}. A mathematically rigorous proof that all correlations decay exponentially has been provided in \cite{MRS04}. Presumably a non-zero density of dissipative sites could be a sufficient to break criticality, but to our knowledge, this possibility has not been investigated. In any case, the field theory emerging from the dissipative sandpile model must be massive, with mass(es) inversely proportional to the lattice correlation length(s).

To make all sites dissipative, one can simply add to the toppling matrix of the standard model an integer multiple of the identity matrix, $\Delta \to \Delta(t) = \Delta + t\,{\mathbb I}$ with $t$ an integer, while leaving all non-diagonal entries unchanged. According to the update of the heights after the toppling of site $j$, namely $h_i \to h_i - \Delta_{j,i}(t)$, a toppled site loses $t$ sand grains more than what it used to lose (whether or not the toppled site is on a boundary). That this change makes the correlation functions decay exponentially should be clear, for the following simple reason.

The new toppling matrix $\Delta(t)$ is a massive Laplacian matrix. It is well-known that the inverse Laplacian $\Delta^{-1}(t)$ has a kernel given at large distances by $G_{i_1,i_2}(t) \simeq \frac1{2\pi} K_0\big(|i_1-i_2|\sqrt{t}\big) + \ldots$, and decays exponentially like ${\rm e}^{-r\sqrt{t}}$ at large distances ($K_0$ is the modified Bessel function). Thus all multisite probabilities examined in the earlier sections, for observables like minimal cluster variables, arrow variables, isolated dissipation or boundary heights, will similarly decay exponentially. Though technically less clear for bulk heights equal to 2, 3 or 4, the same decay is expected for the reason explained above: there is a loss of sand each time a site topples, which makes the typical avalanches short-ranged, which in turn induces correlations of heights on local scales only.

To take a concrete examle, let us look at the correlation of two heights 1 in the dissipative model. The technique explained in Section \ref{sec5.3} in terms of defect matrices goes through. At dominant order, the result, which is the off-critical extension of (\ref{11}), reads \cite{MR01}
\be
\sigma_{1,1}(i_1,i_2;t) = -t^2 \frac{\P_1^2}2 \left\{K_0''^2 - K_0 \, K_0'' + \frac 1\pi K_0'^2 + \frac{1+\pi^2}{2\pi^2} K_0^2 \right\} + \ldots
\label{11mass}
\ee
where the argument of the Bessel functions is $r\sqrt{t}$ and $\P_1$ on the r.h.s. is the critical probability; the dots stand for higher orders in $t$. We see that the correlation decays exponentially, with a correlation length proportional to $\xi \sim t^{-1/2}$. 

How do we compute the scaling limit in the massive model ? The general discussion in Section \ref{sec3} suggested that setting $i = \frac{\vec x}\eps$ in the lattice corrrelator and taking the limit over $\eps$ (after multiplying the correlator by a suitable power of $\eps$) yields the field theoretic correlator. This cannot be the right way to proceed in the dissipative model. Because the correlators decay exponentially, the limit for $\eps$ going to zero of $\exp{(-|\vec x_1-\vec x_2|\sqrt{t}/\eps)}$ vanishes whathever the power of $\eps$ it is multiplied by.

The only way to get a non-trivial limit is to take a double limit: as we take the large distance limit by setting $i = \frac{\vec x}\eps$, we simultaneously take the large correlation length limit by accordingly adjusting the dissipation rate. In the present case, we should take the latter proportional to $\eps^2$: we therefore set $t = M^2\eps^2$, with $M$ playing the role of a mass (inverse correlation length in the continuum field theory).

Looking at the lattice correlator (\ref{11mass}), we see that the factor $t^2$ carries the overall dimension of the fields involved: $t^2$ is proportional to $M^4$, and thus inversely proportional to a distance to the fourth power. It replaces the explicit dependence in $r^4$ in the non-dissipative model. Eventually we find that the scaling limit of the correlator (\ref{11mass}) is
\be
\lim_{\eps \to 0} \; \eps^{-4} \; \sigma_{1,1}\big(\frac{z}\eps,\frac{w}\eps;\eps^2M^2\big) = -M^4 \: \frac{\P_1^2}2 \left\{K_0''^2 - K_0 \, K_0'' + \frac 1\pi K_0'^2 + \frac{1+\pi^2}{2\pi^2} K_0^2 \right\},
\label{11masscal}
\ee
where the argument of the Bessel function is now $M|z-w|$. It is straightforward to check that the $M \to 0$ limit of the previous expression is equal to $-{\P_1^2}/{2|z-w|^4}$, obtained in Section \ref{sec5}.
 
The last question is: the expression above is the correlator of what field and in what field theory ? The most obvious guess turns out to be correct:  let us look in the massive extension of the free symplectic fermion theory. It contains the same two fields as before, which simply acquire a mass through a mass term in the action,
\be
S = \frac 1{\pi} \int {\rm d}z {\rm d}\overline z \; \big(\partial \theta \overline\partial \tilde \theta + \frac{M^2}4 \theta\tilde\theta \big)\,.
\ee
The 2-point correlators of the two fundamental fields are now given by
\be
\langle \theta(z,\overline z) \theta(w,\overline w) \rangle = \langle \tilde\theta(z,\overline z) \tilde\theta(w,\overline w) \rangle = 0, \quad \langle \theta(z,\overline z) \tilde\theta(w,\overline w) \rangle = K_0(M|z-w|).
\ee
Using Wick's theorem, it is a simple matter to check that the following local field,
\be
h_1(z,\overline z;M) = -\P_1 \Big[ \partial \theta \, \overline\partial \tilde \theta  + \overline \partial \theta \, \partial \tilde \theta + \frac{M^2}{2\pi} \: \theta\tilde\theta \Big],
\ee
has a 2-point correlator \footnote{The insertion by hand of the dissipation field $\omega(\infty)$ at infinity in the field theoretic correlator is not required in the dissipative model, as dissipation is present everywhere in the bulk.} in the massive fermionic theory that is precisely given by (\ref{11masscal}). The 3- and 4-point correlators of the same field have been checked to reproduce the corresponding lattice results. The field $h_1(z,\overline z;M)$ is therefore what the height 1 variable in the dissipative sandpile model converges to in the scaling limit.

Similar correlators have been computed for many minimal clusters in \cite{MR01}, with an unexpectedly simple result. The field describing the minimal cluster variable $S$ in the dissipative model appears to be simply given by
\be
h_S(z,\overline z;M) = h_S(z,\overline z) - \P_S \, N_S \, \frac{M^2}{2\pi} \: \theta\tilde\theta,
\ee
where $\P_S$ is the probability of $S$ in the non-dissipative model, and $N_S$ is the size of the cluster $S$. The field $h_S(z,\overline z)$ is still given by (\ref{real}) in terms in the (now massive) fermions.

Likewise the mixed 2-point correlators for all boundary heights on open and closed boundaries have been explicitly evaluated in the dissipative model \cite{PR05b}. For them too, it is found that the boundary fields given in Section \ref{sec6.5} get additional terms proportional to $M^2 \,\theta\tilde\theta$. 

The nature of the higher height fields remains elusive but is definitely worth investigating as it would add a most valuable and crucial element of understanding of the sandpile model.

\subsection{Aspects of universality}
\label{7.2}

Universality is the statement that the large distance properties of statistical models should only depend on some gross features of the way they are defined; microscopic details which become invisible from large distances should not matter. The statement is admittedly not very precise, but in concrete instances, leads to an expected robustness with respect to local modifications. In sandpile models, these would include the precise way sand is deterministically redistributed among neighbours (provided some form of isotropy is preserved), or, to a certain extent, the specific graph or lattice on which the model is defined. Features that do matter are a substantial introduction of dissipation, as we have seen in the previous section, a directed redistribution of sand after toppling \cite{DR89}, a dynamics with stochastic toppling rules \cite{Ma91}, the formulation of the model on a hierarchical geometric structure like the Bethe lattice \cite{DM90}, and of course a change of dimensionality of the underlying lattice.

Very early on, universality with respect to the planar lattice on which the sandpile is being formulated has been tested via a renormalization group approach \cite{PP97,LH02} and numerical simulations \cite{HL03}. More recently, exact calculations of height correlations have been carried out on the honeycomb and triangular lattices.

In \cite{ADMR10}, all calculations of height 1 correlations presented in the previous sections have been worked out on the hexagonal lattice (in the non-dissipative model). These include the 2-, 3- and 4-site probabilities for heights 1 in general positions, in the bulk and on open and closed boundaries, as well as 1-site probabilities on the UHP, again for both types of boundary conditions. The results show that, although the subdominant contributions differ from those on the square lattice, the dominant terms are exactly identical, up to normalizations. The same distinctive features are found, like the fact that the 3-site bulk correlation vanishes in the scaling limit (the dominant term in the lattice result has dimension $-7$ instead of $-8$), and the change of sign for the UHP 1-site probabilities when changing the boundary condition from open to closed (see Section \ref{sec6.3}). Up to normalization, the field identifications of the height 1 variable in the bulk and on open and closed boundaries have been confirmed.

The results have been extended to higher heights on the honeycomb lattice, and to all heights on the triangular lattice \cite{PR18}. Interestingly, these two regular lattices have coordination numbers different from the square lattice, with the consequence that the height variables take in each case a different number of values : four for the square lattice, three for the honeycomb lattice and six for the triangular lattice. This naturally raises the question of which height variables scale to logarithmic fields, and which do not.

The calculations have been carried out by using the technique developed in \cite{KW15}, already used on the square lattice. The 1-site probabilities on the infinite honeycomb lattice are all rational,
\be
\P_1 = \frac1{12}, \qquad \P_2 = \frac7{24}, \qquad \P_3 = \frac58,
\ee
while those on the infinite triangular lattice are somewhat more complicated, like
\be
\P_6 = \frac{1175}{864}-\frac{365}{144\sqrt{3}\pi}-\frac{289}{12\pi^2}+\frac{30\sqrt{3}}{\pi^3}+\frac{45}{\pi^4}-\frac{54\sqrt{3}}{\pi^5}\simeq 0.286,
\ee
and very similar expressions for $\P_{1 \le a \le 5}$.

Concerning the nature of the height variables in the scaling limit, the results confirm what the reader has probably already suspected: far from boundaries, the height 1 variable becomes a primary field with conformal weights $(h,\overline h)=(1,1)$, while each of the higher heights scales to a logarithmic partner of the height 1, exactly like on the square lattice. On boundaries, all height fields are non-logarithmic. Moreover, all computed correlations \footnote{Some of the calculations done on the square lattice could not be worked out. For instance, we could not find a  proper method of images to compute the Green matrix on the triangular half-plane with the closed boundary condition, and therefore could not investigate that boundary condition.} exhibit the same bulk and boundary behaviours as on the square lattice. Thus for what concerns the type of the underlying lattice, universality has been explicitly and successfully verified.

\section{Conformal summary}
\label{sec8}

This last section is more specifically oriented towards conformal aspects of the sandpile model. We will summarize what we believe is currently known of the conformal picture, and discuss some of the most peculiar and not so well understood issues. We will almost exclusively discuss the non-chiral bulk fields, but before coming to those, we briefly comment on the chiral boundary fields encountered so far.

The boundary fields have been somewhat less investigated than the bulk fields. We have encountered two types of boundary fields, those arising from boundary observables and the boundary condition changing fields. In the first class, we have considered the height fields on open and closed boundaries and the dissipation field. Except for the dissipation on a closed boundary, none of them is logarithmic and no evidence of a logarithmic partner has been found. All can be expressed as local fields in the symplectic fermions. 

In the second class, we found primary fields of weight $-\frac 18$ and $\frac 38$, which are both standard fields in a $c=-2$ CFT. Due to the values of their conformal weight, they cannot be local in the symplectic fermions but are naturally accomodated \footnote{Very much like the spin field of the Ising model belongs naturally to the free Majorana fermionc theory with $c=\frac 12$, despite being non-local in the fermions.} in the symplectic fermion theory \cite{GK99}. The status of the other boundary condition changing fields related to the windy boundary conditions is uncertain, and should be further investigated before their exact nature can be reliably stated. 

Thus overall the boundary fields raise no particular questions. They are fairly simple fields which fit well within the symplectic theory. From this point of view the bulk fields are somehow more intriguing.

Most of the bulk fields we have encountered seem to have a realization in terms of symplectic fermions, by which we mean that the fermionic expressions reproduce the known correlators. A few have not been realized in this way so far, namely the height variables $h_{a\ge 2}$ not equal to 1, logarithmic partners of the height 1 field $h_1$, as well as the two fields $\rho$ and $\overline \rho$, to which they transform under $L_1$ and $\overline L_1$ respectively. 

Although we have not given any physical interpretation of $\rho$ and $\overline\rho$, they appear to be related to the derivatives of the dissipation field $\omega$ \cite{PR17},
\be
\rho = \delta \, \overline L_{-1} \omega, \qquad \overline\rho = \delta \, L_{-1} \omega ,
\ee
where $\delta$ is a constant which may depend on the lattice considered, and equal to $\delta = \frac{\pi \P_1}2$ on the square lattice. In addition, the primary field $h_1$ may be consistently identified as being proportional to the derivatives of $\rho$ and $\overline\rho$,
\be
L_{-1} \rho = \overline L_{-1} \overline\rho = \beta \lambda h_1, \qquad \beta=\frac12,
\ee
where $\lambda$ is defined from $L_0 \, h_2 = \overline L_0 \, h_2 = h_2 + \lambda h_1$ and depends on the normalizations of $h_1, h_2$. Combining these relations with the previous ones yields the somewhat surprising result that the height 1 field is proportional to the Laplacian of the dissipation field, $h_1 \sim \partial\overline\partial \omega$. The correlator (\ref{3om}) confirms this: applying $\partial_1 \overline\partial_1 \partial_2 \overline\partial_2$ on it indeed yields a multiple of $1/|z_1-z_2|^4$, itself proportional to $\langle h_1(z_1,\overline z_1)h_1(z_2,\overline z_2) \omega(\infty)\rangle$.

From these observations, it follows that all bulk fields encountered so far, namely
\be
h_{a>1},\: h_1, \: \rho, \: \overline\rho, \: \rho_\to, \: \rho_\uparrow, \: \phi_S, \: \phi_\leftrightarrow, \: \phi_\updownarrow, \: \omega, \: {\mathbb I},
\ee
belong to the same conformal representation, as they are all related to each other by the action of Virasoro modes $L_n$ or $\overline L_n$. Indeed $\rho_\to$ and $\rho_\uparrow$ are not quasi-primary and transform to a multiple of $\mathbb I$ under $L_1$ or $\overline L_1$, while $\phi_S, \, \phi_\leftrightarrow$ and $\phi_\updownarrow$ are a linear combinations of $h_1$ and the chiral and antichiral stress-energy tensors $T$ and $\overline T$. In fact in terms of fermions, all these fields, except $h_{a>1}$, are proportional to or are linear combinations of $\mathbb I$, $\theta\tilde\theta$, $\theta\partial\tilde\theta$, $\theta\overline\partial\tilde\theta$, $\partial\theta\overline\partial\tilde\theta$, $\overline\partial\theta\partial\tilde\theta$, $\partial\theta\partial\tilde\theta$ and $\overline\partial\theta\overline\partial\tilde\theta$. Clearly the main question is: do the fields $h_{a>1}$ also have a realization in terms of symplectic fermions ?

In the symplectic fermion theory, the conformal representation which contains the fields quoted above is larger because it contains many other fields, like $\theta\partial\theta$ or $\partial\theta\overline\partial\theta$, which have not yet been found in the sandpile model. Among other peculiarities, the fermionic theory also contains four logarithmic pairs $(\phi^{\alpha\beta},\psi^{\alpha\beta})$ of weight $(1,1)$, given by $\phi^{\alpha\beta} = \partial\theta^\alpha\overline\partial\theta^{\beta}$ for the primary fields, and $\psi^{\alpha\beta} = \theta\tilde\theta \, \partial\theta^\alpha\overline\partial\theta^{\beta}$ for their logarithmic partners, where $\theta^{\alpha}$ and $\theta^{\beta}$ are independently either $\theta$ or $\tilde\theta$, see \cite{GK99} for more details.

Conformal representations of the above type are called staggered modules, and have been first studied in \cite{Ro96} in their chiral version. As far as we know, it has been first noticed in \cite{GK96} for the case of modules containing rank 2 Jordan blocks, that these representations are characterized by an intrinsic complex parameter $\beta$, known as a logarithmic coupling, an indecomposability parameter or a beta-invariant. The parameter $\beta$ is crucial because it specifies the equivalence class of such representations, whose general structure was further studied in \cite{KR09} in the rank 2 case. The non-chiral staggered modules are far less understood and documented, and reflect the difficulty to formulate a consistent and local logarithmic CFT; see however \cite{DF08} and \cite{Ri12}. It is nonetheless believed that the parameter $\beta$ present in the chiral representations plays the same role of equivalence class label in the non-chiral ones, even if the latter may have more than one such label. 

Concentrating on the action of the chiral Virasoro modes, the parameter $\beta$ arises when we consider the triangular relations satisfied by a generic logarithmic pair $(\phi,\psi)$ of weights $(1,1)$ and the associated $\rho$,

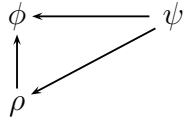
\begin{center}
\begin{pspicture}[shift=0](1,-1)(1.1,0.0)
\psset{xunit=0.8cm}
\psset{yunit=0.8cm}
\rput(0,0){$\phi$}
\rput(2.6,0){$\psi$}
\rput(0,-1.5){$\rho$}
\psline[linecolor=black,linewidth=0.7pt]{->}(2.2,0)(0.25,0)
\psline[linecolor=black,linewidth=0.7pt]{->}(2.3,-0.2)(0.23,-1.3)
\psline[linecolor=black,linewidth=0.7pt]{->}(0,-1.2)(0,-0.3)
\end{pspicture}
\end{center}

The arrows coming out of $\psi$ indicate the actions $(L_0 - 1)\psi = \lambda \phi$ and $L_1\psi = \rho$. It is important to note that if the normalization of $\psi$ is fixed, those of $\lambda\phi$ and $\rho$ are fixed as well (the value of $\lambda$ depends on the way $\phi$ is normalized). The vertical arrow indicates that $L_{-1}\rho$ is proportional to $\lambda\phi$, 
\be
L_{-1}\rho = \beta (\lambda\phi),
\label{beta}
\ee 
the proportionality factor $\beta$ being intrinsic to the representation as all normalizations have already been fixed. In addition, these relations are invariant under the change $\psi \to \psi + \alpha\phi$ because the field $\phi$ is primary ($L_1\phi=0$), and so they do not depend on which logarithmic partner is considered.

To answer the above question thus amounts to check whether the sandpile representation and the symplectic representation have the same value of $\beta$. The value of $\beta$ in the sandpile model has been given above: if the pair $(\phi,\psi)$ is chosen to be $(h_1,h_2)$, with the same normalization as the height variables on the square lattice, in which case $\lambda=-\frac12$, then one finds $\beta=\frac12$ \cite{JPR06}.

The field $h_1$ has been already identified in terms of fermions, and yields a natural choice for the primary field $\phi$ on the symplectic side,
\be
\phi_\theta = - \P_1 \big(\partial \theta \, \overline\partial \tilde \theta  + \overline \partial \theta \, \partial \tilde \theta\big).
\ee
As mentioned earlier, the lattice results in the scaling limit are consistent with $h_1$ being degenerate at level 2, namely $(L_{-1}^2 - 2 L_{-2}^{})h_1 = 0$, see Section \ref{sec5.5}. The same equation is satisfied by $\phi_\theta$.

The only candidate for the logarithmic partner of $\phi_\theta$ is proportional to $\theta \tilde\theta \, (\partial \theta \, \overline\partial \tilde \theta  + \overline \partial \theta \, \partial \tilde \theta)$ up to an irrelevant multiple of $\phi_\theta$. By computing its conformal transformations via its OPE with the chiral stress-energy tensor, one finds that the following normalization, 
\be
\psi_\theta = -\P_1 \, \theta \tilde\theta \, (\partial \theta \, \overline\partial \tilde \theta  + \overline \partial \theta \, \partial \tilde \theta),
\ee
satisfies $L_0 \psi_\theta = \psi_\theta - \frac 12 \phi_\theta$, for the same value $\lambda = -\frac12$. The same OPE reveal in addition that 
\be
\rho_\theta = L_1 \psi_\theta = -\frac{\P_1}2 \, \big(\theta \, \overline\partial \tilde \theta + \overline  \partial\theta \, \tilde\theta \big), 
\ee
from which, upon using $\partial \overline\partial \theta = \partial \overline\partial \tilde \theta = 0$, one obtains
\be
L_{-1} \rho_\theta = \partial \rho_\theta = -\frac{\P_1}2 \big(\partial \theta \, \overline\partial \tilde \theta  + \overline \partial \theta \, \partial \tilde \theta\big) = \frac 12 \phi_\theta.
\ee
Comparing with (\ref{beta}), the value of the logarithmic coupling is found to be $\beta_\theta = -1$ in the fermionic realization. As a consequence, the symplectic fermion theory cannot accomodate the height fields $h_{a>1}$, and therefore {\it does not appear to be the correct CFT to describe the scaling limit of the sandpile model}. 

As one might suspect, the value of $\beta$ has strong consequences on correlation functions involving $\psi$. A detailed comparison between $\beta = \frac 12$ versus the fermionic realization $\beta_\theta = -1$ has been made in \cite{JPR06}; it was shown in particular that the correlations with a trial field $h_2$ corresponding to a value $\beta=-1$ do not match the lattice results \footnote{As an example, the correlator $\langle h_2(z,\overline z) \rangle_{\rm mix}$ displayed in (\ref{h2mix}), and corresponding to the four-point function $\langle \phi^{\rm op,cl}(x_1) \, \phi^{\rm cl,op}(x_2) \, h_2(z,\overline z) \rangle$, can be computed upon assuming that $h_2$ is a logarithmic partner of $h_1$ carrying a generic value of $\beta \neq 0$. Its general form is given in \cite{Ru13}.}. On general grounds, this can also be understood from the fact that the value of $\beta$ determines the singular descendant of $\psi$, which, if set to zero, yields a $\beta$-dependent differential equation satisfied by any correlator containing $\psi$. In the present case, the singular logarithmic field is a combination of a descendant of $\psi$ at level 5 and a descendant of $\rho$ at level 6, with the following explicit dependence on $\beta$ \cite{KR09},
\bea
&& \hspace{-7mm} \xi = \big(L_{-1}^3 - 8 L_{-2}^{}L_{-1}^{} + 12 L_{-3}^{}\big)\big(L_{-1}^2 - 2 L_{-2}^{}\big) \psi - {\textstyle \frac 1\beta} \Big[\!-\!{\textstyle \frac{16}3}(\beta+1) L_{-2}^2 L_{-1}^2 + {\textstyle \frac 43} (14\beta+5) L_{-3}^{} L_{-2}^{} L_{-1}^{} \nonumber\\
&& \hspace{1cm} -6\beta L_{-3}^2 -6(\beta-2) L_{-4}^{} L_{-1}^2 + 8 \beta L_{-4}^{} L_{-2}^{} - {\textstyle \frac 23} (5\beta+2) L_{-5}^{} L_{-1}^{} + 4\beta L_{-6} \Big] \rho.
\eea
Using the relations $L_1 \psi = \rho, \, (L_0-1)\psi = \lambda \phi$, as well as the degeneracy condition $(L_{-1}^2 - 2 L_{-2}^{})\phi = 0$ (and the value $c=-2$), one can verify that the field $\xi$ satisfies $L_1 \xi = L_2 \xi = 0$ provided the identity $L_{-1}\rho = \beta \lambda \phi$ holds. A rather convincing confirmation for the value of $\beta=\frac 12$ in the sandpile model is therefore to check that the various correlators involving $h_2$ indeed satisfy the condition $\xi=0$ for $\beta=\frac 12$. It has been done for the correlator (\ref{h2mix}).

The situation seems therefore to be the following. The sandpile model contains a conformal logarithmic representation whose structure is very similar to the one appearing in the symplectic fermion theory, but which is nevertheless inequivalent to it. As far as the logarithmic partner $\psi$ is not brought in, the two representations look the same; this explains why some of the fields can be realized in terms of symplectic fermions. However the fermionic theory does not contain the $\beta=\frac 12$ representation found in the sandpile model, from which one concludes that it does not describe its scaling limit. 

To characterize the CFT that does describe the sandpile model, even if a Lagrangian realization of it cannot be found, remains an enormous challenge. At the moment, this looks to be an extremely ambitious question in view of the (very) small number of fields which have been successfully identified.


\end{document}